\documentclass[article]{siamart190516}

    \usepackage{amsfonts} % For AMS Beautification
    \usepackage{fullpage} % For More Realistic Page Usage
    \usepackage{float}
    \usepackage{graphicx}
    \usepackage{sidecap}
    \usepackage{physics}
    \usepackage{multirow}
    \usepackage{breakcites}

\renewcommand{\vec}[1]{\underline{\mathbf{#1}}}
\newcommand{\eps}{\varepsilon}
\newcommand*\red{}
\newcommand*\blue{}
\begin{document}
  
    \title{Neuromechanical Mechanisms of Gait Adaptation in \textit{C. elegans}: Relative Roles of Neural and Mechanical Coupling}
    
    % Insert your name below
    \author{Carter L. Johnson, Timothy J. Lewis, and Robert D. Guy}

    % Insert a specific date below if appropriate
    \date{Draft Date: \today}
    
    \maketitle
 
    \begin{center}\bf Abstract \end{center}

Understanding principles of neurolocomotion requires the synthesis of neural activity, sensory feedback, and biomechanics.
The nematode \textit{C. elegans} is an ideal model organism for studying locomotion in an integrated neuromechanical setting because its {\red neural circuit has a well-characterized modular structure and its undulatory forward swimming gait adapts to the surrounding fluid with a shorter wavelength in higher viscosity environments. This adaptive behavior emerges from the neural modules interacting through a combination of mechanical forces, neuronal coupling, and sensory feedback mechanisms. However, the relative contributions of these coupling modes to gait adaptation are not understood.}
% Local oscillations driven by proprioceptive inputs are sufficient to produce the undulatory waveform and gait adaptation.  
% However, it is not entirely understood how gait adaptation emerges from the various effects of mechanical forces, neuronal coupling, and sensory feedback mechanisms.
Here, an integrated neuromechanical model of \textit{C. elegans} forward locomotion is developed and analyzed. 
{\red The model consists of repeated neuromechanical modules that are coupled through the mechanics of the body, short-range proprioception, and gap-junctions.}
The model captures the experimentally observed gait adaptation over a wide range of {\red mechanical} parameters, provided that the muscle response to input from the nervous system is faster than the body response to changes in internal and external forces.  {\red The modularity of the model allows the use of} 
the theory of weakly coupled oscillators to identify the relative roles of body mechanics, gap-junctional coupling, and proprioceptive coupling in coordinating the undulatory gait.  
The analysis shows that the wavelength of body undulations is set by the relative strengths of these three coupling forms.  {\red In a low-viscosity fluid environment, the competition between gap-junctions and proprioception produces a long wavelength undulation, which is only achieved in the model with sufficiently strong gap-junctional coupling.}
The experimentally observed decrease in wavelength in response to increasing fluid viscosity is the result of an increase in the relative strength of mechanical coupling, which promotes a short wavelength.

%-------------------------------------------------------------------
\section{Introduction}
%-------------------------------------------------------------------

The central goal of neuroethology is to understand how an organism's body and nervous system interact with its environment to produce behaviors such as locomotion.  Model organisms have been used to study the complex interactions between the nervous system, body mechanics, and environmental dynamics in generating and coordinating locomotion \cite{doi:10.1137/S0036144504445133,Marder_1996}.  Some studies of locomotion in model organisms highlight feedforward control of locomotion, where the nervous system drives motor activity and sensory feedback plays only a modulatory role; {\red these include legged locomotion in cockroaches and swimming} in lamprey, crayfish, and leeches \cite{Cohen_1992,fuchs2011intersegmental,doi:10.1137/S0036144504445133,Mullins:2011aa,Sigvardt:1996aa,Skinner:1997aa,Skinner:1998aa,Tytell_2010,Zhang_2014}.
{\red However, other systems such as stick insect locomotion are better understood as integrated neuromechanical systems because sensory feedback is essential to coordinating movements \cite{Borgmann_2007,Ludwar_2005,Pearson_2004}.}  This sensory feedback is necessary for navigating more complex environments and can often lead to gait adaptation \cite{ayali2015comparative}.  The nematode \textit{C. elegans} is an ideal model organism for studying locomotion in an integrated neuromechanical setting 
because of its relatively simple and fully-described nervous system \cite{White:1986aa}, limited stereotypical locomotive behavior \cite{Nigon:2017aa}, dependence on sensory feedback for forward locomotion \cite{Schafer_2006,Wen:2012aa}, and undulatory gait that adapts to different fluid environments \cite{Berri_2009,Fang-Yen:2010aa,Sznitman2010}.

\textit{C. elegans} locomote forward using alternating dorsal and ventral body bends that propogate from anterior to posterior.   
 The properties of this undulatory gait adapt to fluid environments of different viscosities: higher external fluid viscosities result in slower undulations of shorter wavelengths \cite{Berri_2009,Fang-Yen:2010aa,Sznitman2010}. 
 In water, \textit{C. elegans} swim with a relatively long wavelength and relatively fast undulation frequency (roughly 1.5 bodylengths and 1.8 Hz) \cite{Fang-Yen:2010aa}.  On agar, \textit{C. elegans} crawl with a short wavelength and slow undulation frequency (0.65 bodylengths and 0.3 Hz) \cite{Fang-Yen:2010aa}. 
Previously, it was thought that these were two distinct gaits (swimming vs. crawling).  However, recent experiments have shown that
the wavelength and frequency of swimming in highly viscous fluids resemble crawling on agar surfaces \cite{Berri_2009,Fang-Yen:2010aa}, and instead of a sharp swim/crawl transition, there is a smooth transition between the two gaits as the fluid viscosity of the environment is varied \cite{Berri_2009,Fang-Yen:2010aa,Sznitman2010}.   
% How this adaptation in gait emerges from the interactions between the external environment, mechanical forces, and internal sensory feedback mechanisms is not understood.
% {\red The relative roles of the external environment, mechanical forces, and internal sensory feedback mechanisms on shaping this gait adaptation trend is not entirely understood.}

%------------ Circuitry/hypotheses of proposed mechanisms of coordination ----
There are several hypotheses for how the undulatory gait is generated and coordinated \cite{Gjorgjieva_2014}; however, it is generally agreed that proprioception plays a key role 
\cite{Boyle:2012aa,Niebur_1991,Wen:2012aa}.  One hypothesis is that there is a central pattern-generating (CPG) neural unit in the head that initiates the propagation of the bending wave --- higher fluid viscosities slow the propagation and shorten the wavelength \cite{Wen:2012aa}. Another hypothesis is that the ventral nerve cord consists of a network of neural modules that are capable of either (i) intrinsic neural oscillations \cite{Olivares_2018} or (ii) \textit{neuromechanical} oscillations (i.e., involving an entire feedback loop from neural to muscular to body mechanics and back through proprioception) \cite{Boyle:2012aa,Bryden_2008}.  
Recent experiments support the presence of multiple neural or neuromechanical oscillators \cite{Fouad:2018aa} {\red and the important role of proprioception in coordinating the oscillations \cite{Ji2020.06.22.164939}.}

% and gait adaptation has been demonstrated in computational models consisting of a chain of neuromechanical oscillators \cite{Boyle:2012aa,Denham_2018}. 

{\red Previous neuromechanical models consisting of a chain of neuromechanical oscillators (i.e., local oscillations driven by proprioceptive inputs) have been able to reproduce the undulatory waveform and gait adaptation \cite{Boyle:2012aa,Denham_2018}. 
% Fang-Yen et al. \cite{Fang-Yen:2010aa} and Sznitman et al. 2010 \cite{Sznitman2010} used a low-amplitude viscoelastic beam driven by a wave of muscle activity to describe the body mechanics. 
 % Wen et al. \cite{Wen:2012aa} used a similar mechanical model, but with a minimal neural mechanism to drive local oscillations in the body and proprioceptive input from anterior regions to propogate the undulatory wave.   In these models, gait adaptation emerges as external fluid viscosity slows down the propagation of the undulatory wave \cite{Fang-Yen:2010aa,Sznitman2010,Wen:2012aa}.
% On the other hand, 
% They assumed bistable B neurons, and use a basic repeating neural motif first introduced by Chalfie et al. \cite{chalfie1985neural} and used in neuromechanical models \cite{Bryden_2008,edos_1990,Niebur_1991,NIEBUR199351}.  
Boyle et al. \cite{Boyle:2012aa} showed that a mechanistic neuromechanical model driven by proprioception could reproduce the continuous swim-crawl transition.
Denham et al. \cite{Denham_2018} examined a similar neuromechanical model and  through computational explorations showed how the mechanical properties and neural parameters affect gait adaptation in proprioceptively-driven control.
% Denham et al. \cite{Denham_2018} examined a simpler neuromechanical model and further explored how mechanical and neural parameters contribute to the wavelength. 
% Boyle et al. \cite{Boyle:2012aa} found that increasing the range of proprioception can increase the model wavelength, and Denham et al. \cite{Denham_2018} found that lowering the neural sensitivity to stretch can also increase the wavelength.
 % However, these models only examined the interplay between body mechanics and long-range proprioception on wavelength adaptation.  
 % Short-range proprioception and gap-junctions between neurons can also account for the coordination of the undulatory waveform, as shown by a similar neuromechanical model by Izquierdo and Beer \cite{Izquierdo_2018}.  This model only looked at crawling on agar, so wavelength adaptation was not examined. 
% Different models' predictions and assumptions have been validated experimentally: the role of proprioception \cite{Wen:2012aa}, bistable B-type neurons \cite{Mellem:2008aa}, the emergence of distributed local oscillations \cite{Fouad:2018aa}, and the proprioception-induced relaxation-oscillation mechanism \cite{Ji2020.06.22.164939}. 
% A similar model by Izquierdo and Beer \cite{Izquierdo_2018} found that gap-junctional coupling and short-range proprioception could produce coordinated undulations, though gait adaptation was not examined.
However, the relative contributions to gait adaptation of each of the individual coupling modes (mechanical, gap-junctional, and proprioceptive coupling) are not well understood.}

Here, we introduce a neuromechanical model of the \textit{C. elegans} forward locomotion system {\red that consists of a chain of neuromechanical oscillators coupled by gap-junctions and nearest-neighbor proprioception}.  We use our model to systematically analyze the role of body mechanics, {\red gap-junctional} coupling, and proprioceptive coupling in gait adaptation. 
The model captures the experimentally observed {\red wavelength-trend of gait} adaptation over a wide range of parameters, provided that the muscle response to input from the nervous system is faster than the body response to changes in force.  {\red We also find that sufficiently strong gap-junctional coupling is essential to achieving the long-wavelengths in our model.}
The modular structure of our model allows the use of the theory of weakly coupled oscillators to further dissect out the mechanisms underlying gait adaptation.  Specifically, we assess the influence of each coupling modality (mechanical, {\red gap-junctional}, and proprioceptive). We find that {\red proprioceptive} coupling leads to a posteriorly-directed traveling wave with a characteristic wavelength. {\red Gap-junctional} coupling promotes synchronous activity (long wavelength), and mechanical coupling promotes a high spatial frequency (short wavelength).  The wavelength of the undulatory waveform is set by the relative strengths of these three coupling forms. As the external fluid viscosity increases, the mechanical coupling strength increases and the wavelength decreases, resulting in the observed {\red wavelength trend of gait adaptation.}

%-------------------------------------------------------------------
\section{Neuromechanical Model}\label{chapter2}
%-------------------------------------------------------------------

\begin{figure}
\centering\includegraphics[width=.9\textwidth]{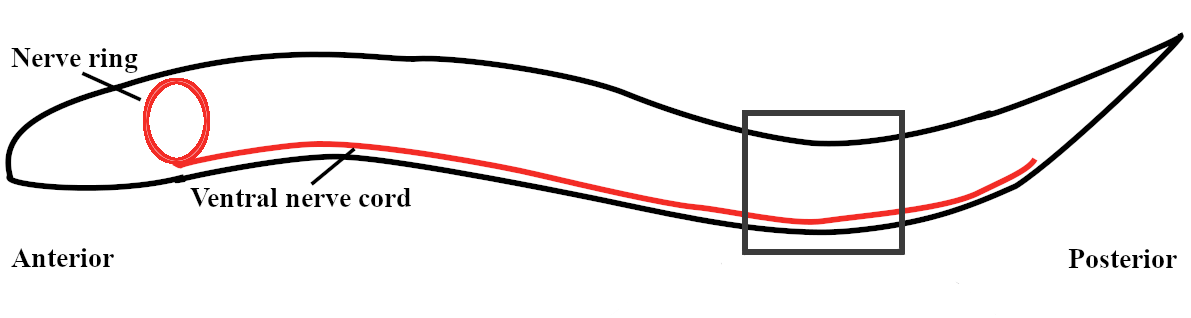}
\centering\includegraphics[width=.9\textwidth]{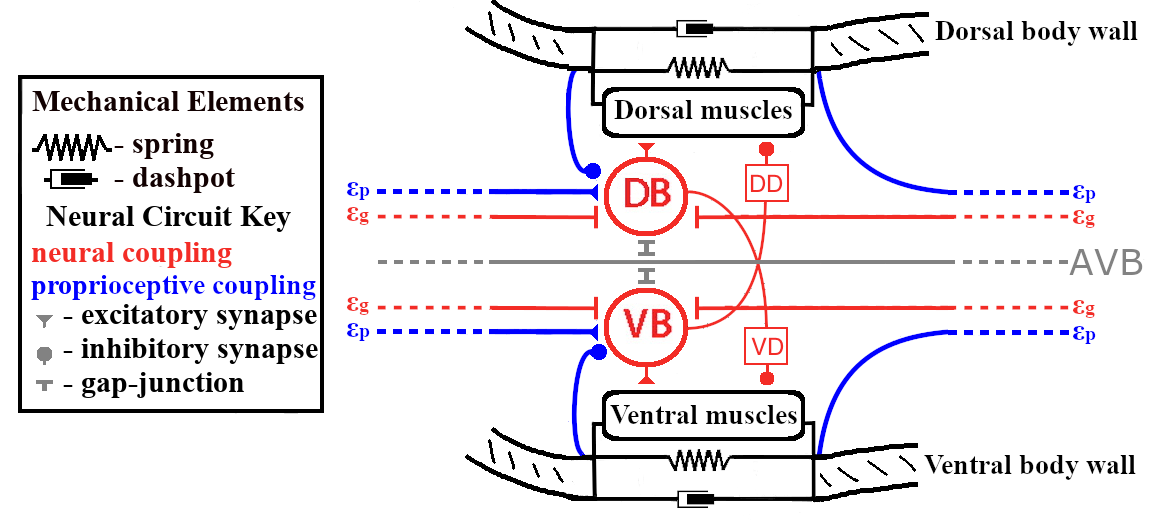}
% \centering\includegraphics[width=.9\textwidth]{wiring_diagram_module_only.png}
\caption{The highlighted schematic here depicts the repeating neuromechanical module: a 4-motorneuron functional unit consisting of DB, VB, DD, and VD-class neurons, the post-synaptic muscles, and corresponding body wall region.  The dorsal B-class (ventral B-class) neurons are excitatory and synapse onto the ipsilateral muscles and contralateral D-class neurons.  The dorsal D-class (ventral D-class) neurons are inhibitory and synapse onto the dorsal (ventral) muscles.  The B-class motorneurons also receive proprioceptive feedback from the local body segment (inhibitory) and anterior segments (excitatory).  The interneuron AVB is connected to VB and DB via gap-junctions, and the VB (DB) neurons are also coupled via gap-junctions with their nearest neighbors of the same class.  The body wall is modeled as a viscoelastic material connected to a contractile muscle.
}
\label{wiring_diag}
\end{figure}

The neuromechanical model developed here describes the motor circuit, body-wall muscles, and the resulting body shapes of \textit{C. elegans}. The body description is derived from a continuous centerline-approximation of an active viscoelastic beam, whereas the muscles and neural subcircuits are discrete in nature.  {\red The mechanical model is similar to that used in \cite{Fang-Yen:2010aa,Sznitman2010,Wen:2012aa}.}
The model for the motor circuit uses the repeated motif of Haspel and O'Donovan \cite{Haspel14611}: 6 modules of roughly 12 motor neurons and 12 muscle cells, of these 12 repeated motor neurons roughly 6 {\blue(2 ventral B-class, 2 ventral D-class, 1 dorsal B-class, and 1 dorsal D-class neurons)} are responsible for forward locomotion. {\blue In our model, we collapse the pair of VBs into a single compartment and the pair of VDs into a single compartment. We use a simplified version of Haspel and O'Donavan's connectivity scheme \cite{Haspel14611}, detailed in Section \ref{neural_module_subsubsection} and shown in Figure \ref{wiring_diag}.  Previous models present various simplified circuits \cite{Bryden_2008,Boyle:2012aa,chalfie1985neural,Denham_2018,Izquierdo_2018,NIEBUR199351}, and our model is most similar to \cite{Boyle:2012aa}, except that we omit the VD-to-VB inhibition.}
The model also includes proprioception: the B-class motor neurons respond to bending in the local and {\red nearest} anterior regions of the body \cite{Wen:2012aa,Zhen:2015aa}. 

A schematic of our neuromechanical model is shown in Figure \ref{wiring_diag}, which highlights the modular structure of the neural circuit, body-wall muscles, and the corresponding body region. Within each module, the motor subcircuit drives the body-wall muscles, which in turn apply contractile forces to bend the corresponding body region.  The body mechanics then feed back into the neural circuit through proprioceptive feedback, which translates body-wall length changes into neural signals.  This structure allows each module to function, in isolation, as a neuromechanical oscillator, and it suggests that the full body functions as a system of coupled neuromechanical oscillators.  
%{\red The coupling between two neuromechanical modules is shown in Figure \ref{wiring_diag_2modules_coupling}.}

\subsection{Model Development}\label{sect_model_dev}
    
\subsubsection{Body Mechanics} 
The nematode body is modeled as an active viscoelastic beam for small amplitude displacements submerged in {\blue a Newtonian} fluid.  \textit{C. elegans} usually operates in a regime where inertia plays a minor role (i.e., low Re), thus the equation of motion is a balance of internal elastic forces, internal viscous forces, and a fluid drag force described by a local drag coefficient $C_N$ \cite{Fang-Yen:2010aa,Sznitman2010,Wiggins:1998aa}:
\begin{align}
C_N \dot{y} &= -k_b \partial_{xx} \qty(\kappa + \dfrac{\mu_b}{k_b}\dot{\kappa}+M(x,t)),  \label{continuum_bodymechanics_PDE}
\end{align}
where $x$ is the length-wise body coordinate, $t$ is time,  $y(x,t)$ is the displacement in the ventral-dorsal plane, $\kappa(x,t)$ is the curvature, and $M(x,t)$ is the active moment that comes from internal muscle activity. The parameter $k_b$ is the bending modulus, $\mu_b$ is the effective internal viscosity, and the normal drag coefficient $C_N$ is proportional to the external fluid viscosity $\mu_f$ ($C_N = \alpha \mu_f$, see Appendix \ref{appendix_deriv_mech_params}).  The values for these parameters are given in Table \ref{table_1}, and a discussion of how they were selected is provided in Section \ref{sect_param_discuss}.

We consider small amplitude undulations, so that the curvature $\kappa(x,t)$ is approximately the second spatial derivative of the displacements $y(x,t)$:
\begin{equation}
\kappa(x,t) \approx \partial_{xx}y(x,t). \label{kappa_is_yxx}
\end{equation}
Taking two partial derivatives in $x$ of equation \ref{continuum_bodymechanics_PDE} and applying force-free, moment-free boundary conditions, the curvature $\kappa(x,t)$ of the body satisfies
\begin{align}
\alpha \mu_f \dot{\kappa} &= -k_b \partial_{xxxx} \qty(\kappa + \dfrac{\mu_b}{k_b}\dot{\kappa}+M(x,t)),  \label{continuum_bodymechanics_PDE_v2}\\
\kappa(x,t) &+ \dfrac{\mu_b}{k_b}\dot{\kappa}(x,t) +M(x,t) = 0, \  &\text{ for } x = 0, x=L, \label{cont_pde_BC1}\\
\partial_{x} &\qty(\kappa + \dfrac{\mu_b}{k_b}\dot{\kappa}+M(x,t)) = 0,\  &\text{ for } x = 0, x=L, \label{cont_pde_BC2}
\end{align}
where $x=0$ is the head and $x=L$ is the tail ($L$ is the body length). Note that in equations \ref{continuum_bodymechanics_PDE_v2}-\ref{cont_pde_BC2}, a positive curvature $\kappa(x,t)$ represents a bend towards the dorsal side.  The active moment $M(x,t)$ comes from internal muscle activity, which will be defined below. 

 \subsubsection{Muscles}
 The body is driven by six modules of roughly 6 ventral and 6 dorsal muscle cells, that apply contractile forces to either the dorsal or ventral side \cite{Haspel14611,Zhen:2015aa}. These muscle modules split the body into six distinct regions of length $\ell = L/6$.  Each ventral/dorsal muscle group applies a contractile force as a function of its activity level $A(t)$.  The ventral and dorsal ($k=V,D$) muscle activities $A_{k,j}$ in the $j^\text{th}$ module are given by
\begin{equation}
\tau_m \dot{A}_{k,j} = - A_{k,j} + I_M(k,j), \label{muscle_module_single}\\
\end{equation}
where $\tau_m$ is the timescale of muscle activation and $I_M(k,j)$ is the input from the $j^\text{th}$ neural module (described below). 
The tension $\sigma(A(t))$ generated by the muscle is only contractile ($\sigma\geq0$) and saturates at some peak force $c_{m}$:
 \begin{align}
\sigma(A(t)) &= \dfrac{c_m}{2}\qty(\tanh(c_{s}(A(t)-a_0)) + 1), \label{muscle_contract_force} \end{align}
where $c_s, a_0$ define the scale and shift of the nonlinear threshold.
In the $j^{\text{th}}$ module, the dorsal and ventral muscles apply contractile forces to opposite sides of the body, which induces a moment $m_j(t)$ on the centerline from $x_{j-1}=(j-1)\ell$ to $x_j=j\ell$:
\begin{equation}
m_j(t) = \sigma(A_{V,j}(t)) - \sigma(A_{D,j}(t)). \label{single_moment}
\end{equation}

The active moment $M(x,t)$ as a function of the body coordinate $x$ is then given by
\begin{align}
M(x,t) &=  m_j(t) \text{ for } x \in [x_{j-1}, x_j).
\label{internal_moment_full}
\end{align}

\subsubsection{Neural Module}\label{neural_module_subsubsection}
{\red The motor neurons necessary for forward locomotion are the DB (dorsal B-class) and VB (ventral B-class) \cite{Haspel14611,Zhen:2015aa}, and the neurons DD (dorsal D-class) and VD (ventral D-class) have been predicted as necessary for swimming \cite{Boyle:2012aa,DengENEURO.0241-20.2020}.}
{\red Figure \ref{wiring_diag} shows the neural module used in our model. There are 11 VB, 13 VD, 7 DB, and 6 DD neurons in the ventral nerve cord, which corresponds to about 2 VB, 2 VD, 1 DB, and 1 DD neurons in each of the six modules in our model \cite{Haspel14611}. In each module, the pair of VB neurons are connected via gap-junctions and have similar inputs and outputs, so we model each pair of VB neurons as a single entity. We assume the same for each pair of VD neurons. } The neural modules in our model are similar in structure to Boyle et al. \cite{Boyle:2012aa}{\blue, except that we omit the VD-to-VB inhibition}. Each neural module is driven by constant input from the head interneuron AVB \cite{Haspel14611,White:1986aa,Zhen:2015aa}.  
The D-class neurons are assumed to invert excitation from the B-class neurons into inhibition of the contralateral muscles.
The B-class neurons are modeled as bistable, non-spiking elements, in line with recordings of similar motor neurons involved in head-turns \cite{Mellem:2008aa}.  The activities of the ventral and dorsal ($k=V,D$) B-class neurons in the $j^{\text{th}}$ neural module are given by
\begin{align}
\tau_n \dot{V}_{k,j} &= F(V_{k,j}) + P(k,j) + I_{gj}(k,j), \label{nonlinear_bistable}
\end{align}
where 
\begin{align}
F(V_k) &= V_k - V_k^3 + I. \label{non_fun_F}
\end{align}
Here, $\tau_n$ is the timescale of neural activity, and $I$ is the offset from the constant ``on'' input from AVB.
$P(k,j)$ is proprioceptive feedback into the neuron, and $I_{gj}(k,j)$ is gap-junctional (electrical) coupling between neurons, both of which will be described below.
 
The D-class neurons are excited by the ipsilateral B-class neurons and inhibit the contralateral body-wall muscles. This effect is captured by direct inhibition of the muscles by the B-class neurons.  We model the B-class neurons as exciting the ipsilateral muscles and inhibiting the contralateral muscles. The input from the $j^\text{th}$ neural module to the ventral/dorsal muscles is given by
\begin{equation}
I_M(k,j) = \begin{cases} V_{V,j} - V_{D,j}, & \text{ if } k=V\\
V_{D,j} - V_{V,j}, & \text{ if } k=D.
\end{cases} \label{input_to_muscles}
\end{equation}

\subsubsection{Proprioceptive Feedback}
To close the neuromechanical loop, the body segment curvatures feed back into the circuit via proprioceptive processes in the VB and DB neurons.  There are two types of proprioception in this model: local (from the body region covered by the muscles of the module) and nonlocal (from neighboring anterior body regions).

Local proprioceptive feedback acts to reset the neural modules, i.e., switch between dorsal bend commands and ventral bend commands.  
Thus, \textit{local} proprioception is modeled as an excitatory current to the ventral B-class neurons in response to positive average curvature over the local module of length $\ell = L/6$, and an inhibitory current in response to negative average local curvature. The input to the dorsal B-class neurons is the same but with the polarities reversed. This feedback acts to relax the contracted muscles and contract the relaxed muscles.

Nonlocal proprioception promotes a wave of activity that propogates from anterior to posterior. The anatomical structures underlying proprioception are unknown \cite{Zhen:2015aa}, however, the evidence in Wen et al. \cite{Wen:2012aa} suggests that proprioceptive information affects the B-class motorneurons and is propagated posteriorly. {\red To be consistent with this directionality,} in our model positive nonlocal \textit{anterior} segment curvature yields a weak inhibitory current to the ventral B-class neurons and a weak excitatory current to the dorsal B-class neurons.  Negative nonlocal anterior segment curvature yields similar currents with the polarities reversed to each side. {\red 
% This is chosen to be consistent with both the directionality suggested by Wen et al. \cite{Wen:2012aa} and the directionality of the traveling wave. 
This same directionality was shown to produce locomotion in the earlier neuromechanical model of Izquierdo and Beer \cite{Izquierdo_2018}.}  This is similar to the assumptions of Boyle et al. \cite{Boyle:2012aa}, but {\blue with the opposite} directionality and sign of nonlocal proprioception.  

The proprioceptive feedback to the ventral and dorsal B-class neurons in the $j^{\text{th}}$ neural module ($j=1,\dots,6$) of length $\ell = L/6$ is modeled by
\begin{align}
P(V,j) &= + c_p \dfrac{1}{\ell}\int_{(j-1)\ell}^{j \ell}\kappa(x,t) \dd x - \eps_p \dfrac{1}{\ell} \int_{(j-2)\ell}^{(j-1) \ell}\kappa(x,t) \dd x, \label{prop_feedback_V}\\
P(D,j) &= - c_p \dfrac{1}{\ell} \int_{(j-1)\ell}^{j \ell} \kappa(x,t) \dd x + \eps_p \dfrac{1}{\ell} \int_{(j-2)\ell}^{(j-1) \ell} \kappa(x,t) \dd x,  \label{prop_feedback_D}
\end{align}
where $c_p$ is the strength of \textit{local} proprioception, $\eps_p$ is the strength of \textit{nonlocal} anterior proprioception, and $\kappa(x,t) = 0$ for $x\notin[0,L]$ for notational simplicity.

\subsubsection{Gap-Junctional Coupling} The B-class neurons are also connected via gap-junction synapses to neighboring B neurons of the same type (ventral/dorsal) \cite{Haspel14611,White:1986aa,Zhen:2015aa}. The gap-junctions are modeled as symmetric ohmic resistors with constant conductance, so that the gap-junctional coupling to the ventral and dorsal ($k=V,D$) B-class neurons in the $j^{\text{th}}$ neural module are described by

\begin{align}
I_{gj}(k,j) = \eps_g (V_{k,j-1}-V_{k,j}) + \eps_g(V_{k,j+1}-V_{k,j}), \label{gap_junction_coup}
\end{align}
where $\eps_g$ is the strength of gap-junction coupling and $V_{k,0} = V_{k,7} = 0$ for notational simplicity.

%-------------------------------------------------------------------
\subsection{Model Discretization for Numerical Simulation}\label{sect_discret_and_sim}

To simulate the model described in Section \ref{sect_model_dev}, the body is discretized into six modules in correspondence with the six neuromuscular modules, so that there are six discrete body segment curvatures.
The 4th-order difference operator $D_4$ is used to approximate the 4th spatial derivative with zero-force, zero-torque boundary conditions:
\begin{equation}\dfrac{1}{\ell^4}D_4 = \dfrac{1}{\ell^4}\qty(\begin{matrix}
7 &   -4  &   1  &      &    &   \\
    -4  &   6 &   -4 &    1  &     &   \\
     1  &  -4  &   6  &  -4   &  1   &   \\
  &   1   & -4 &    6  &  -4  &   1\\
& &    1 &   -4   &  6 &   -4 \\
       &      &    &   1   & -4   &  7 \\
\end{matrix}).\label{4th_diff_op}\end{equation}
Discretizing equations \ref{continuum_bodymechanics_PDE_v2}-\ref{cont_pde_BC2} and using \ref{single_moment}-\ref{internal_moment_full} yields a linear differential equation for the vector of body segment curvatures $\vec{\kappa}$:
\begin{equation}
\qty(\alpha \mu_f I_6 + \dfrac{\mu_b}{\ell^4} D_4) \dot{\vec{\kappa}} = -\dfrac{k_b}{\ell^4} D_4 \qty(\vec{\kappa} + \sigma(\vec{A}_V) - \sigma(\vec{A}_D)), \label{full_kappa}
\end{equation}
where $I_6$ is the $6\times6$ identity matrix.

In this discretization, the neural and muscle activity dynamics of all the modules are given by
\begin{align}
\tau_m \dot{\vec{A}}_V &= - \vec{A}_V + \vec{V}_V - \vec{V}_D, \label{full_AV}\\
\tau_m \dot{\vec{A}}_D &= - \vec{A}_D + \vec{V}_D - \vec{V}_V, \label{full_AD} \\
\tau_n \dot{\vec{V}}_V &= F(\vec{V}_V) + c_{p}\vec{\kappa} -  \eps_p W_p \vec{\kappa} + \eps_g W_g \vec{V}_V, \label{full_VV}\\
\tau_n \dot{\vec{V}}_D &= F(\vec{V}_D) - c_{p}\vec{\kappa} + \eps_p W_p \vec{\kappa} + \eps_g W_g \vec{V}_V, \label{full_VD}
\end{align}
where each vector entry (e.g., $A_{V,j}$) is the corresponding activity of the $j^\text{th}$ module.
In equations \ref{full_VV} and \ref{full_VD}, $W_p$ is the nonlocal proprioceptive connectivity matrix (equation \ref{w_p_matrix}), which comes from discretizing equation \ref{prop_feedback_V}, and $W_g$ is the gap-junction connectivity matrix (equation \ref{w_g_matrix}), which comes from discretizing equation \ref{gap_junction_coup}:

\noindent\begin{minipage}{0.5\textwidth}
\begin{equation}
W_p = \qty(\begin{matrix}
0 &  &  & &  \\
1 & 0 & &  & \\
& & \ddots & \ddots &  \\
& & & 1 & 0
\end{matrix})\label{w_p_matrix},\end{equation}\end{minipage}
\begin{minipage}{0.4\textwidth}
\begin{equation} W_g = \qty(\begin{matrix}
-1 & 1 & &   \\
1 & -2 & 1    & \\
& \ddots & \ddots & \ddots   \\
 & & 1 & -1  \\
\end{matrix}).\label{w_g_matrix}\end{equation}\end{minipage}  \\

\noindent  A numerical solution to the system of differential equations \ref{full_kappa}-\ref{full_VD} is generated using the ode23 method in MATLAB.

%-------------------------------------------------------------------
\subsection{Parameter Discussion}\label{sect_param_discuss}

Some parameters in the model are well-constrained by experimental data, while others are not.
Quantities that are directly measurable include the body length $L = 1$ mm, average body radius $R=40$ $\mu$m, cuticle width $r_{c} = 0.5$ $\mu$m, and wavelength $\lambda/L$ and frequency $f$ in fluids of various viscosities $\mu_f$.   The timescales in the system are less certain. The range 50-200 ms is used for the muscle activation timescale $\tau_m$, which is the range of measurements of peak muscle force generation time in Milligan et al. (1997) \cite{Milligan2425}. As with previous models \cite{Boyle:2012aa,Denham_2018,Izquierdo_2018}, the neural activity is chosen to be the fastest process in the model, but while Boyle et al. \cite{Boyle:2012aa} considered the B-neurons as instantaneous switches, here the neural activity timescale is set at $\tau_n = 10$ ms.

The internal viscosity $\mu_b$ and Young's modulus $E$ have been estimated across several orders of magnitude \cite{Backholm_2013,Fang-Yen:2010aa,Sznitman2010}, so caution is exercised in using one set of parameters from one source over another.  Of more importance in the model is the mechanical timescale 
\begin{equation}
\tau_b = \dfrac{\mu_b}{k_b}, \label{mech_timescale}
\end{equation} 
which is the timescale of relaxation in an inviscid fluid. In equation \ref{mech_timescale}, $k_b$ is the bending modulus, which is derived from the Young's modulus $E$ and the geometry of the cuticle in Appendix \ref{appendix_deriv_mech_params} following previous modeling procedures \cite{Cohen_2014,Sznitman2010}.   {\red In Appendix \ref{appendix_deriv_mech_params}, we also compare the range of mechanical parameters used here with previous models.}
Given the range of mechanical parameters reported in the literature, the mechanical timescale could be as small as $\tau_b = 1$ ms or as large as $\tau_b = 5$ s.  The role of this timescale is explored in Section \ref{sect_timescales_param_study}.  

The electrophysiological details of the internal neural circuit are largely unknown, thus all the feedback and coupling strengths $c_p, c_m, \varepsilon_p, \varepsilon_g$, the parameters of the nonlinear functions $F(V)$ and $\sigma(A)$ are not well constrained.   The feedback strengths $c_m=10$, $c_p=1$ and parameters of the nonlinear functions $F(V)$ ($a=1, I = 0$) and $\sigma(A)$ ($c_s = 1, a_0 = 2$) were chosen so that the neuromechanical oscillator robustly gives the correct frequency ($\sim 1.76 Hz$) in a low-viscosity environment. The values for the coupling parameters $\varepsilon_p$ and $\varepsilon_g$, on the other hand, are explored in the next section.

\begin{table}
\centering
\caption{Range of parameters explored and sources. See Section \ref{sect_param_discuss} for more details and Appendix \ref{appendix_deriv_mech_params} for derivations.}
\begin{tabular}{ |c|c|c|c| } 
\hline
Parameter & Name & Range of values & References\\
\hline
$L$ & Body length & 1 mm & \cite{White:1986aa}\\
\hline
$R$ & Average body radius & 40 $\mu$m & \cite{cohen_ranner_2017} \\
\hline
$r_{\text{cuticle}}$ & Cuticle width & 0.5 $\mu$m & \cite{cohen_ranner_2017}\\
\hline
E & Young's modulus & $3.77$ kPa - $1.3\times10^{4}$ kPa & \cite{Backholm_2013,Fang-Yen:2010aa,Sznitman2010}\\
\hline
$I_c$ & Second moment of area of cuticle & $2.0 \times 10^{-7} (\text{mm})^4$ & \cite{cohen_ranner_2017}\\
\hline $k_b$ &  Bending modulus &  $7.53 \times 10^{-10} - 2.6\times 10^{-6}$ N$\cdot$(mm)$^2$ & \cite{Backholm_2013,Fang-Yen:2010aa,Sznitman2010}\\ 
\hline
$\mu_b$ & Body viscosity & $2 \times 10^{-11} - 1.3\times10^{-7}$ N(mm)$^2$s & \cite{Backholm_2013,Fang-Yen:2010aa,Sznitman2010}\\ 
\hline
$\mu_f$ & Fluid viscosity & $1 - 2.8 \times 10^4$ mPa$\cdot$s & \cite{Fang-Yen:2010aa}\\
\hline
$C_N$ & Normal drag coefficient & $3.4 \mu_f$ & \cite{cohen_ranner_2017,Fang-Yen:2010aa} \\
\hline
$\tau_b$ & Mechanical timescale $\tau_b = \mu_b/k_b$& 1 ms - 5 s & \cite{Backholm_2013,Fang-Yen:2010aa,Sznitman2010}\\
\hline
$\tau_m$ & Muscle activation timescale & 50-200 ms & \cite{Milligan2425}\\
\hline
\end{tabular}
\label{table_1}
\end{table}

%-------------------------------------------------------------------
\section{Model Results}\label{sect_results}

\textit{C. elegans} locomote forward using alternating dorsal and ventral body bends that propogate in the form of a traveling wave from anterior to posterior.   The spatial wavelength of this traveling wave changes in response to changes in the fluid viscosity \cite{Berri_2009,Fang-Yen:2010aa,Sznitman2010}.  In this section, we show that our model captures this gait adaptation for a wide range of mechanical and neural parameters,  provided that the muscle response to input from the nervous system is faster than the body response to changes in internal and external forces.  {\red We also show that sufficiently strong gap-junctional coupling is essential to achieving the long-wavelengths in our model.}

\subsection{Model Captures Gait Adaptation}\label{sect_capt_gait_Adapt}

\begin{figure}
\centering\includegraphics[width=.7\textwidth]{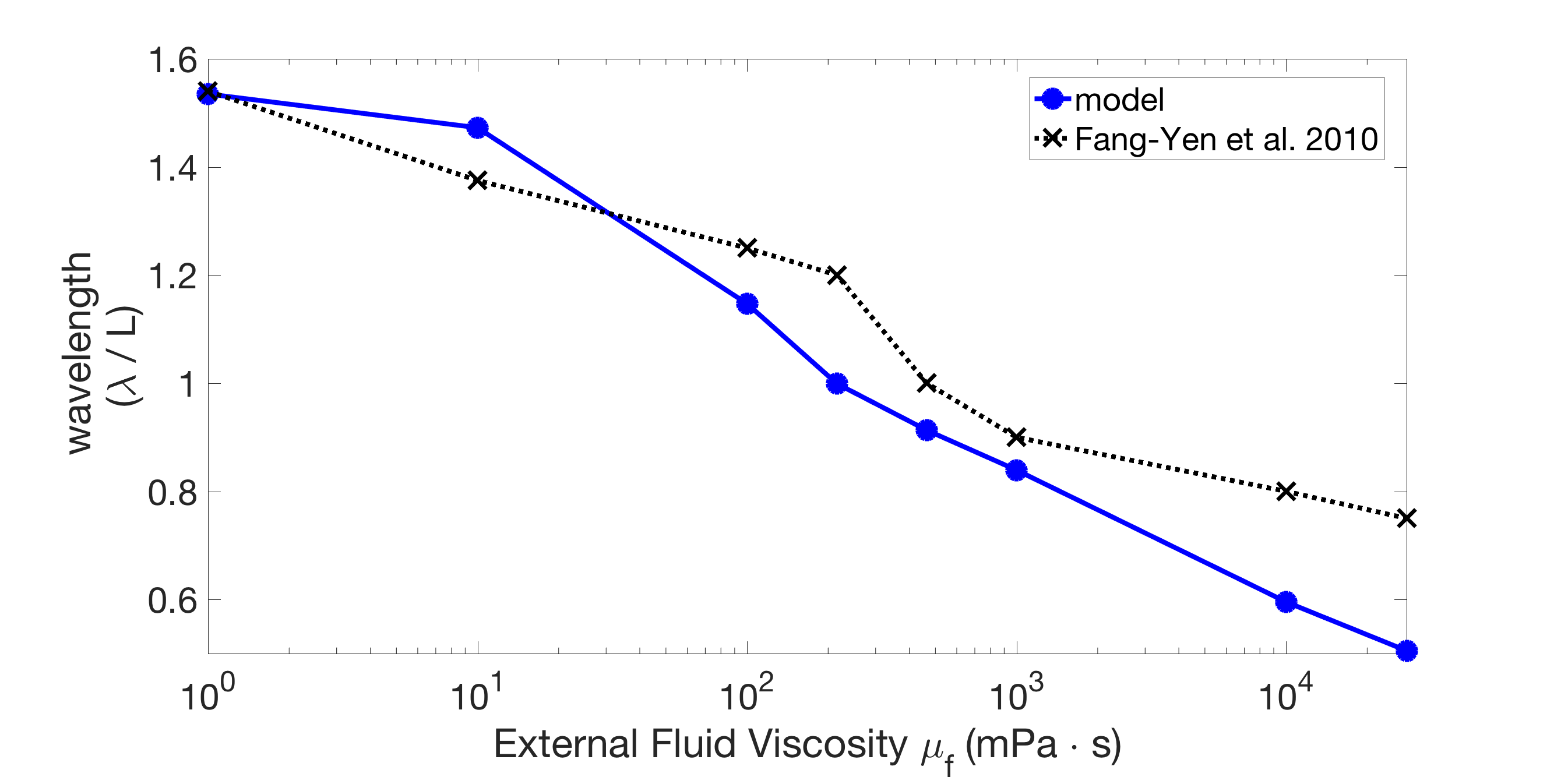}
\includegraphics[width=.15\textwidth]{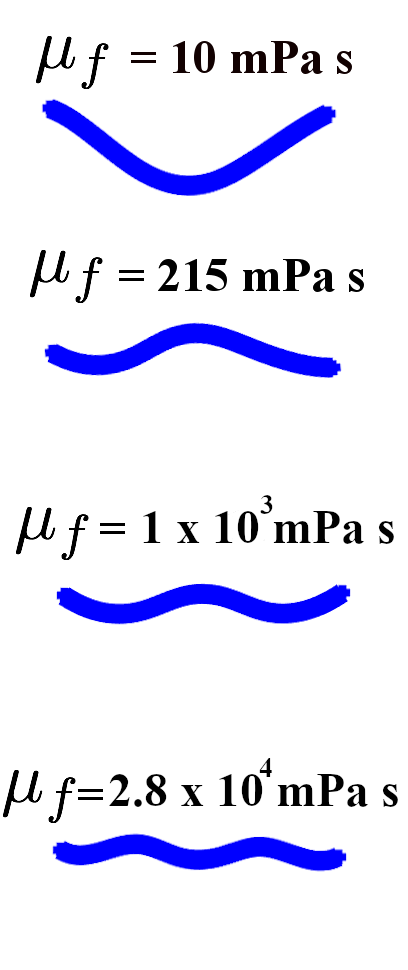}
\caption{The model captures the quantitative trend of gait modulation seen in experiments such as \cite{Fang-Yen:2010aa}. Here, $\tau_b = 0.5$ s, $\tau_m = 0.1$ s, and $\mu_b = 1.3\times 10^{-7}$ N(mm)$^2$s. In water ($\mu_f = 1$ mPa s) the wavelength is roughly 1.5 bodylengths, and increasing the fluid viscosity $\mu_f$ smoothly reduces the wavelength down to roughly 0.75 bodylengths in the most viscous case ($\mu_f = 28 $ Pa s).  
}
\label{result1_lambdavgamma_fullonly}
\end{figure}

We fit the model to match the wavelength and frequency in water, and then ran it in different fluid environments. Our model captures the quantitative effect of external fluid viscosity on the body wavelength seen in experiments and previous models. Figure \ref{result1_lambdavgamma_fullonly} shows an example of the wavelength trend of the model for fixed body parameters $\tau_b = 500$ ms, $\tau_m = 50 $ms (the wavelengths were computed from the model output as described in Appendix \ref{appendix_defining_wvln}). Figure \ref{result1_lambdavgamma_fullonly} also shows that the model wavelengths are in close quantitative agreement with the experimentally-measured wavelengths of Fang-Yen et al. \cite{Fang-Yen:2010aa}.
In water ($\mu_f = 1$ mPa$\cdot$s) the wavelength is roughly 1.5 bodylengths, and increasing the fluid viscosity $\mu_f$ smoothly reduces the wavelength down to roughly 0.75 bodylengths in the most viscous case ($\mu_f = 2.8\times 10^4$ mPa$\cdot$s).
This wavelength trend is similar to what has been observed in other experiments \cite{Berri_2009,Sznitman2010}, and in Section \ref{sect_timescales_param_study},  we show that our model captures this trend robustly over a wide range of parameters.

The undulation frequency also changes in response to changes in the fluid viscosity \cite{Berri_2009,Fang-Yen:2010aa,Sznitman2010}.  In Fang-Yen et al. \cite{Fang-Yen:2010aa}, the frequency decreases from 1.7 Hz to 0.30 Hz as fluid viscosity increases from 1 mPa s to $2.8 \times 10^4$ mPa s.  Our model also exhibits a decrease in frequency as fluid viscosity $\mu_f$ increases, but not of the same magnitude (1.7 Hz - 1.6 Hz).  Discussion of this discrepancy is given in Section \ref{sect_discussion}.

\subsection{Parameter Study Highlights Importance of Timescale Ordering in Capturing Gait Adaptation}\label{sect_timescales_param_study}

\begin{figure}
\centering\includegraphics[width=\textwidth]{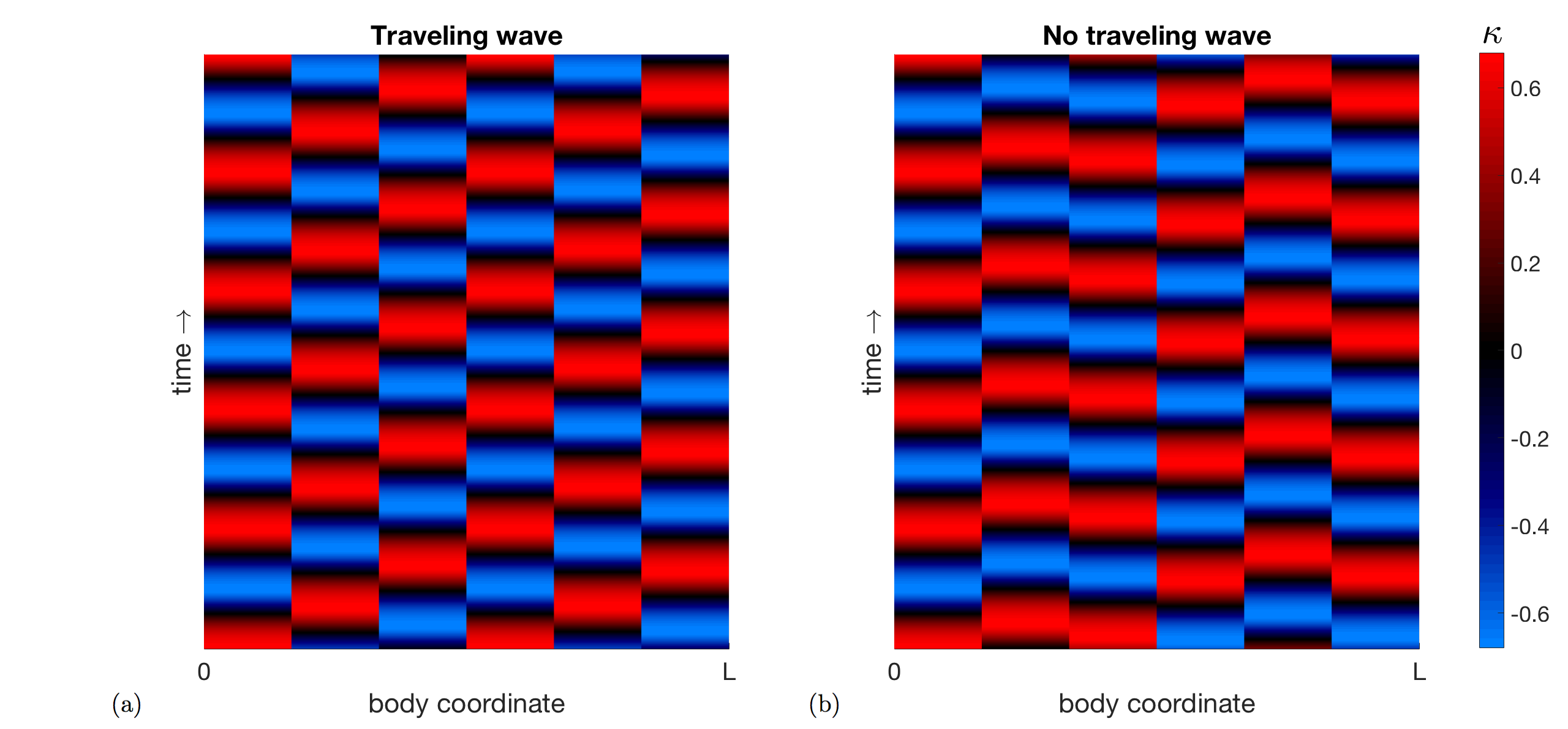}
\caption{Sample model curvature kymographs (curvature vs. time) for various parameter regimes. For some parameter regimes, the gait adaptation trend generally held and there was a traveling wave at all $\mu_f$ values; (a) gives an example of this behavior for $\tau_b =$ 0.51 s, $\mu_b = 1.3 \times 10^{-7}$ N(mm$^2$) s, and $\mu_f$ = 28 Pa s.  For other parameter regimes, high enough external fluid viscosity $\mu_f$ resulted in a loss of the traveling waveform; (b) gives an example of this behavior for $\tau_b =$ 0.51 s, $\mu_b =  1.5 \times 10^{-9}$ N(mm$^2$) s, and $\mu_f$ = 28 Pa s. }
\label{sample_kymos}
\end{figure}

We performed a parameter study to show that the model robustly captures gait adaptation {\red as the fluid viscosity $\mu_f$ is varied.  
The mechanical parameters $\tau_b$, $\mu_b$, the proprioceptive coupling strength $\eps_p$, and the muscle timescale $\tau_m$ were varied, while the other parameters of the model were held fixed, including the gap-junctional coupling strength $\eps_g = 0.0134$. (For more extensive parameter explorations, see \cite{johnson2020neuromechanical}.)}
%The model worm locomotes forward via a posteriorly-directed traveling wave of body deformations.  
For some parameter regimes, the body deformations were traveling waves for all fluid viscosities $\mu_f$, but this was not the case for other parameter regimes.  Figure \ref{sample_kymos} shows kymographs of the body curvature that demonstrate two typical cases exhibited by the model.
% one a traveling wave (a), and the other not (b).
% {\red as the fluid viscosity $\mu_f$ is varied.  All the parameters of the model were examined in \cite{johnson2020neuromechanical}, and we found that the key parameters that effect gait adaptation are the mechanical parameters $\tau_b$, $\mu_b$, the coupling strengths $\eps_p$, $\eps_g$, and the muscle timescale $\tau_m$. In this parameter study, we vary only four parameters ($\tau_b$, $\mu_b$, $\eps_p,$ and $\tau_m$), while the other parameters of the model were held fixed. In \cite{johnson2020neuromechanical}, we also found that increasing the gap-junctional coupling strength $\eps_g$ shifts the wavelength adaptation to higher viscosities $\mu_f$ in \cite{johnson2020neuromechanical}, but in this parameter study we fix it at $\eps_g = 0.0134$. }

\begin{figure}
\centering\includegraphics[width=.5\textwidth]{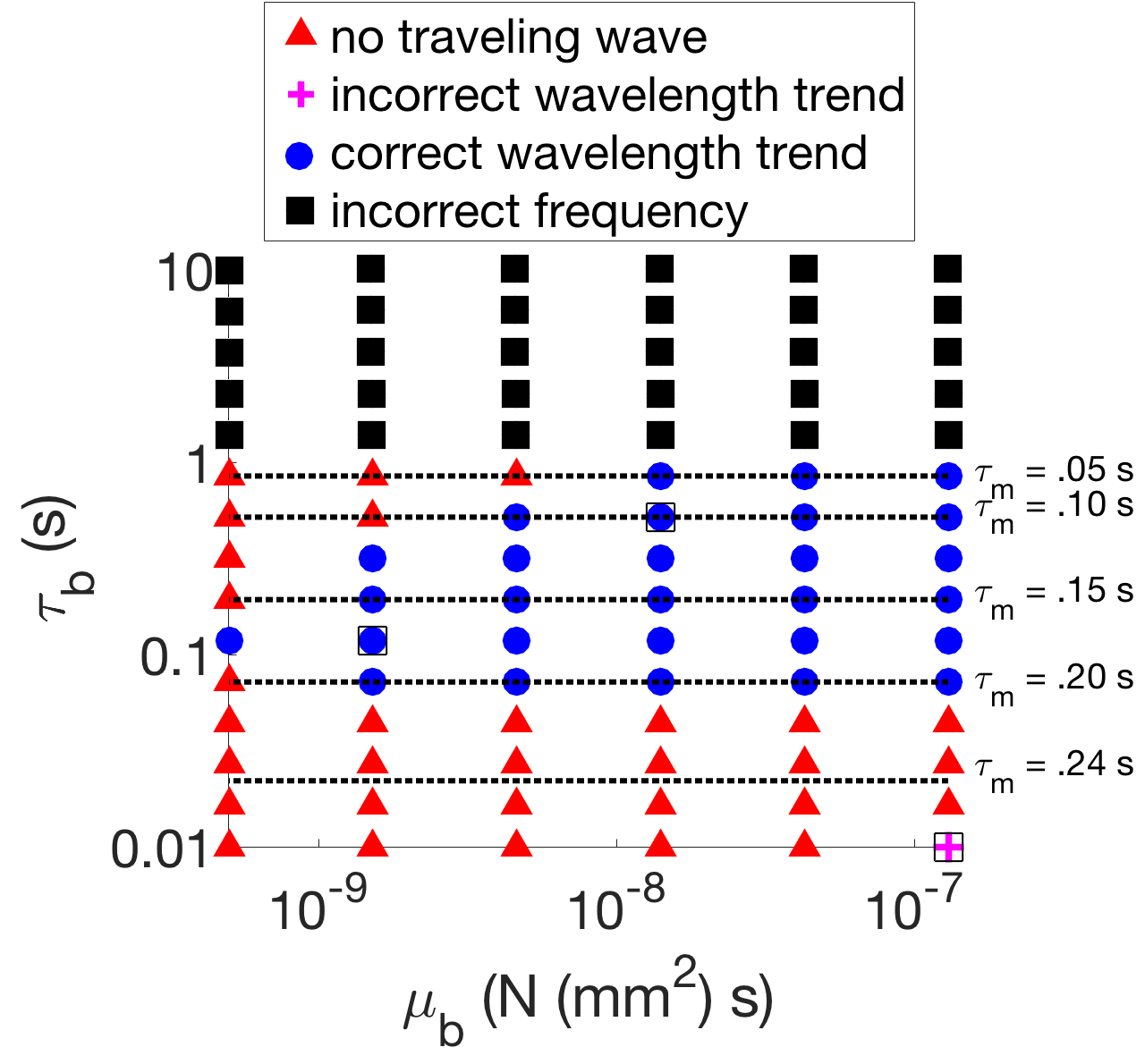}
\caption{Classification of the model behavior for different mechanical parameters $\mu_b$ and $\tau_b$. For each parameter combination $(\mu_b, \tau_b)$, the muscle timescale $\tau_m$ was fit to match the undulation frequency in water ($\tau_m$ contours shown in black dashes).  Boxed markers indicate parameter combinations which have the wavelength trend illustrated in Figure \ref{sample_runs_vs_gamma}.}
\label{parameter_Sweep_wvln_trends}
\end{figure}

The model parameters {\red$\tau_b$, $\mu_b$, $\eps_p,$ and $\tau_m$} were systematically varied to characterize the model behavior.  For a given body timescale $\tau_b$ and body viscosity $\mu_b$, the muscle activity timescale $\tau_m$ was selected in the range 50-250 ms to match the undulation frequency (1.7 Hz) in water ($\mu_f = 1 $ mPa s) {\red within 1\%}.  Next, the proprioceptive coupling strength $\varepsilon_p$ was selected to match the wavelength (1.5 bodylengths) in water {\red within 1\%}.
%, and the gap-junctional coupling strength was fixed at $\eps_g = 0.0134$. 
{\red We are able to separate these effects and perform these one-parameter searches due to the weakly coupled nature of our model.}
The model was then run in different fluid viscosity $\mu_f$ environments and the emergent coordination trend is reported in Figure \ref{parameter_Sweep_wvln_trends}.  The model behavior was classified {\red exclusively} as either: (1) not a traveling wave for all fluid environments, (2) incorrect wavelength trend, (3) qualitatively correct wavelength trend, or (4) incorrect frequency in water.

\begin{figure}
\centering(a)\includegraphics[width=0.3\textwidth]{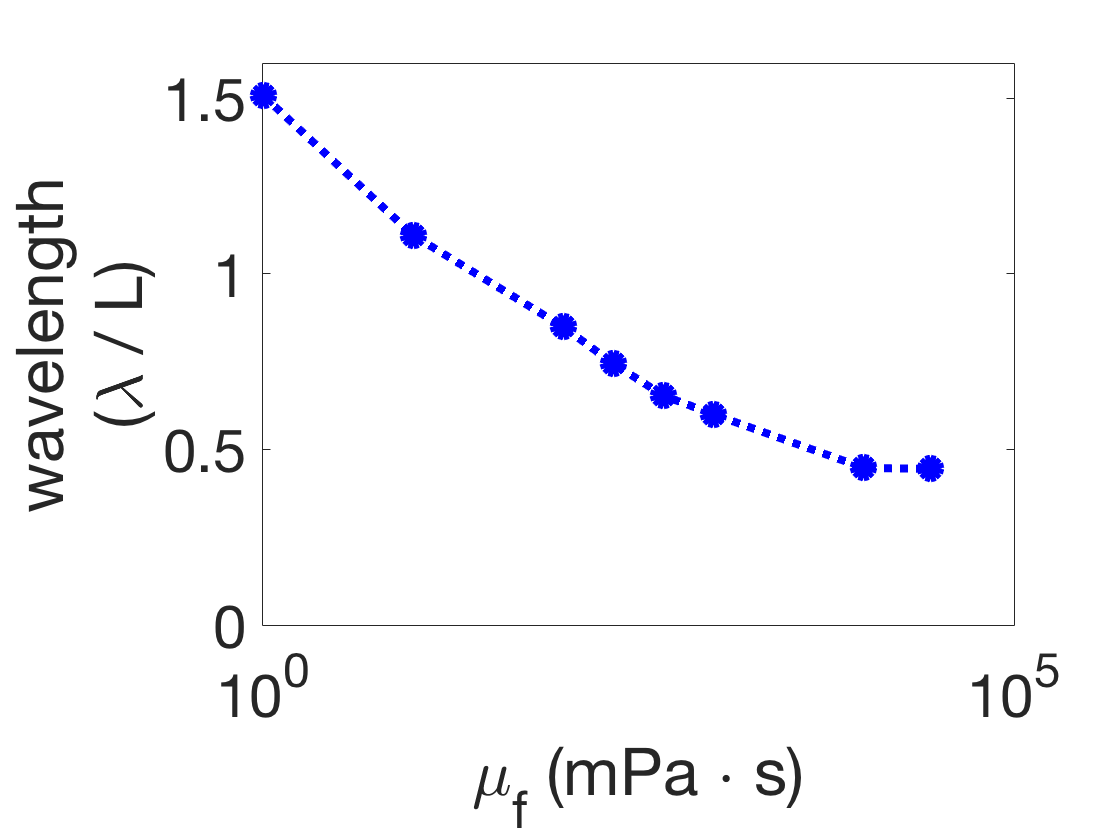}
(b)\includegraphics[width=0.3\textwidth]{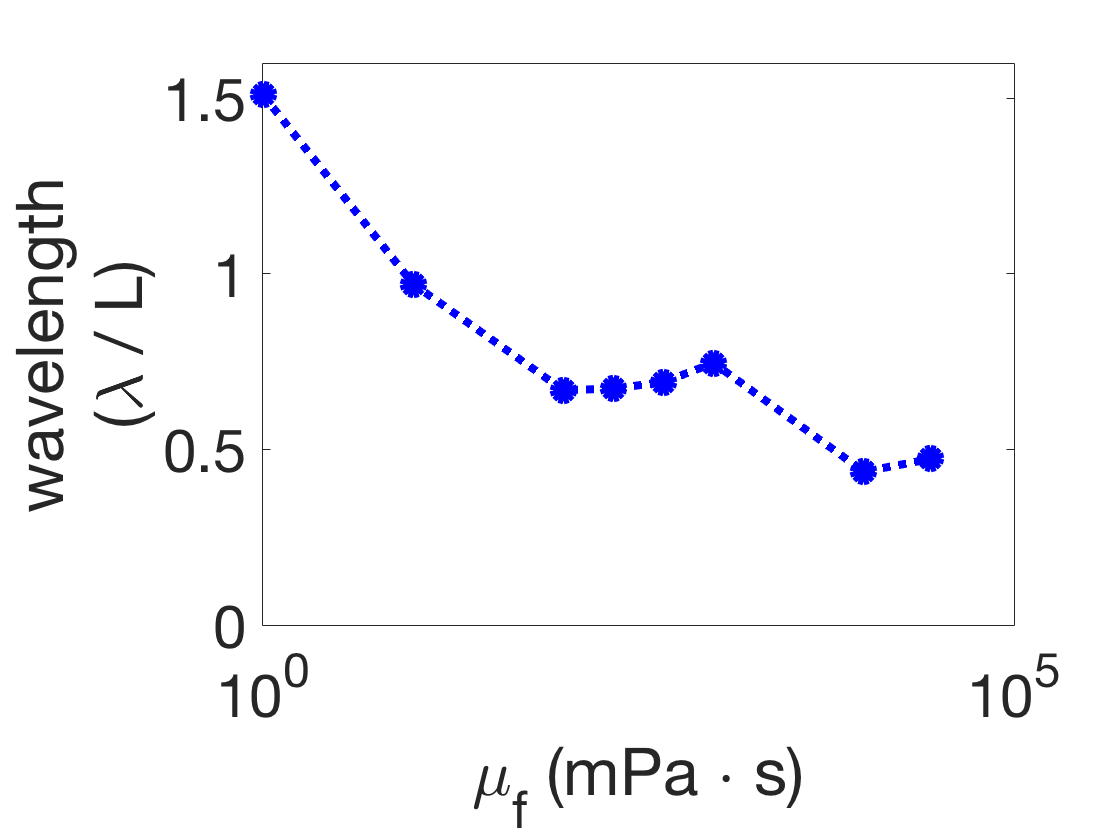}
(c)\includegraphics[width=0.3\textwidth]{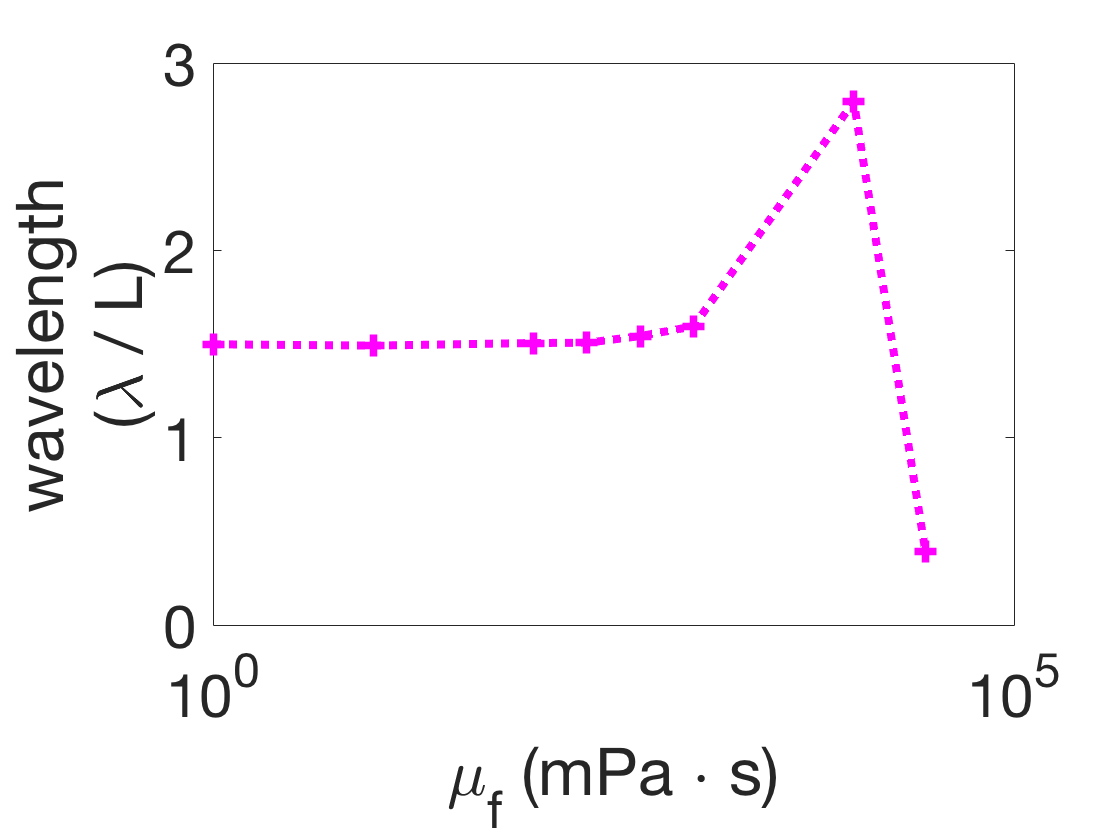}
\caption{Model wavelength vs. external fluid viscosity $\mu_f$ for various parameter regimes (the boxed markers in Figure \ref{parameter_Sweep_wvln_trends}).
(a) and (b) show examples of the qualitatively correct wavelength trend, while (c) shows an incorrect trend.  (a) has $\tau_b =$ 0.51 s , $\mu_b =1.4\times 10^{-8}$ N(mm$^2$)s, (b) has $\tau_b =$ 0.12 s , $\mu_b =1.5\times 10^{-9}$ N(mm$^2$)s, and (c) has $\tau_b =$ 0.01 s , $\mu_b =1.3\times 10^{-7}$ N(mm$^2$)s.  } 
\label{sample_runs_vs_gamma}
\end{figure}

There is no traveling wave (red triangles) if, for any viscosity $\mu_f$, the difference between the minimum and maximum pairwise-phase difference is greater than or equal to 0.5, because this indicates that there is no consistent directionality to the phase differences in the body.  A range of observed wavelength trends in various parameter regimes (the boxed markers in Figure \ref{parameter_Sweep_wvln_trends}) are illustrated in Figure \ref{sample_runs_vs_gamma}. Figure \ref{sample_runs_vs_gamma}(a) and (b) show examples of the qualitatively correct wavelength trend (blue circles), while (c) shows the the incorrect trend, which was only obtained at a single parameter combination. The wavelength trend is incorrect because the wavelengths increased dramatically as the fluid viscosity increased, as opposed to generally decreasing.

A few key observations can be made from Figure \ref{parameter_Sweep_wvln_trends}.  First, if the mechanical timescale $\tau_b$ is too large, then the frequency in water cannot be obtained (see the black squares in Figure \ref{parameter_Sweep_wvln_trends}).
  Second, if the mechanical timescale $\tau_b$ is too small, then there will not be a traveling wave for all fluid viscosities $\mu_f$.
  This suggests that while the body stiffness $k_b$ and body viscosity $\mu_b$ have been estimated across several orders of magnitude in various experiments and models, the effective mechanical body timescale $\tau_b = \mu_b/k_b$ lies within the relatively narrow range $0.07-1$ s.
  
In order to match the frequency, the muscle timescale $\tau_m$ must be inversely related to $\tau_b$.  When the body timescale $\tau_b$ is increased, the muscle timescale $\tau_m$ must decrease to compensate.  The frequency in water cannot be obtained for $\tau_b$ too large since it would require decreasing the muscle activity timescale $\tau_m$ below physiological limits.  Similarly, when the body timescale $\tau_b$ is decreased, the muscle timescale $\tau_m$ must be increased to compensate for the frequency. For $\tau_b$ too small, there is not a traveling wave for all fluid viscosities $\mu_f$; this occurs soon after $\tau_b < \tau_m$.  
% In Figure \ref{parameter_Sweep_wvln_trends}, this can be seen in the large red no-traveling-wave region corresponding to $\tau_b = 0.07$ s and $\tau_m = 0.20$ s.
This suggests that the relative ordering of the timescales $\tau_b, \tau_m, \tau_n$ is key to the coordination. Generally, the mechanical timescale $\tau_b$ must be the largest, the muscle activity timescale $\tau_m$ intermediate, and the neural timescale $\tau_n$ the shortest.  The mechanism by which this timescale ordering affects coordination is explained in Section \ref{sect_two_osc_analysis}.

Remarkably, whenever there is a traveling wave in this systematic parameter search, it almost always has the qualitatively correct wavelength trend.  This wavelength trend is consistent with gait adaptation across several orders of magnitude of the mechanical parameters.
 % {\red Furthermore, we found in Section \ref{section_role_of_Gapjns} and \cite{johnson2020neuromechanical} that increasing the gap-junctional coupling strength $\eps_g$ does not change this trend, but instead merely shifts the wavelength adaptation to higher viscosities $\mu_f$ (see Figure \ref{role_of_gapjns_figure}(a).}

{\red\subsection{Gap-Junctions are Necessary for Long Wavelengths in the Model}\label{section_role_of_Gapjns}

To determine the role of gap-junctions in our neuromechanical model, we investigate how changing the gap-junctional coupling strength affects the wavelength trend and overall coordination.  
Other models have achieved accurate wavelength adaptation without gap-junctional coupling, albeit with proprioceptive ranges longer than nearest-neighbor \cite{Boyle:2012aa,Denham_2018}, so we assess the degree to which gap-junctional coupling is necessary in our model.  
We find that in our model, sufficiently strong gap-junctional coupling is necessary to obtain the long wavelengths in low-viscosity environments (e.g. water).

We vary the gap-junctional coupling strength in our model over several orders of magnitude ($\eps_g =0$ and $\eps_g \in [10^{-4},10^{-1}]$). We fix the parameters $\tau_b=0.5$ s, $\tau_m=0.1$ s, and $\mu_b = 1.3 \times 10^{-7}$ N$(mm)^2$s and find the proprioceptive coupling strength $\eps_p$ to match the wavelength in water (approximately 1.5 bodylengths). 
Figure \ref{role_of_gapjns_figure}(a) shows the resulting wavelength trend of the model for different gap-junctional coupling strengths $\eps_g$. For sufficiently strong gap-junctional coupling strengths ($\eps_g \geq 0.005$), the wavelength in water was matched and the full range of wavelength adaptation was found (circle-markers in Figure \ref{role_of_gapjns_figure}(a)).  For weaker gap-junctional coupling strengths ($\eps_g < 0.005$), the wavelength in water was not matched (plus-markers, Figure \ref{role_of_gapjns_figure}(a)). We further show that the model with weak gap-junctional coupling strength ($\eps_g < 0.005$) cannot match the wavelength in water for any proprioceptive coupling strength. Figure \ref{role_of_gapjns_figure}(b) shows the model wavelengths in water as a function of proprioceptive coupling strength $\eps_p$ for different gap-junctional coupling strengths $\eps_g$ and reiterates that the model cannot match the long wavelength without sufficiently strong gap-junctional coupling. 
Our neuromechanical model thus relies on gap-junctional coupling between neural modules to achieve the long wavelengths in low-viscosity environments.  

% Other models achieve this via longer-ranged proprioception \cite{Boyle:2012aa,Denham_2018}, however in the absence of gap-junctional coupling in our model ($\eps_g = 0$), short-range (nearest-neighbor) anterior proprioception is not sufficient.

% At larger gap-junctional coupling strenths $\eps_g$, the steep transition from long wavelengths (1.5 bodylengths) to shorter wavelengths occurs at higher fluid viscosities $\mu_f$. The experimentally-observed wavelength trend of Fang-Yen et al. \cite{Fang-Yen:2010aa} (shown in X’s and black-dotted lines in Figure \ref{role_of_gapjns_figure}(a)) is more closely matched by $\eps_g \in [0.01, 0.05]$. In the previous parameter study, we used $\eps_g = 0.0134$, which is near this range of gap-junctional coupling strengths. 

\begin{figure}
\centering(a)\includegraphics[width=0.75\textwidth]{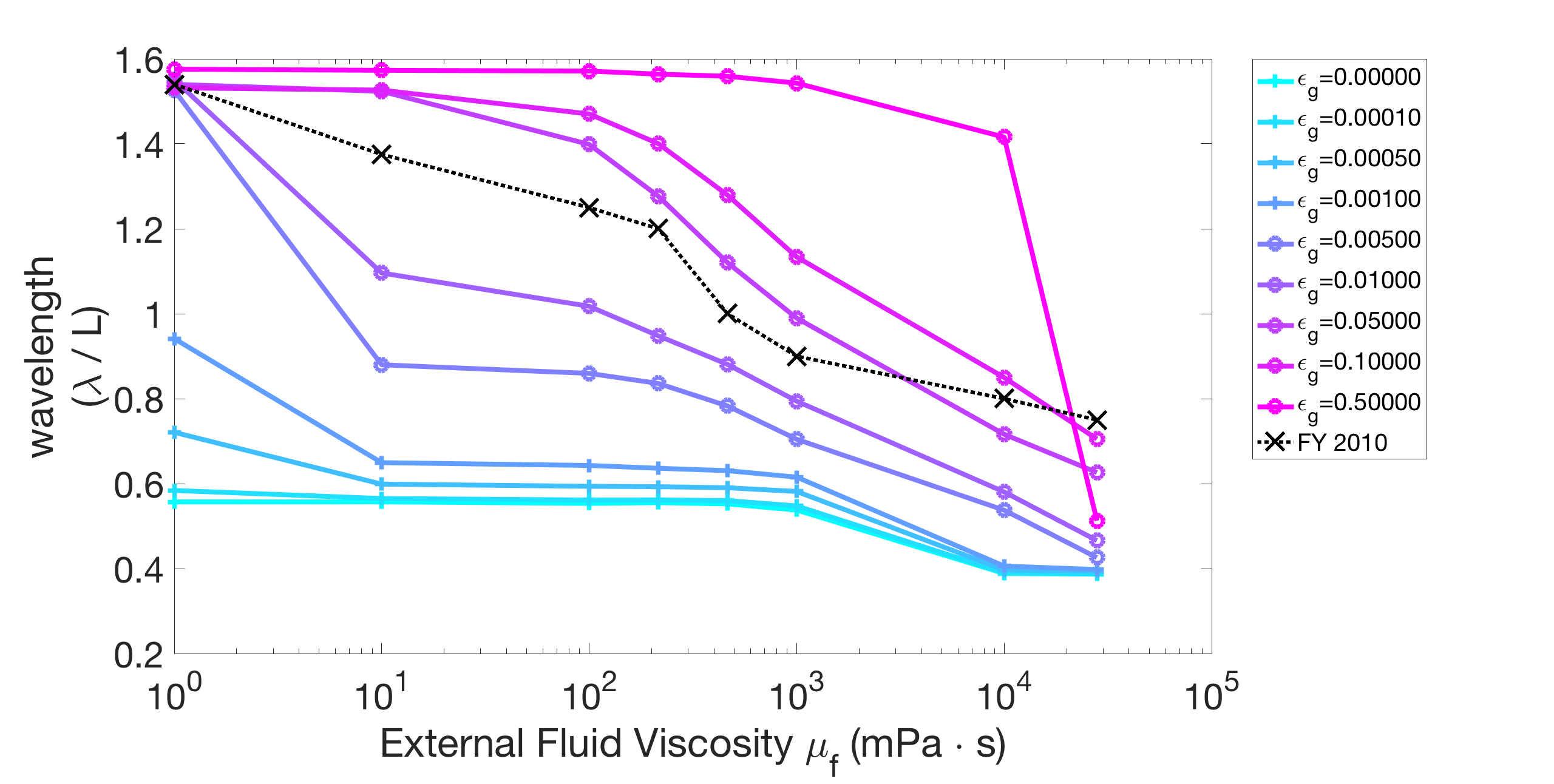}
\centering(b)\includegraphics[width=0.75\textwidth]{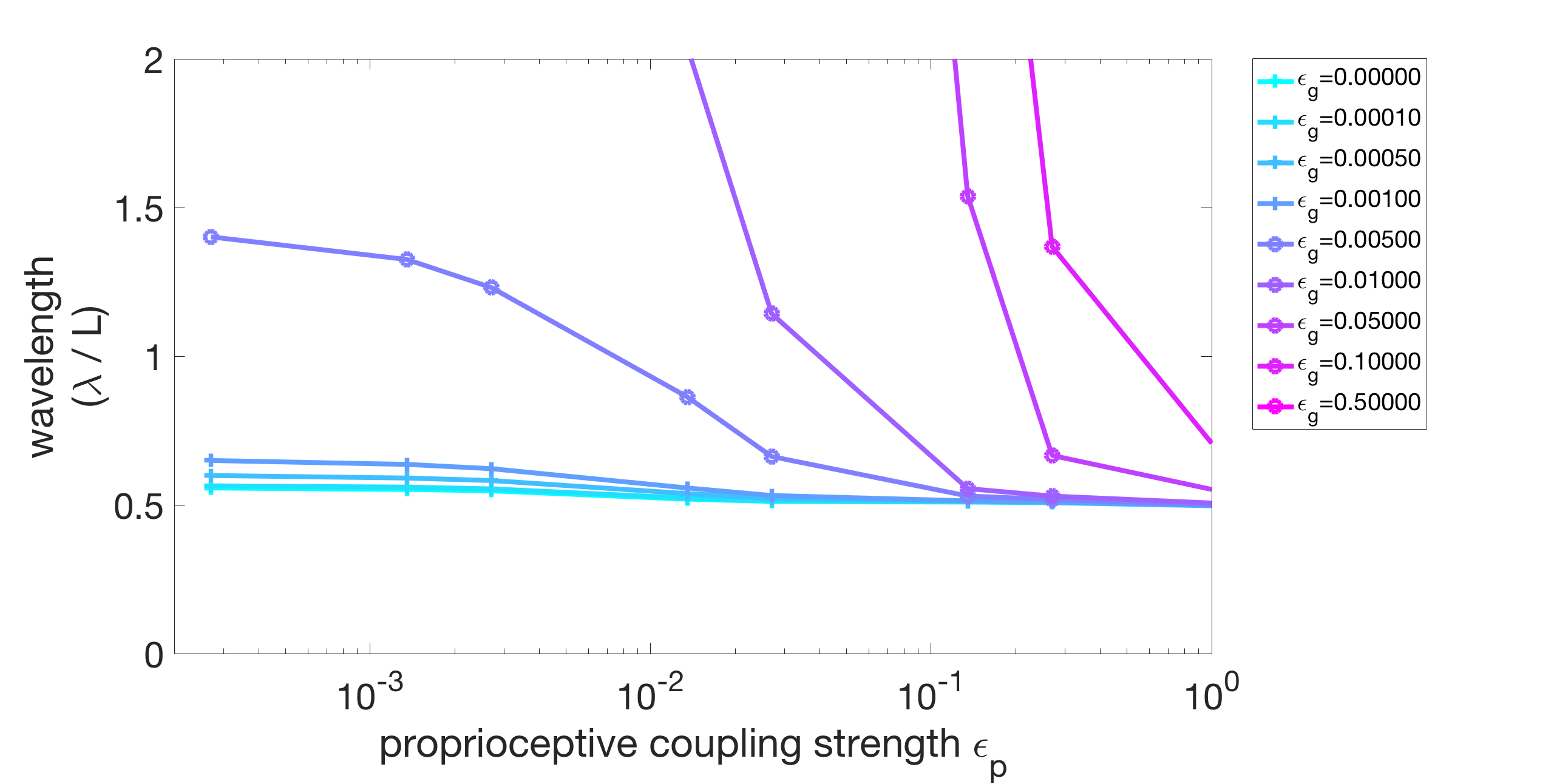}
\caption{(a) The resulting wavelength trend of the model (wavelength vs. external fluid viscosity $\mu_f$) for different gap-junctional coupling strengths $\eps_g$. In each run, the proprioceptive coupling strength $\eps_p$ was (attempted to) fit to match the wavelength in water (1.5 bodylengths). Circle-markers indicate successful fits, plus-markers indicate unsuccessful fits, and X-markers indicate data from Fang-Yen et al. \cite{Fang-Yen:2010aa}. Note that for $\eps_g < 0.005$, the wavelength in water was not matched. 
(b) The model wavelength in water ($\mu_f = 1$mPa s) as a function of proprioceptive coupling strength $\eps_p$ for different gap-junctional coupling strengths $\eps_g$. Note that for $\eps_g < 0.005$, the wavelength in water ($\lambda = 1.5$) is not matched over a wide range of proprioceptive coupling strengths $\eps_p$. }
\label{role_of_gapjns_figure}
\end{figure}
}

%---------------------------------
\section{The Neuromechanical Model as a Network of Coupled Oscillators: Insight Into Mechanisms Underlying Gait Adaptation}

The neuromechanical model is able to robustly capture the quantitative trend of gait adaptation across a wide range of parameters. In this section, the modular structure of the model will be exploited to uncover the fundamental mechanisms underlying gait adaptation.  
{\red In Section \ref{modules_are_oscillators}, we show that the isolated neuromechanical modules are oscillators. (Note that from the standard theory of weakly coupled oscillators point of view the modules would be considered ``intrinsic oscillators'', whereas from a neurolocomotion perspective, the modules are ``extrinsic oscillators'', i.e., {\blue local }proprioceptive feedback is required to generate the oscillations.)}  These modules form a network of coupled oscillators with three forms  of coupling: mechanical (through the body and external fluid), proprioceptive, and gap-junctional.  Furthermore, this coupling is relatively weak, and thus the theory of weakly coupled oscillators \cite{Kopell_1986,Schwemmer2012} can be applied to identify the coordinating effects of each coupling modality.
% The undulation exhibited by the model is a traveling wave of dorsal-ventral bending from the first module to the last.  The phase differences between the bends of each module define the wavelength. The neuromechanical model (equations \ref{full_kappa}-\ref{full_VD}) can be thought of as a system of coupled neuromechanical oscillators, where each local module oscillates on its own and the coupling between modules coordinates their phases. 
%
% The theory of weakly coupled oscillators is used to break down the mechanisms of coordination into the oscillator properties and the coupling dynamics.  
% For the parameter regimes where the model captures the correct wavelength trend (the blue parameter regimes in Figure \ref{parameter_Sweep_wvln_trends} (left)), it can be explained \textit{how} the model achieves this coordination. 
We demonstrate that the competition between mechanical coupling and neural coupling provides an explicit mechanism for gait adaptation.

\subsection{Isolated Neuromechanical Modules are Oscillators}\label{modules_are_oscillators}
\begin{figure}
\centering(a)\includegraphics[width=0.3\textwidth]{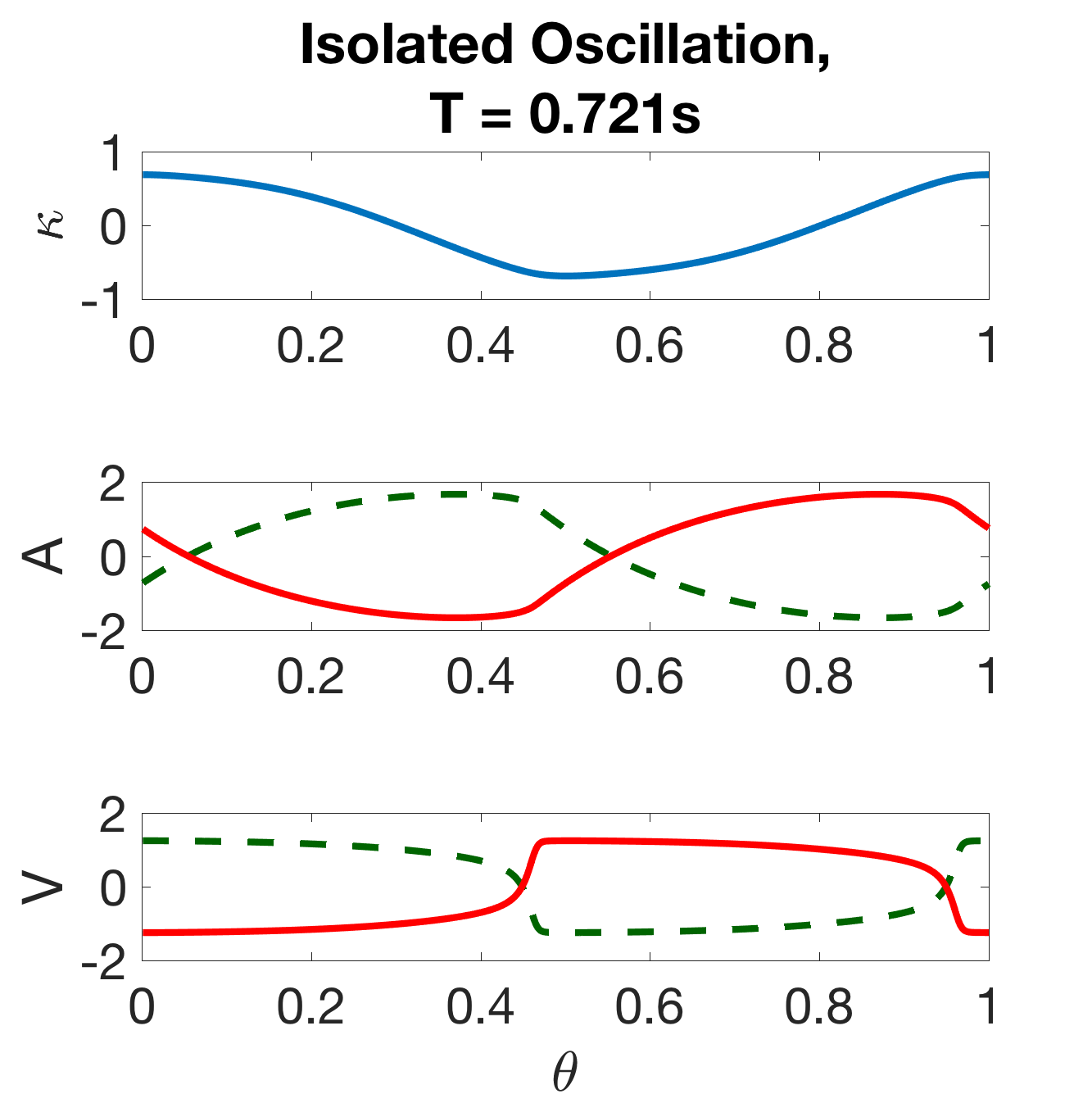}
(b)\includegraphics[width=0.3\textwidth]{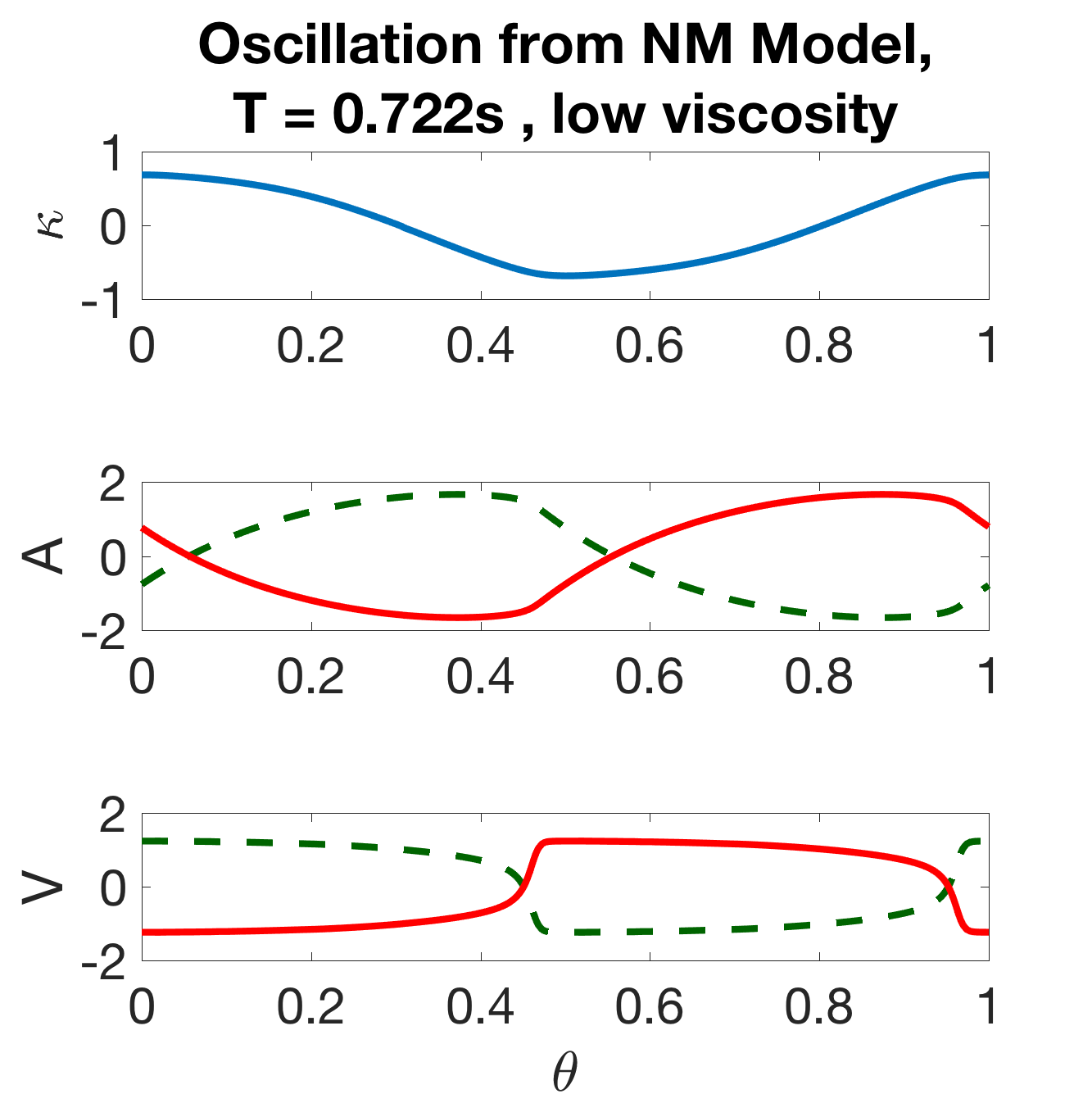}
(c)\includegraphics[width=0.3\textwidth]{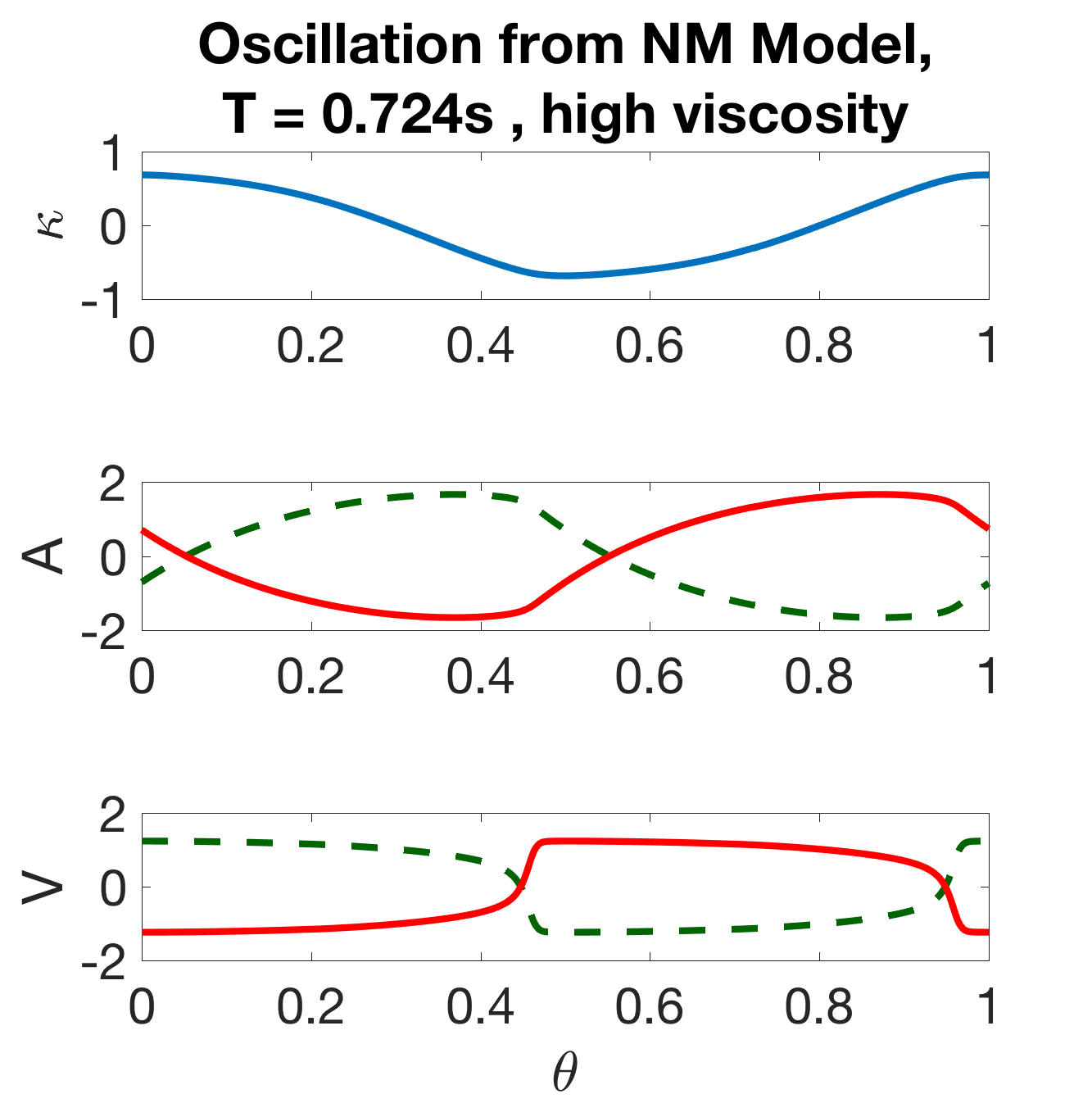}
\caption{The period and amplitude of the oscillations in $\kappa,A_V,A_D,V_D,V_V$ are all relatively similar for (a) the single, isolated neuromechanical module, (b) the single module in the full neuromechanical model at low viscosity ($\mu_f = 1$mPa s), and (c) the single module in the full neuromechanical model at high viscosity ($\mu_f = 2.8\times10^4$ mPa s).  Ventral neural/muscle activities are given in green dashed lines, dorsal neural/muscle activities are given in red solid lines.}
\label{single_oscillation_3versions}
\end{figure}

A single, isolated neuromechanical module is defined as a neural subcircuit, the corresponding muscles and body section, and local proprioceptive feedback (without coupling through the body or neural circuitry).  The dynamics for this isolated module are governed by
\begin{align}
\dot{\kappa} &= - \dfrac{1}{\tau_b}\qty(\kappa + \sigma(A_V)-\sigma(A_D)) ,\label{single_osc_kappa}\\
\dot{A}_V &= \dfrac{1}{\tau_m} \qty(- A_V + V_V - V_D) \label{single_osc_AV},\\
\dot{A}_D &= \dfrac{1}{\tau_m}\qty(- A_D + V_D - V_V) \label{single_osc_AD},\\
\dot{V}_V &= \dfrac{1}{\tau_n }\qty(F({V}_V) + c_{p}\kappa) \label{single_osc_VV}, \\
\dot{V}_D &= \dfrac{1}{\tau_n}\qty(F({V}_D)- c_{p}\kappa) \label{single_osc_VD}.
\end{align}
Note that this is the model described in Section \ref{chapter2}, omitting the intermodular coupling.
The isolated modules exhibit robust oscillations over a wide range of parameters, and a single period of the module is shown for each state-variable in Figure \ref{single_oscillation_3versions}(a). Thus, the neuromechanical modules are oscillators, wherein each B-class neuron promotes either a dorsal or ventral bend and the local proprioceptive feedback acts to switch the bistable B neurons from one state to the other.  %{\red These are \textit{extrinsic} oscillators in the sense that they require a full neuromechanical loop with proprioceptive feedback in order to generate oscillations.} 
The basic cycle of the oscillator is as follows: when activated, the ventral B-class neuron ($V_V$) excites the ventral muscles which build up activity ($A_V$) to induce a ventral bend (negative $\kappa$); when the curvature $\kappa$ is sufficiently large, the local proprioceptive feedback deactivates the ventral B-class neuron and activates the dorsal B-class neuron, and the cycle continues towards a dorsal bend.  

The system of six identical, uncoupled neuromechanical oscillators is described by 
\begin{equation}
\dot{\vec{X}}_j = S(\vec{X}_j), \ j = 1,\dots,6
\end{equation}  
where \begin{equation}
\vec{X}_j = \qty[
\kappa_j,
A_{V,j},
A_{D,j},
V_{V,j},
V_{D,j}]^T,
\end{equation}
and $S(\vec{X})$ is given by equations \ref{single_osc_kappa}-\ref{single_osc_VD}.
The oscillations correspond to a $T$-periodic limit cycle $\vec{X}^{LC}(t)$ in $(\kappa, A_V, A_D, V_V, V_D)$-state-space.
This limit cycle can be parametrized by phase 
\begin{equation}
\theta_j = \qty(\omega t + \theta_j^0) \text{ mod 1}
\end{equation}
with the initial phase $\theta_j^0 \in [0,1)$.  As $\theta_j$ increases at a constant rate $\omega = 1/T$, $\vec{X}^{LC}(\theta_j)$ traces out the limit cycle through state-space and the state of each oscillator on the limit cycle is given by 
\begin{equation}
\vec{X}_j(t) = \vec{X}^{LC}(\theta_j),
\end{equation}
where the only distinguishing feature between the oscillators is their unique phase $\theta_j$.  Figure \ref{single_oscillation_3versions}(a) shows the components of $\vec{X}^{LC}(\theta)$.

\subsection{Network of Coupled Oscillators}

Rearranging equations \ref{full_kappa}-\ref{w_g_matrix}, the neuromechanical model can be written as a network of coupled oscillators:
\begin{equation}
\dot{\vec{X}}_j = S(\vec{X}_j) + C_j(\vec{X}_1, \dots, \vec{X}_6), \ \ j=1,\dots,6 \label{totally_separated_model}
\end{equation}
where $C_j(\vec{X}_1, \dots, \vec{X}_6)$ describes the coupling dynamics from all the modules to the $j^{th}$ module through gap-junctions, nonlocal proprioception, and body mechanics:
\begin{equation}
C_j(\vec{X}_1,\dots,\vec{X}_6) = \qty[\begin{array}{c}
\eps_m \sum_{k=1}^6(D_4^{-1})_{jk}\ \dot{\kappa}_k,\\
0,\\
0,\\
\frac{1}{\tau_n} \sum_{k=1}^6 \eps_p (W_p)_{jk}\ \kappa_k + \eps_g(W_g)_{jk}\ V_{V,k},\\
\frac{1}{\tau_n} \sum_{k=1}^6 - \eps_p (W_p)_{jk}\ \kappa_k + \eps_g (W_g)_{jk}\ V_{D,k}\\
\end{array}].
\end{equation}
 The parameter $\eps_m = \alpha \mu_f \ell^4/\mu_b$ is the effective mechanical coupling strength.  %This form suggests that the neuromechanical model can be analyzed as a chain of intrinsic, isolated oscillators ``plus'' coupling between the modules.

{\red The oscillations of the isolated module (equations \ref{single_osc_kappa}-\ref{single_osc_VD}) are indistinguishable (on the scale of Figure \ref{single_oscillation_3versions}) from the oscillations of a module within the fully-coupled network (equations \ref{full_kappa}-\ref{full_VD}) at both low and high external fluid viscosity $\mu_f$.
%Furthermore, Figure \ref{single_oscillation_3versions}(b,c) show that the change in oscillator period between the low and high viscosity cases is $\Delta T/T < 0.003$, so the change in frequency as external fluid viscosity $\mu_f$ is varied is small.  
That is, the dynamics of the coupled module never deviate substantially from the limit cycle of the uncoupled module. This indicates that the intrinsic dynamics of the module dominate over the influence of coupling on any given cycle, which implies that the coupling is ``weak''.  However, small changes in the phase of a module due to coupling can accumulate over many cycles to significantly influence the phase differences between the modules.}
% Therefore the dynamics in the coupled modules are collapsed to the limit cycle and the coupling dynamics are considered ``weak'' relative to the intrinsic oscillator dynamics.}

Because the coupling is weak {\red (as defined above)}, the theory of weakly coupled oscillators can be applied (see \cite{Schwemmer2012} for details).
The coupling only alters the phase of the oscillators on their respective limit cycles and the effect on amplitude is negligible, therefore the phase completely describes the state of a neuromechanical module.  Equation \ref{totally_separated_model} can be reduced to the so-called phase equations, a set of differential equations describing the evolution of the phases of each oscillator:
\begin{equation}
\dot{\theta}_j = \omega_j + \sum_{k=1}^6 \eps_m (D_4^{-1})_{jk} H_m (\theta_k-\theta_j) + \eps_g (W_g)_{jk} H_g(\theta_k-\theta_j) + \eps_p  (W_p)_{jk}  H_p(\theta_k-\theta_j) ,\label{weak_coupling_ThetaJ}
\end{equation}
where $\theta_j$ is the phase of the $j^{th}$ oscillator, $\omega$ is the intrinsic frequency, and $H(\phi)$ are the interaction functions that describe the change in frequency (resulting from either mechanical, proprioceptive, or gap-junction coupling) as a function of the phase difference $\phi = \theta_k - \theta_j$ of a given pair of oscillators:
\begin{align}
H_m(\phi) &= -\dfrac{1}{T} \int_0^T Z_\kappa(t) \dot{\vec{\kappa}}^{LC}(t - \phi) \dd t, \label{mech_interact_fn}\\
H_p(\phi) &= \dfrac{1}{\tau_n}\dfrac{1}{T} \int_0^T Z_{V_V}(t) \vec{\kappa}^{LC}(t - \phi) -  Z_{V_D}(t) \vec{\kappa}^{LC}(t - \phi) \dd t, \label{prop_interact_fn} \\
H_g(\phi) &= \dfrac{1}{\tau_n}\dfrac{1}{T} \int_0^T Z_{V_V}(t) \qty(\vec{V_V}^{LC}(t - \phi) - \vec{V_V}^{LC}(t)) + Z_{V_D}(t) \qty(\vec{V_D}^{LC}(t - \phi) - \vec{V_D}^{LC}(t))\dd t.  \label{gap_interact_fn}
\end{align}
Here, $Z_\kappa(t)$, $Z_{V_V}(t)$, $Z_{V_D}(t)$ are the $T-$periodic phase response functions to perturbations in the corresponding state variable. 

The coupling modalities define the \textit{structure} of the interaction functions, through the state variables that are coupled, as well as the coupling topology (the connectivity matrices $D_4^{-1}$, $W_g$, and $W_p$ in equation \ref{weak_coupling_ThetaJ}).  
Note that there is a separate H-function for each of the three coupling modalities and these three coupling modalities add linearly to produce the full interaction of the modules. Therefore, the relative contributions of the various coupling types can be analyzed independently through varying the different coupling strengths: fluid viscosity $\mu_f$ (through $\eps_m$) for mechanical, $\eps_p$ for proprioceptive, and $\eps_g$ for gap-junctional.

\subsection{Two Oscillator Analysis Explains the Coordination Mechanism}\label{sect_two_osc_analysis}

Analyzing a pair of two coupled oscillators gives considerable insight into the coordination that each coupling modality produces separately and the mechanisms of coordination.  With only two oscillators, the phase model reduces to
\begin{align}
\dot{\theta}_1 &= \omega+ \eps_m \sum_{j=1}^2 (D_4^{-1})_{1j} H_m (\theta_j-\theta_1) + \eps_g H_g(\theta_2-\theta_1), \label{pair_osc_theta1}\\
\dot{\theta}_2 &= \omega + \eps_m \sum_{j=1}^2 (D_4^{-1})_{2j} H_m (\theta_j-\theta_2) + \eps_p H_p(\theta_{1}-\theta_2) + \eps_g H_g(\theta_1-\theta_2). \label{pair_osc_theta2}
\end{align}
In the two oscillator case, the matrix $D_4^{-1}$ is symmetric, so $(D_4^{-1})_{12} = (D_4^{-1})_{21} = d_{12}$. By defining 
\begin{equation}\phi = \theta_2-\theta_1,\end{equation}
and subtracting equation \ref{pair_osc_theta1} from equation \ref{pair_osc_theta2}, the dynamics of the two oscillator system can be described by a single differential equation for the phase difference between the two oscillators:\\
\begin{align}
\dot{\phi} &= \eps_m d_{12} G_m (\phi) + \eps_p G_p(\phi) + \eps_g G_g(\phi) = G(\phi), \label{pair_osc_phi}
\end{align}
where $G_m(\phi) = H_m(-\phi) - H_m(\phi)$, $G_p(\phi) = H_p(-\phi)$, and $G_g(\phi) = H_g(-\phi) - H_g(\phi)$ are the pair-wise interaction functions, or \textit{G-functions} of the pair.  The stable phase-locked state of the system $\phi^*$ is given by $G(\phi^*)=0,$ $G'(\phi^*)<0$.

\subsubsection{Each Coupling Modality Promotes a Different Coordination Outcome}  
\begin{figure}
\centering(a)\includegraphics[width=.3\textwidth]{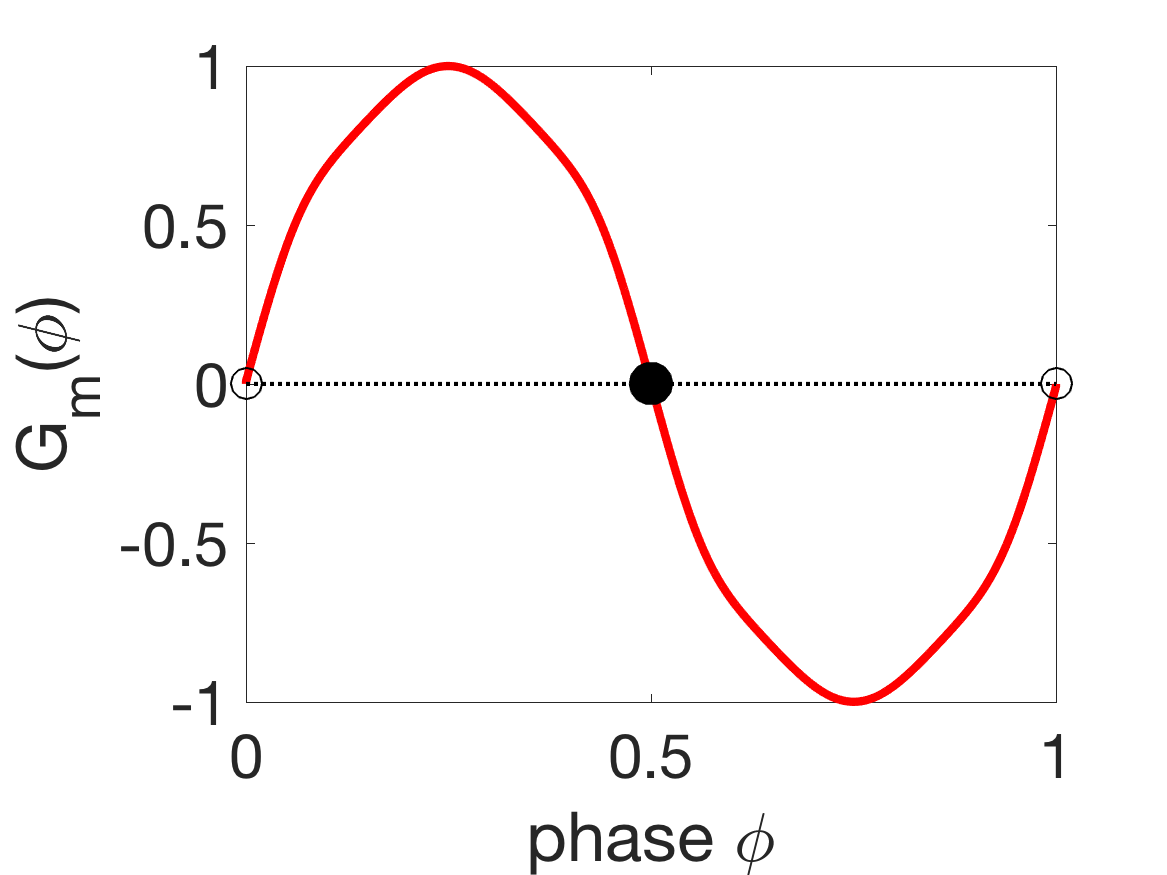}
(b)\includegraphics[width=.3\textwidth]{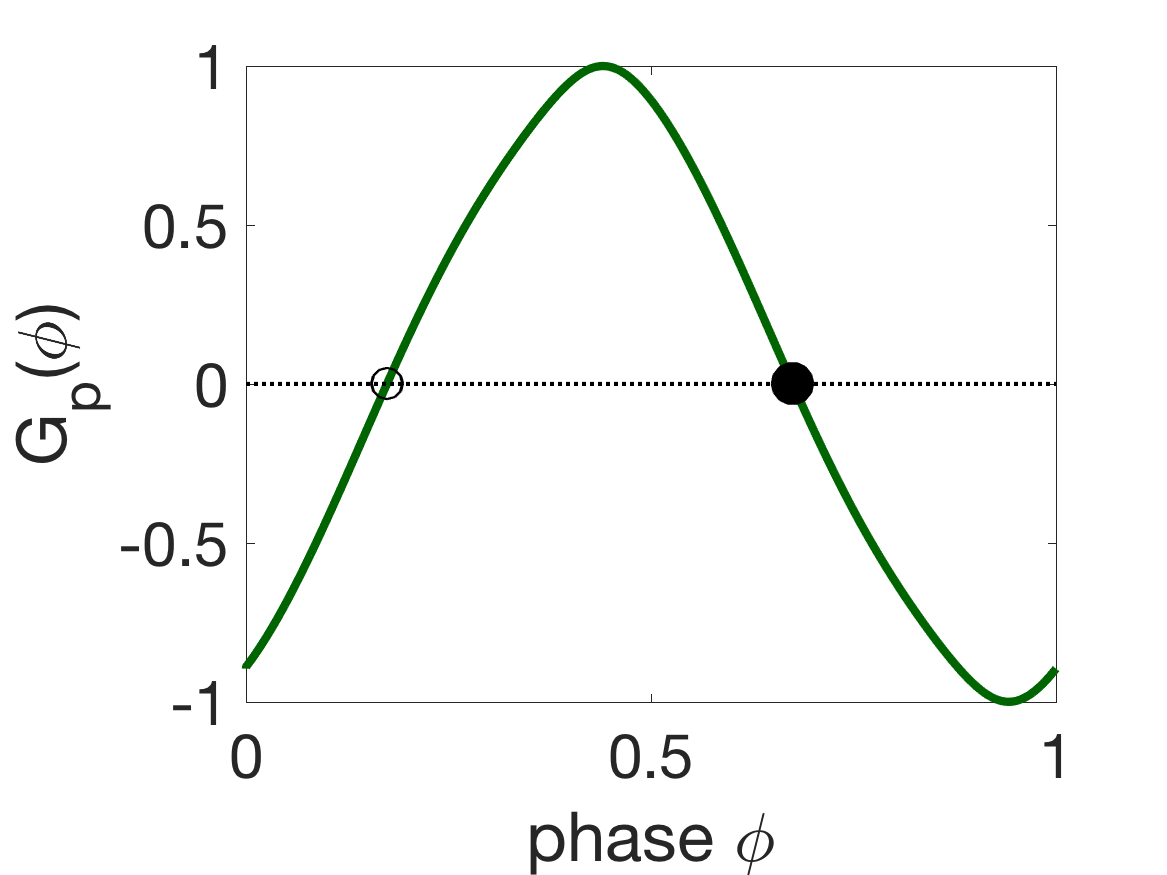}
(c)\includegraphics[width=.3\textwidth]{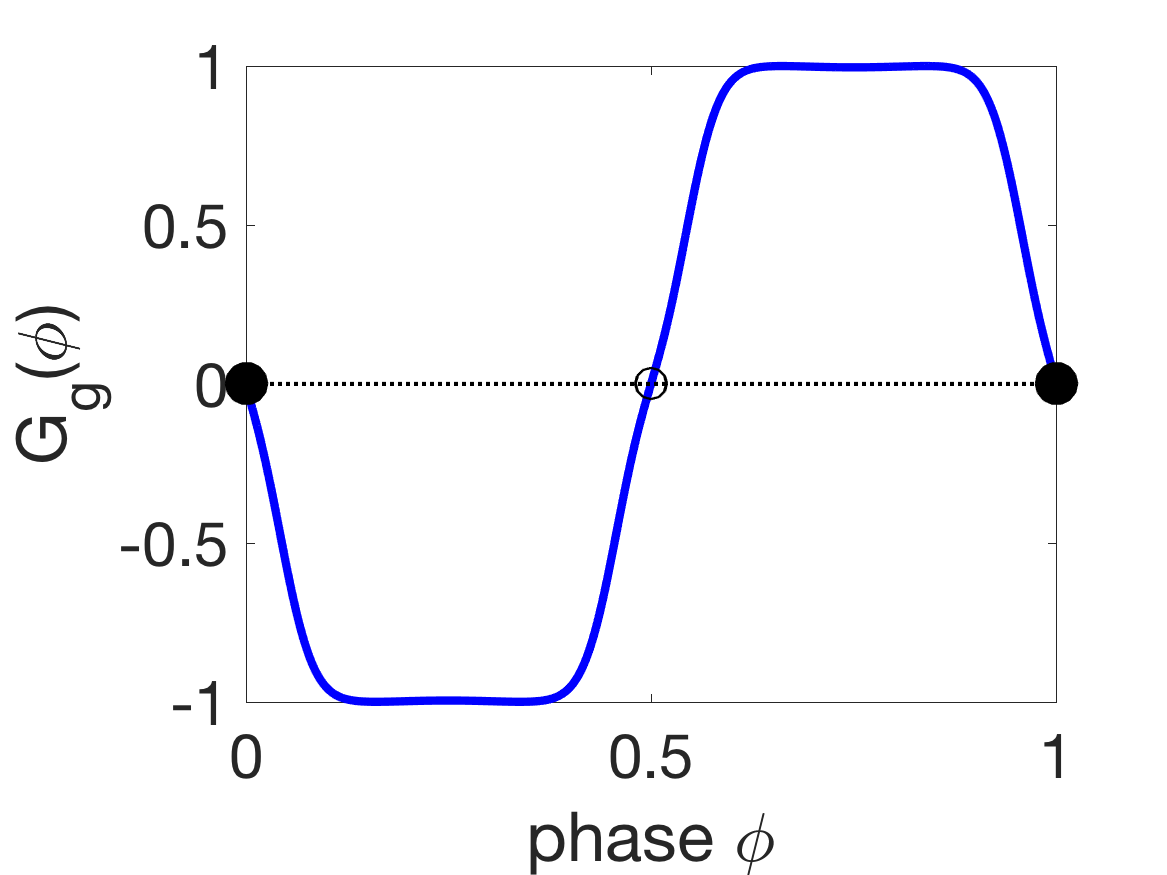}
\caption{Each coupling modality promotes a different coordination outcome in a pair of coupled neuromechanical oscillators based on the stable zero of the corresponding G-function: (a) mechanical coupling promotes antiphase since $G_m(0.5) = 0$ and $G_m'(0.5)<0$; (b) proprioceptive coupling promotes a phase-wave since $G_p(.75) = 0$ and $G_p'(.75)<0$; and (c) gap-junctional coupling promotes synchrony since $G_g(1) = 0$ and $G_g'(1)<0$.}
\label{gfns_plot}
\end{figure}

% Each coupling modality alone promotes a different coordination outcome. 
Figure \ref{gfns_plot} shows the G-functions and corresponding phase-locked states of the different coupling modalities.
For mechanical coupling alone, i.e., $\eps_p = \eps_g =0$,  the stable phase-locked state is anti-phase ($\phi^* = 0.5$), since $G(0.5)=0$ and $G'(0.5)<0$ (Figure \ref{gfns_plot}(a)). Similarly, for proprioceptive coupling alone, the stable state is an intermediate phase-difference ($\phi^*\approx 0.75$,  Figure \ref{gfns_plot}(b)), so the first oscillator leads the second (front-to-back).  For gap-junctional coupling alone, the stable state is synchrony ($\phi^* = 0$, Figure \ref{gfns_plot}(c)). 

The coordination outcome with all three coupling mechanisms present corresponds to the zero of the G-function (equation \ref{pair_osc_phi}), which is a weighted sum of the three individual G-functions.  Thus, coordination can be examined in the context of this weighted sum as the three coupling strengths are varied: external fluid viscosity $\mu_f$ for mechanical coupling, proprioceptive coupling strength $\eps_p$, and gap-junction coupling strength $\eps_g$.

\subsubsection{Neural Coupling Sets the Low-Viscosity Wavelength}

\begin{figure}
\centering(a)\includegraphics[width=.45\textwidth]{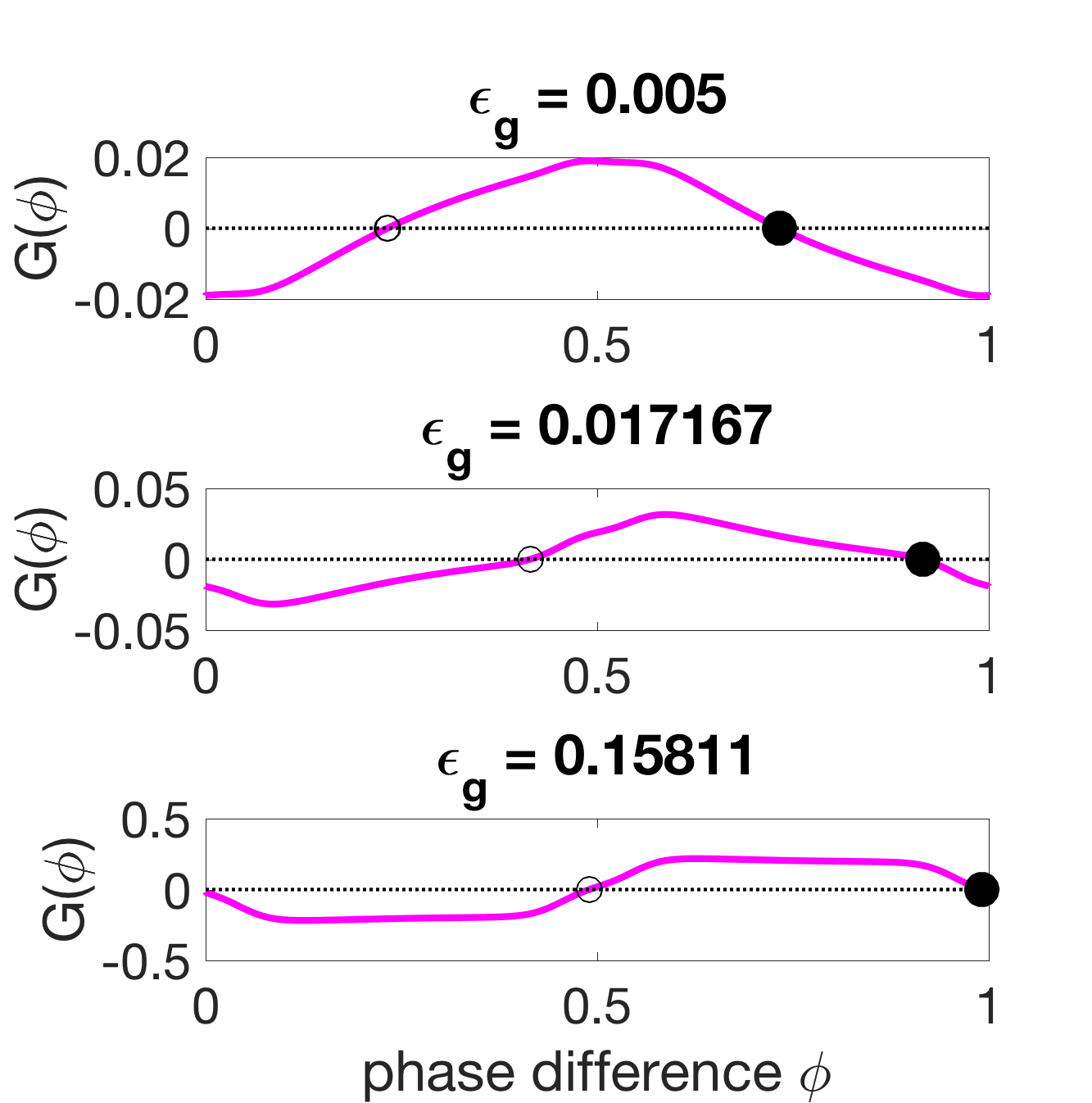}
\centering(b)\includegraphics[width=.45\textwidth]{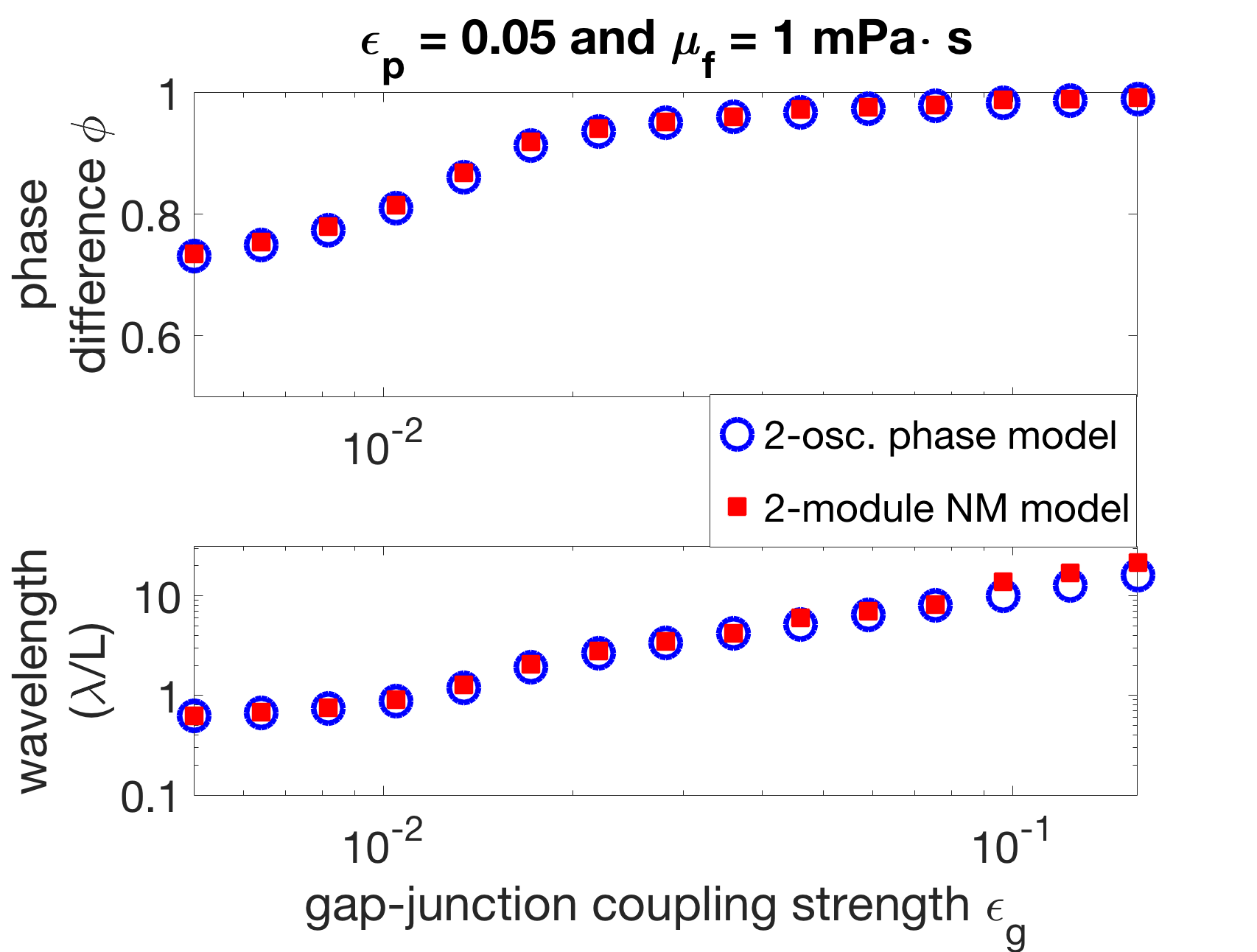}
\caption{In the low-viscosity limit, the stable phase-locked states of the pair of neuromechanical oscillators is set by the competition between proprioceptive and gap-junctional coupling.  (a) The linear combination of the G-functions given by equation \ref{pair_osc_phi} for $\varepsilon_p = 0.05$, $\mu_f = 1$ mPa$\cdot$s, and various $\eps_g$.  Note that as the gap-junctional coupling strength $\eps_g$ increases, the stable phase-locked phase difference $\phi^*$ moves from roughly $\phi^*=0.75$ towards $\phi^* = 1$. (b) The stable phase-locked states of the pair can be tuned by varying the two forms of neural coupling: proprioceptive and gap-junctional.  When proprioceptive coupling dominates, the stable phase-locked state is a phase difference of roughly $\phi^* = 0.75$, and when gap-junctional coupling dominates, the stable phase-locked state is synchrony $\phi^* = 1$.  The resulting wavelength in the body, if the pair-wise phase difference was constant in the six-oscillator model, can be tuned by varying the two forms of neural coupling: proprioceptive and gap-junctional.  When proprioceptive coupling dominates, the wavelength is roughly 0.75 bodylengths, and when gap-junctional coupling dominated, the wavelength is infinite, since each oscillator pair is in perfect synchrony and thus the body is a standing wave.
}
\label{neural_coupling_plots}
\end{figure}

The stable phase difference $\phi^*$ of the pair of the neuromechanical oscillators can be used to define a wavelength in the full body (for details see Appendix \ref{appendix_defining_wvln}): 
\begin{equation}
\dfrac{\lambda}{L}  = \dfrac{1}{6\qty(1-\phi^*)}.
\label{wvln_per_bodylength_defining_wvln}
\end{equation}

In the low external fluid viscosity case ($\mu_f=1$ mPa$\cdot$s), 
setting $\eps_p=0.05, \eps_g = 0.01$ as in Section \ref{sect_timescales_param_study} provides a good approximation of the experimentally observed wavelength for the mechanical parameters $k_b = 2.6\times 10^{-7}$ N (mm)$^2$, $\mu_b = 1.3\times 10^{-7}$ N (mm)$^2$ s.  For these parameters, the relative sizes of the G-functions in equation \ref{pair_osc_phi} are
\begin{align}
\eps_m d_{12}\max|G_m (\phi)| &= 3.532 \times 10^{-5},\\
\eps_p \max |G_p(\phi)| &= 2.016, \\
\eps_g \max |G_g(\phi)| &= 1.259.
\end{align}
Thus, at low viscosity, mechanical coupling is almost negligible compared to neural coupling, so the coordination is determined by proprioceptive and gap-junctional coupling.

How the wavelength is set in this low-viscosity case can be examined by varying the neural coupling strengths.  Figure \ref{neural_coupling_plots}(a) shows that as the gap-junctional coupling strength $\eps_g$ is increased relative to the proprioceptive coupling strength, the phase-locked states move from close to the zeros of $G_p(\phi)$ towards the zeros of $G_g(\phi)$.
% Figure \ref{neural_coupling_plots}A and B show how the stable phase-locked state changes as $\eps_g$ is varied, and Figure \ref{neural_coupling_plots}C shows the resulting wavelengths (according to equation \ref{wavelength_calc}).  
Figure \ref{neural_coupling_plots}(b) shows that when proprioceptive coupling dominates, the stable phase-locked state corresponds to a phase difference of roughly $\phi^* \approx 0.75$ and corresponds to a wavelength of 0.75 bodylengths according to equation \ref{wvln_per_bodylength_defining_wvln}. When gap-junctional coupling dominates,  the stable phase-locked state is close to synchrony $\phi^*\approx 1$, which corresponds to an infinite wavelength in the full body if this phase difference was constant. In this gap-junction-dominated case, each pair is in perfect synchrony and the body exhibits a standing wave.

To assess the predictive power of the two-oscillator phase model, a simulation of the neuromechanical model with only two modules was performed alongside the phase model. Figure \ref{neural_coupling_plots}(b) shows that the two-oscillator phase model is quantitatively accurate when compared to the phase differences and wavelengths derived from this two-module simulation. Thus, neural coupling sets the low-viscosity wavelength in the two-module neuromechanical model as well.

\subsubsection{Competition Between Mechanical and Neural Coupling Provides a Mechanism for Gait Adaptation}

\begin{figure}
\centering(a)\includegraphics[width=.45\textwidth]{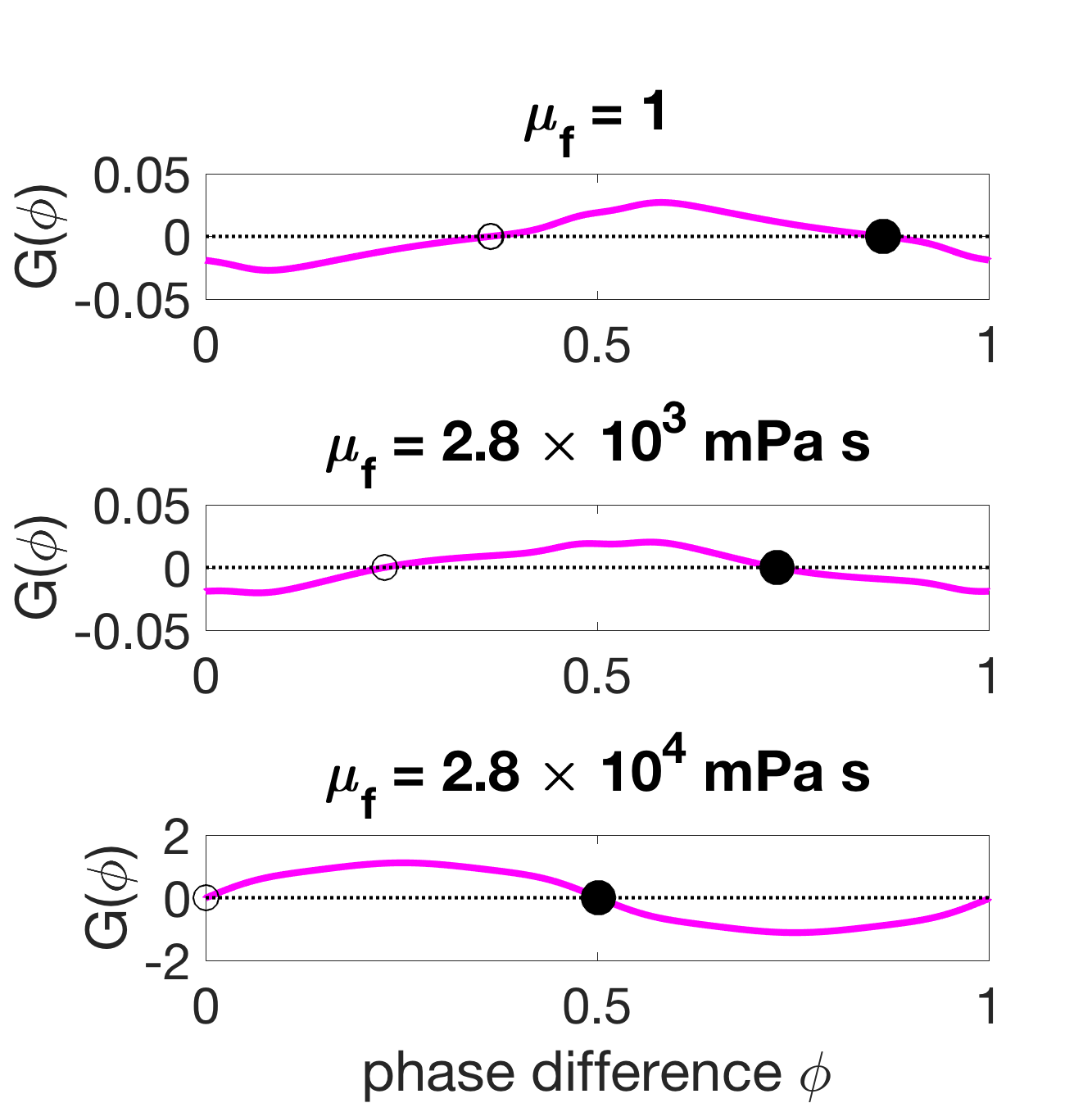}
\centering(b)\includegraphics[width=.45\textwidth]{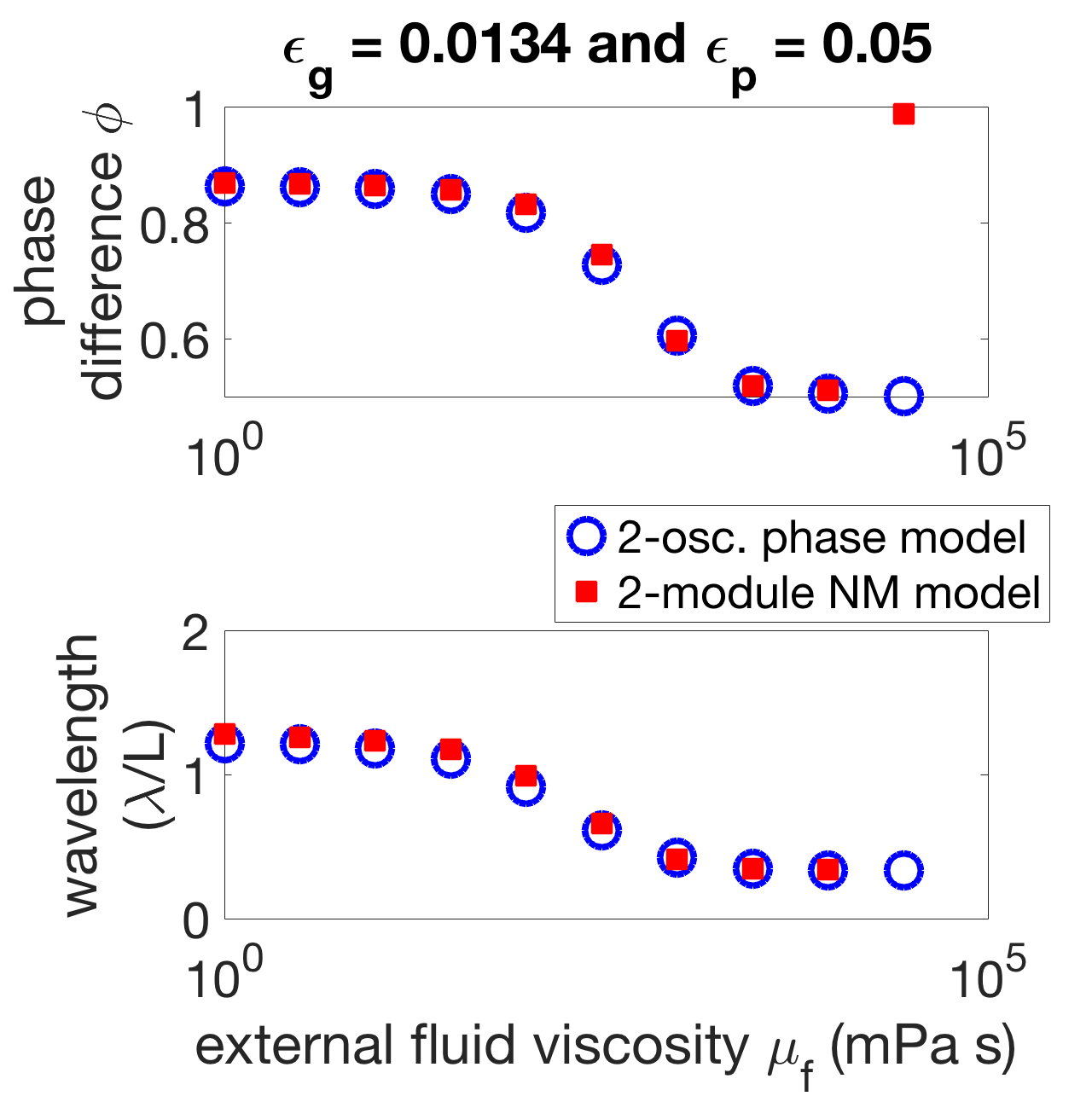}
\caption{Gait adaptation is a result of the competition between mechanical and neural coupling in the pair of neuromechanical oscillators.  (a) The linear combination of the G-functions given by equation \ref{pair_osc_phi} for $\varepsilon_p = 0.05$, $\varepsilon_g = 0.0134$, and various $\mu_f$.  Note that as $\mu_f$ increases, the strength of mechanical coupling increases and the stable phase-locked phase difference $\phi^*$ moves from roughly $\phi^*=0.8$ towards $\phi^* = 0.5$. (b) When neural coupling dominates, the stable phase-locked state is a phase difference of roughly $\phi^* = 0.88$, and when mechanical coupling dominates, the stable phase-locked state is antiphase, i.e., $\phi^* = 0.5$ phase difference. The resulting wavelength in the body, if the pair-wise phase difference was constant in the six-oscillator model, is set by the competition between the mechanical and neural coupling.  When neural coupling dominates, the wavelength is roughly 1.5 bodylengths, and when mechanical coupling dominates, the wavelength is roughly 0.45 bodylengths.
}
\label{mech_coupling_plots}
\end{figure}

To examine the effect of mechanical coupling in the two-oscillator phase model, the neural coupling parameters are fixed to $\eps_p = 0.05$ and $\eps_g = 0.0134$ so that the wavelength in the low-viscosity case is roughly 1.5 bodylengths.  The strength of mechanical coupling is increased in equation \ref{pair_osc_phi} by increasing the external fluid viscosity $\mu_f$. Figure \ref{mech_coupling_plots}(a) shows that as the strength of mechanical coupling is increased, the phase-locked states move from close to the zeros set by $\eps_p G_p(\phi) + \eps_g G_g(\phi)$ towards the zeros of $G_m(\phi)$.
 Figure \ref{mech_coupling_plots}(b) shows how the stable phase-locked state changes as a function of the mechanical coupling strength $\mu_f$.
When neural coupling dominates, the stable phase-locked state is a phase difference of roughly $\phi^*\approx 0.89$, and when mechanical coupling dominates, the stable phase-locked state is antiphase $\phi^* = 0.5$.
Similarly, Figure \ref{mech_coupling_plots}(b) shows that when neural coupling dominates the resulting wavelength (according to equation \ref{wvln_per_bodylength_defining_wvln}) is roughly 1.5 bodylengths, and when mechanical coupling dominates the wavelength is roughly 0.45 bodylengths.

This analysis shows that gait adaptation is a result of competition between mechanical and neural coupling.  The decrease in wavelength as external viscosity $\mu_f$ increases is explained by the increased strength in mechanical coupling and its associated coordination outcome, antiphase. The two-oscillator phase model is quantitatively accurate when compared to phase differences derived from the neuromechanical model with two modules, as shown in Figure \ref{mech_coupling_plots}(b).  Thus, this suggests that the mechanism underlying the behavior in the two-module neuromechanical model is the same as the mechanism of the phase model outlined here.  However, note that the phase difference at the highest fluid viscosity ($\mu_f = 2.8\times 10^4$ mPa s) is different between the two-oscillator phase model and the full two-module neuromechanical model.  This indicates the limit of weak coupling, as the phase reduction is not able to capture the transition to synchrony seen in the two-module neuromechanical model. However, weak coupling holds in the two-oscillator case for the rest of the viscosities $\mu_f$ considered.  Furthermore, this transition to synchrony is not seen in the six-module neuromechanical model.

\subsubsection{Phase Reduction Gives Insight into Timescale Ordering}

\begin{figure}
\centering(a)\includegraphics[width=.45\textwidth]{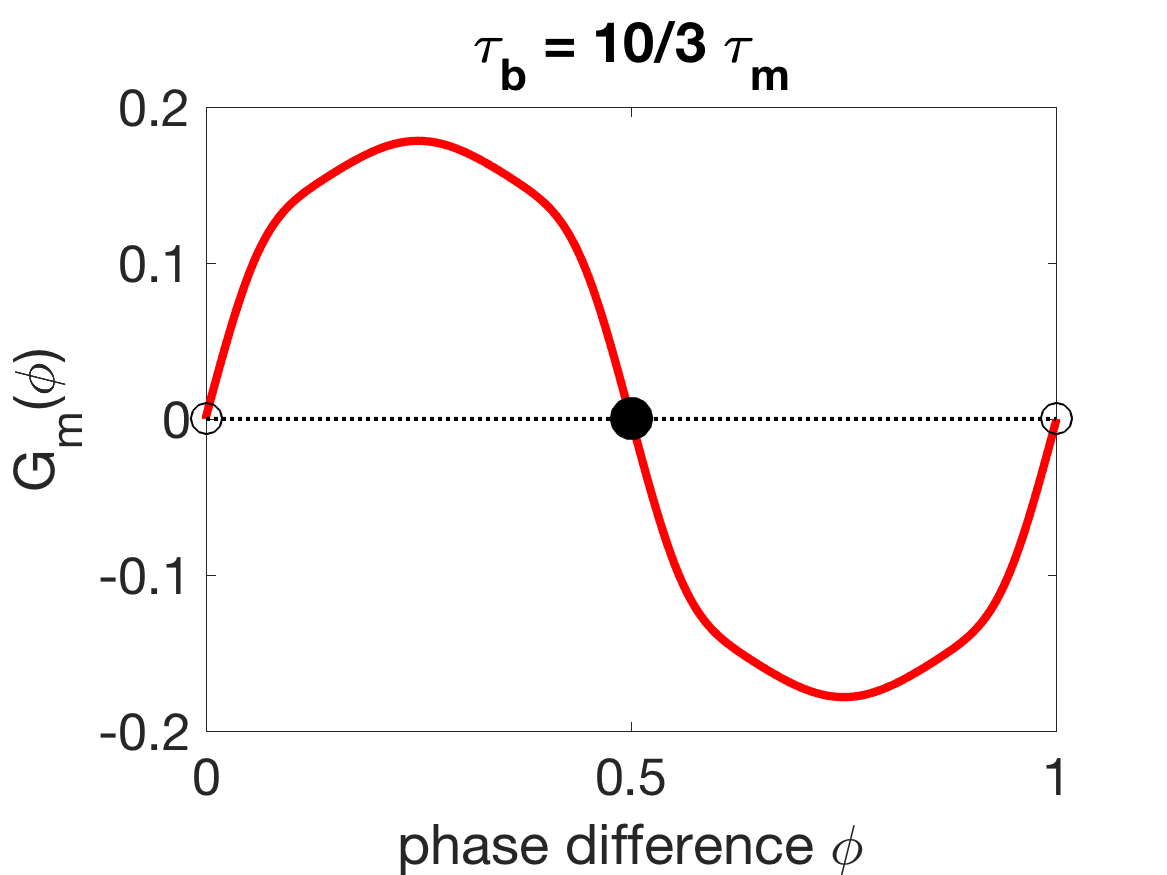}
\centering(b)\includegraphics[width=.45\textwidth]{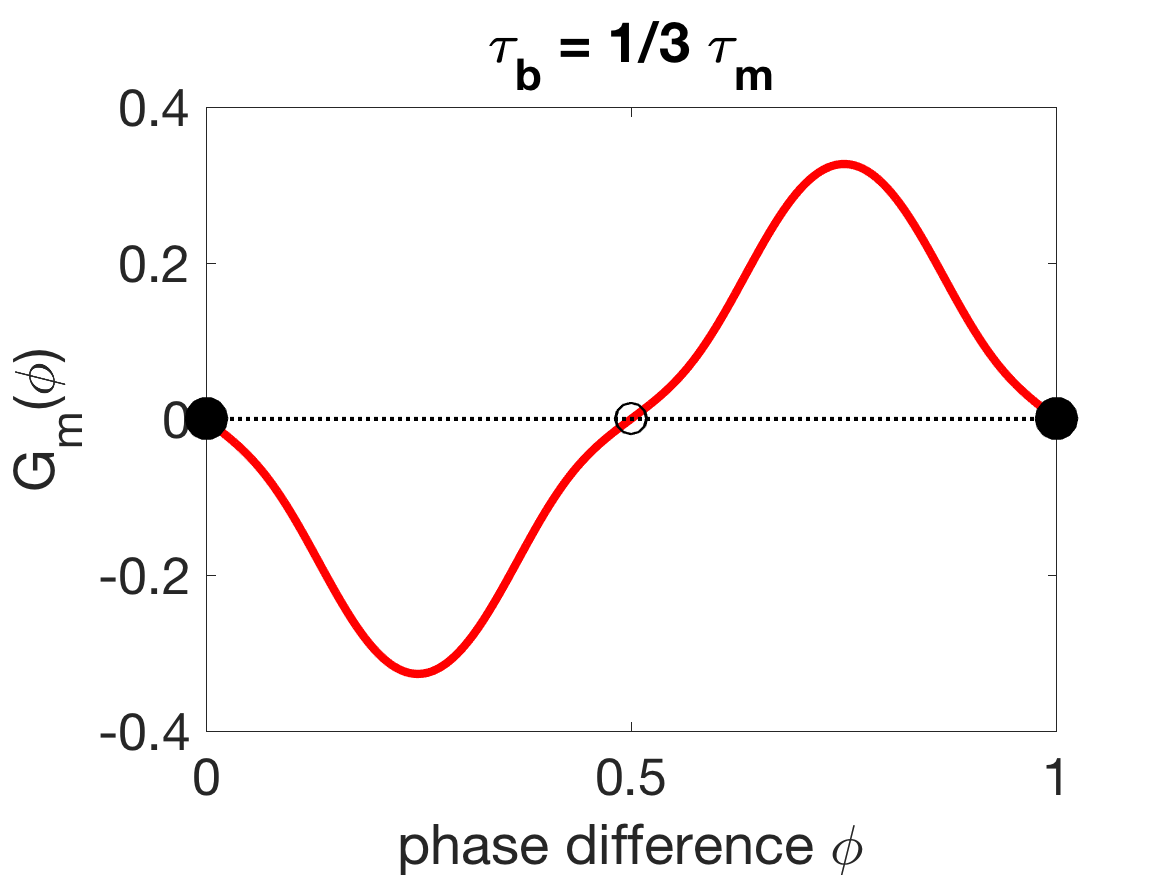}\\
\centering(c)\includegraphics[width=.45\textwidth]{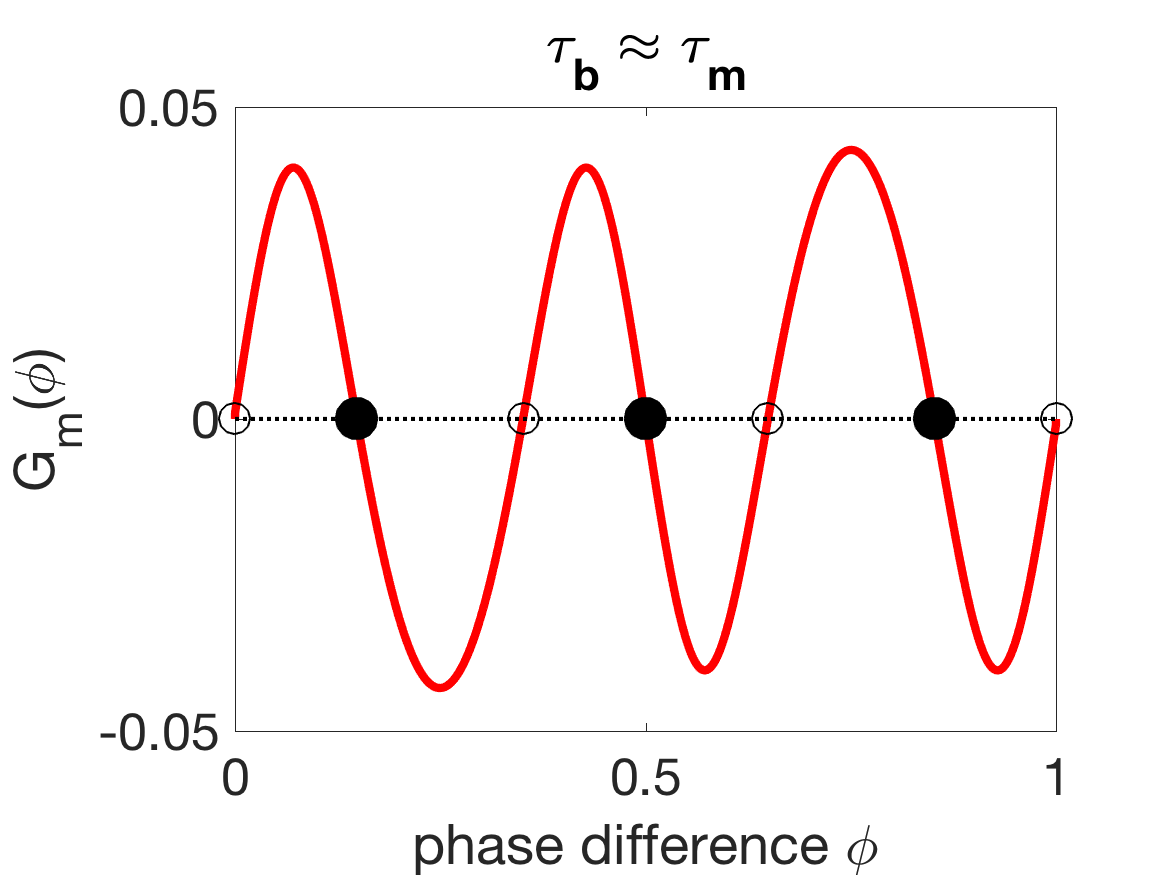}
\centering(d)\includegraphics[width=.45\textwidth]{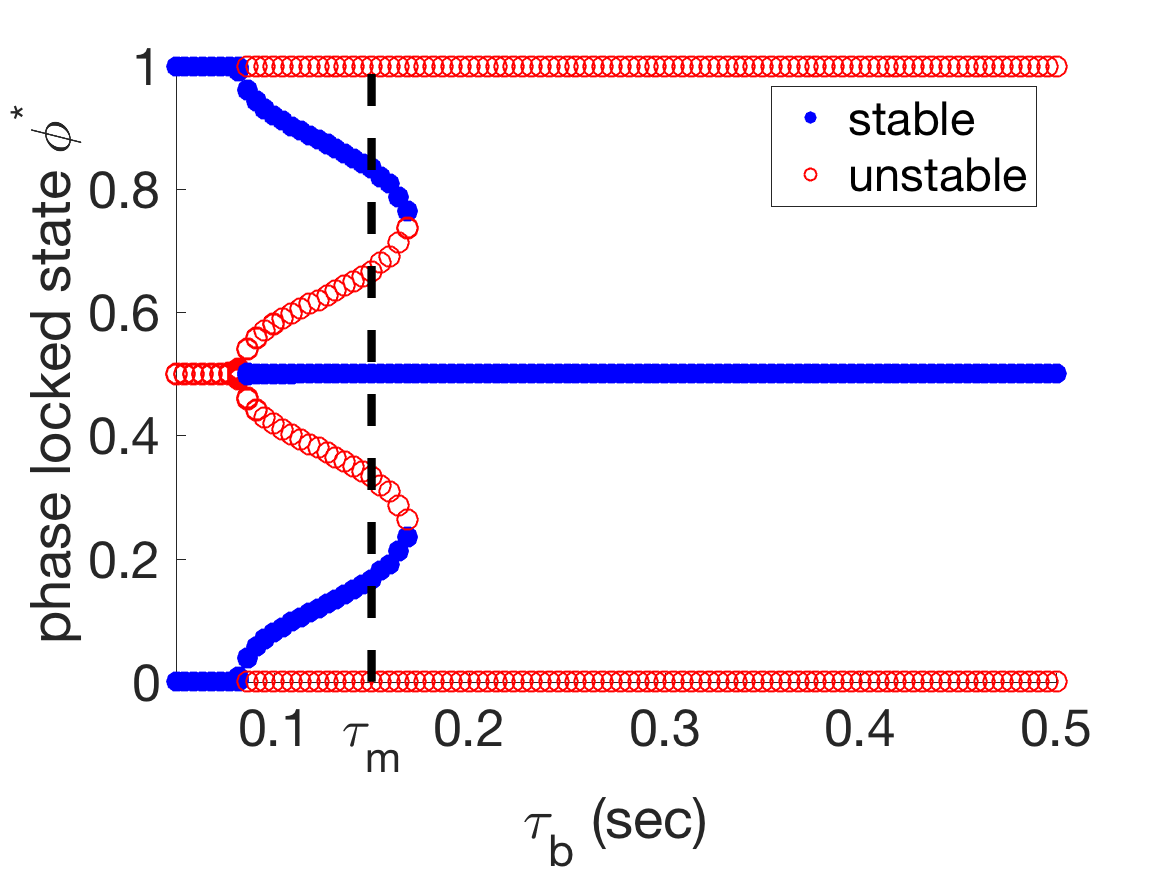}
\caption{(a) Mechanical G-function $G_m(\phi)$ for the pair of neuromechanical oscillators when $\tau_b$ is sufficiently larger than $\tau_m$ ($\tau_b = 0.5$ s, $\tau_m = 0.15$ s). Note the stable phase-locked state is antiphase since $G_m(0.5)=0$ and $G_m'(0.5)<0$. (b) Mechanical G-function $G_m(\phi)$ for the pair when $\tau_b$ is sufficiently smaller than $\tau_m$ ($\tau_b = 0.05$ s, $\tau_m = 0.15$ s).  Note the stable phase-locked state is synchrony since $G_m(1)=0$ and $G_m'(1)<0$, while antiphase is unstable since $G_m'(0.5)>0$. (c) Mechanical G-function $G_m(\phi)$ for the pair when $\tau_b \approx \tau_m$ ($\tau_b = 0.14$ s, $\tau_m = 0.15$ s). (d) Bifurcation diagram for the phase-locked states $\phi^*$ of the mechanical G-function vs. $\tau_b$, for $\tau_m = .15$ s.}
\label{compare_g_fns_tm_tf}
\end{figure}

The phase reduction also explains why generally $\tau_b$ must be larger than $\tau_m$ in order to obtain the correct coordination trend (as described in Section \ref{sect_timescales_param_study}).
The results in the previous subsection indicate that it is important for mechanical coupling to promote antiphase in order to get the correct wavelength trend as external viscosity $\mu_f$ is increased.  Figure \ref{compare_g_fns_tm_tf}(a) shows that, when $\tau_b$ is sufficiently larger than $\tau_m$, the stable zero of $G_m$ is 0.5, i.e., the stable phase-locked state is antiphase. 
However, when $\tau_b$ is sufficiently smaller than $\tau_m$, the stable zero of $G_m$ is 0, i.e., the G-function is flipped and mechanical coupling promotes synchrony.  In this case, the wavelength trend as external viscosity $\mu_f$ is increased is incorrect, since increasing the mechanical coupling strength would pull the oscillators towards synchrony, lengthening the wavelength instead of shortening it. 

The shift in the stabilities of the phase-locked states from antiphase to synchrony is somewhat complicated, as Figure \ref{compare_g_fns_tm_tf}(c) shows that $\tau_b \approx \tau_m$ can yield tristable phase-locked states. A series of paired saddle-node bifurcations and paired super- and sub-critical pitchfork bifurcations (Figure \ref{compare_g_fns_tm_tf}D), marks the transition from stable antiphase to tristability to stable synchrony as $\tau_b$ moves below $\tau_m$.   The change in the stability of the antiphase state promoted by mechanical coupling is the cause of the rapid change in coordination in Figure \ref{parameter_Sweep_wvln_trends} as $\tau_b$ becomes sufficiently smaller than $\tau_m$.  
% However, the complicated series of bifurcations hints at why the transition is not uniform in the other explored parameter, $\mu_b$, which possibly relocates these bifurcation points.

\subsection{Mechanism for Gait Adaptation Holds in Six-Oscillator Case}

We simulate the six-oscillator phase model in order to (i) assess the predictive power of the phase model by a quantitative comparison to the full six-module neuromechanical model and (ii) determine whether the mechanism of gait adaptation analyzed in the two-module case extends to the full six-module case.

\begin{figure}
\centering(a)\includegraphics[width=.465\textwidth]{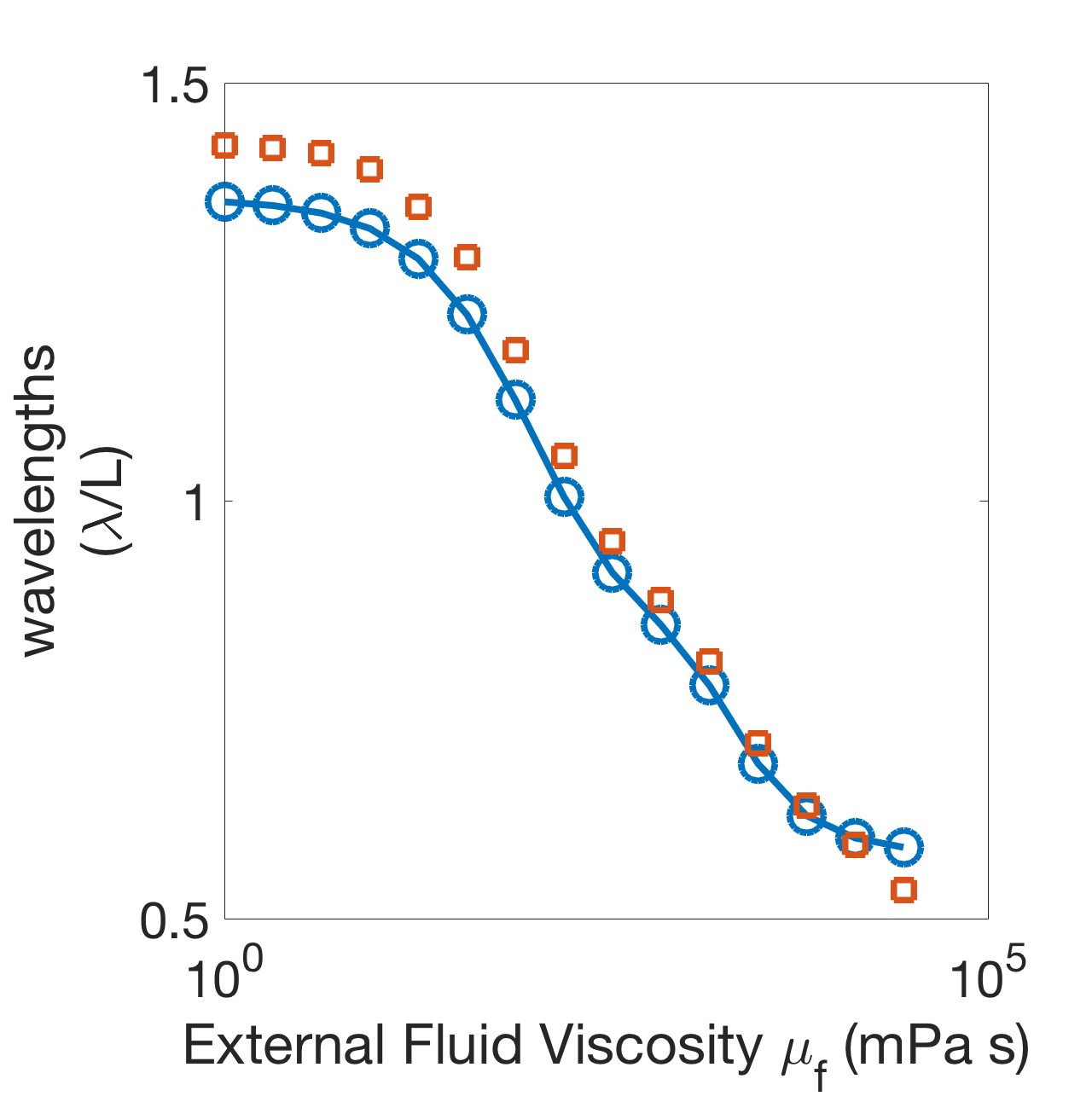}
\centering(b)\includegraphics[width=.465\textwidth]{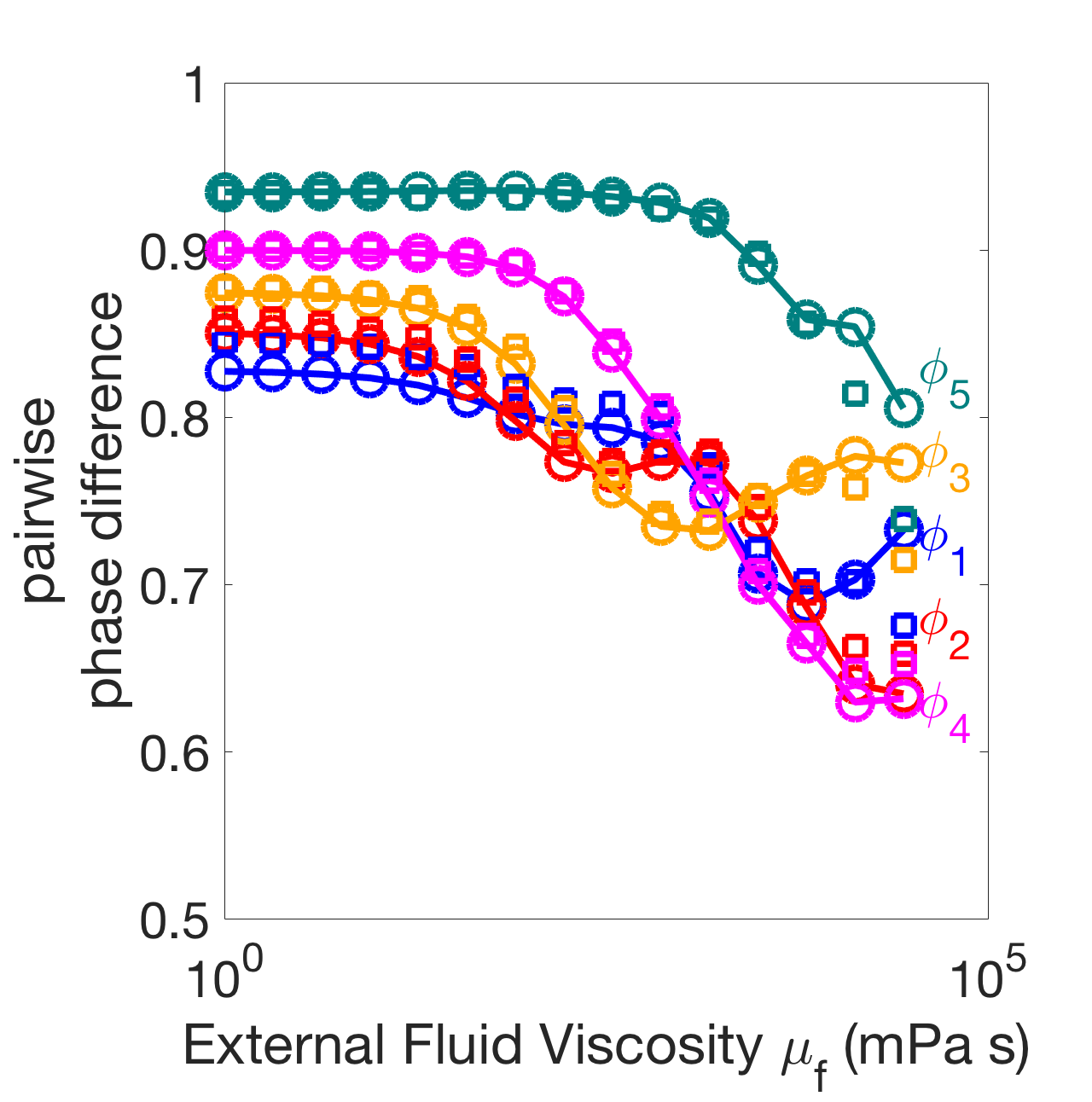}
\caption{(a) The wavelengths generated by the six-oscillator phase model (blue line with circles) and neuromechanical model (red squares) as a function of external fluid viscosity $\mu_f$ for $\varepsilon_p = 0.05$ and $\varepsilon_g = 0.017$. The wavelength is set by the competition between the mechanical and neural coupling. (b)  The phase differences between neighboring oscillator modules in the six-oscillator phase model (lines with circles) and neuromechanical model (squares) as a function of external fluid viscosity $\mu_f$ for $\varepsilon_p = 0.05$ and $\varepsilon_g = 0.017$. Similar to the two-oscillator case, the stable phase differences here are set by the competition between mechanical and neural coupling.  When neural coupling dominates, the stable phase differences are spread out around 0.9, and when mechanical coupling dominates, the stable phase differences move towards antiphase, i.e., closer to 0.5 phase difference, but with strong boundary effects.
}
\label{6box_mech_adapt}
\end{figure}

 Figure \ref{6box_mech_adapt}(a) shows the wavelengths for the six-oscillator phase model (line, circles) and neuromechanical model (crosses) as a function of external fluid viscosity $\mu_f$ for $\varepsilon_p = 0.05$ and $\varepsilon_g = 0.017$ (these coupling strengths were chosen so that the water-wavelength is approximately 1.5).  The wavelengths were computed by equation \ref{wvln_per_bodylength_defining_wvln_nonconstant} in Appendix \ref{appendix_defining_wvln}.  The phase model and the neuromechanical model agree quantitatively even at high $\mu_f$, where the mechanical coupling strength is several orders of magnitude stronger.
 Figure \ref{6box_mech_adapt}(b) shows the stable phase differences between neighboring modules in the six-oscillator phase model (lines, circles) and neuromechanical model (crosses) as a function of external fluid viscosity $\mu_f$.  Again, the phase model and the neuromechanical model are in quantitative agreement.  Furthermore, Figure \ref{6box_mech_adapt}(b) shows that increasing fluid viscosity affects the phase-locked states in the six-oscillator case in a similar way as in the two-oscillator case. When neural coupling dominates at low viscosity, the stable phase differences are spread out near 0.9, and as fluid viscosity increases, the mechanical coupling strength increases and the stable phase differences decrease towards antiphase. {\red However the phase differences do not reach pairwise-antiphase as in the two-oscillator case}.  The large variation between the phase differences across pairs of modules is due to the non-uniformity of coupling matrices $D_4^{-1}$, $W_g,$and $W_p$. The modules in the middle receive stronger mechanical coupling than the modules at the boundaries; the boundary modules receive less gap-junctional coupling because they have one fewer neighboring module; and the first module gets zero nonlocal proprioceptive feedback because it has no anterior neighboring module. 

\begin{figure}
\centering\includegraphics[width=\textwidth]{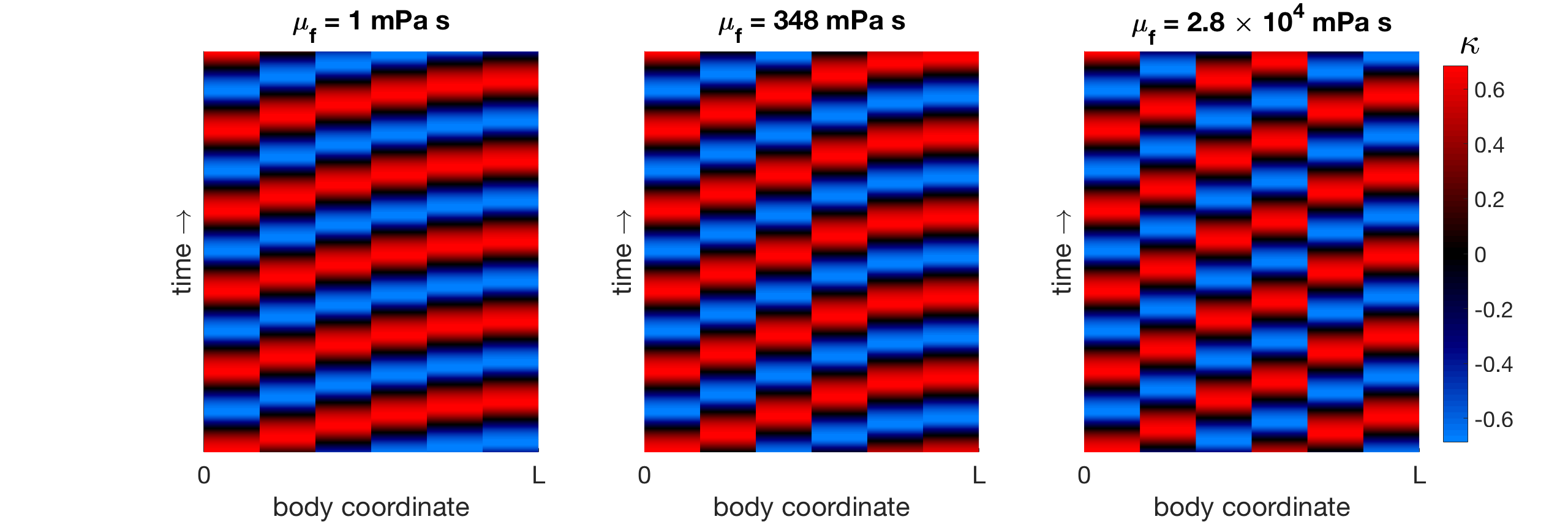}
\centering\includegraphics[width=\textwidth]{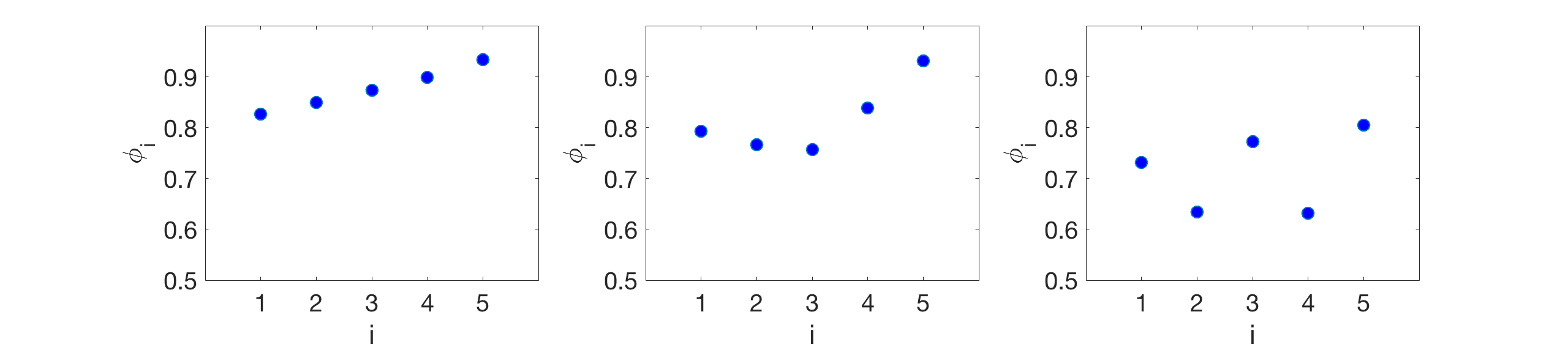}
\caption{\red Curvature kymographs generated from the pairwise phase differences $\phi_i$ of the six-oscillator phase model (from Figure \ref{6box_mech_adapt}(b)) at three selected fluid viscosities (low $\mu_f = 1$ mPa s, medium $\mu_f = 348$ mPa s, and high $\mu_f = 2.8\times10^4$ mPa s). The shortening of the wavelength is a direct consequence of the decreasing pairwise phase differences.
}
\label{kymos_for_6box_mech_adapt}
\end{figure}

The general trend of each phase difference between neighboring modules (decreasing from near-synchrony towards antiphase) underlies the wavelength trend of gait adaptation in Figure \ref{6box_mech_adapt}(a) in both the six-oscillator phase model and the neuromechanical model.  {\red Figure \ref{kymos_for_6box_mech_adapt} shows the curvature kymographs generated by the pairwise phase differences of the phase model (from Figure \ref{6box_mech_adapt}) at three selected fluid viscosities (low $\mu_f = 1$ mPa s, medium $\mu_f = 348$ mPa s, and high $\mu_f = 28$ Pa s).  The shortening of the wavelength seen in the curvature kymographs is a direct consequence of the decreasing pairwise phase differences.}
Thus, the results for the two-oscillator case in Section \ref{sect_two_osc_analysis} extend to the six-oscillator case: the decrease in wavelength in response to increasing fluid viscosity is the result of the corresponding increase in the relative strength of mechanical coupling, which decreases the phase differences between neighboring modules and yields shorter wavelengths.

\section{Discussion}\label{sect_discussion}

The analysis of the neuromechanical model presented here identifies a mechanism for gait adaptation to increasing fluid viscosity in \textit{C. elegans} forward locomotion.  We model the \textit{C. elegans} forward locomotion system as a chain of neuromechanical oscillators coupled by body mechanics, proprioceptive coupling, and gap-junctional coupling. Using the theory of weakly coupled oscillators, we exploit the modular structure of the forward locomotion system to analyze the relative contributions of the various coupling modalities.  We find that proprioceptive coupling between modules leads to a posteriorly-directed traveling wave with a characteristic wavelength. Gap-junction coupling between neural modules promotes synchronous activity (long wavelength), and mechanical coupling promotes a high spatial frequency (short wavelength).  The wavelength of \textit{C. elegans}' undulatory waveform is set by the relative strengths of these three coupling forms. As the external fluid viscosity increases, the mechanical coupling strength increases and therefore wavelength decreases, as observed experimentally.

% This analysis shows that gait adapation results from the competition between mechanical coupling and neural coupling.  The mechanical coupling through the fluid drag and body mechanics tends to pull the oscillator modules into antiphase. As the external fluid viscosity is increased, mechanical coupling is strengthened and the phase-locked outcome shifts towards antiphase between pairs, which results in a shortening of the wavelength of undulation.

By tuning only a few coupling parameters, the model can robustly capture the gait adaptation seen in experiments \cite{Berri_2009,Fang-Yen:2010aa,Sznitman2010} over a wide range of mechanical parameters. The robustness of the model is of particular importance because the experimental measurements of mechanical body parameters vary widely.
 % with values depending on experimental set-up, on the model and interpretation, or even on the activity of the muscles \cite{Denham_2018}.
Our model suggests relationships between the parameters that need to hold in order to get the appropriate coordination and wavelength trend. In particular, the effective mechanical body timescale $\tau_b = \mu_b/k_b$ (the ratio of body viscosity to stiffness) plays a key role.
% that it is the effective mechanical body timescale $\tau_b = \mu_b/k_b$, not necessarily the body stiffness $k_b$ or viscosity $\mu_b$, that plays a key role in coordination.  
Our model yields the correct coordination trend across the entire range of reported mechanical parameters, provided that $\tau_b$ is in the range \mbox{$0.07-1$ s}.
% Our model also points out the importance of timescale ordering in the locomotion system. 
Furthermore, the muscle activity timescale $\tau_m$ must generally be shorter than the effective body mechanics timescale $\tau_b$. In other words, the system must generate contractile forces faster than the body responds, otherwise, there will not be a traveling wave of neuromechanical activity and therefore no effective locomotion for high external fluid viscosities.  
{\red Like previous models \cite{Boyle:2012aa}, the neural timescale $\tau_n$ is assumed to be the smallest, and changing its value will likely not significantly affect the dynamics unless it is increased by an order of magnitude. However, the relative strengths of the neural coupling parameters (proprioceptive coupling strength $\eps_p$ and gap-junctional coupling strength $\eps_g$) are directly responsible for the appropriate coordination in the model.  In particular, the long-wavelengths in the model are only achievable with sufficiently strong gap-junctional coupling.}
% The neural activity being the fastest process was a key assumption to the model of \cite{Boyle:2012aa}; however, the relative scales of the other timescales were not explored.  
%Here, the role of the mechanical body timescale is explained independent of the choice of body stiffness or body viscosity measurement. 
%talk about mechanical parameters being hard to measure - model dependent, interpretation-dependent
% An upper bound on the body mechanics timescale $\tau_b \leq 1$ s is shown to be necessary (in this model) to obtain the correct frequency in water, given physiological ranges for the other timescales.  

%a sentence to start this off to say that our model is similar in structure to the Boyle model
{\red \textit{C. elegans} gait adaptation is marked by a decrease in both the wavelength and  undulation frequency with increasing fluid viscosity \cite{Berri_2009,Fang-Yen:2010aa,Sznitman2010}.
Our model captures the quantitative trend in wavelength but only the qualitative trend in frequency ($1.7-1.6$ Hz as opposed to the range  $1.7-0.3$ Hz given in Fang-Yen et al. \cite{Fang-Yen:2010aa}). Our model is similar in structure to modeling work by Boyle et al. \cite{Boyle:2012aa}  and Denham et al. \cite{Denham_2018}. However, both Boyle et al. \cite{Boyle:2012aa} and Denham et al. \cite{Denham_2018} are able to capture both wavelength and frequency adaptation in their models.  A key difference between our model, Boyle et al. \cite{Boyle:2012aa}, and Denham et al. \cite{Denham_2018} is the role of body viscosity.  In our model, the body viscosity is responsible for setting the timescale and frequency.  Denham et al. \cite{Denham_2018} neglects body viscosity, so the fluid viscosity likely sets the frequency. 
 % Our effective mechanical coupling strength $\eps_m$ is inversely proportional to body viscosity, so the limit of body viscosity $\mu_b \to 0$ gives $\eps_m\to\infty$, which pushes the Denham model out of the weak coupling limit.  Outside of this weak-coupling limit, the external fluid viscosity can directly set the mechanical timescale and frequency adaptation is possible.
  On the other hand, Boyle et al. \cite{Boyle:2012aa} includes an active viscosity component that depends on muscle activation, which likely provides a mechanism for frequency adaptation.  In our model, body viscosity is constant, independent of muscle activation, so including an active viscosity component might be necessary for frequency adaptation.}

%On the other hand, the description of the muscle dynamics and body mechanics are more complex in the Boyle et al. model \cite{Boyle:2012aa}.  Boyle et al. \cite{Boyle:2012aa} also captures gait adaptation, and the large number of parameters and variables of the model allows it to more closely match the wavelengths, amplitudes, and undulation frequencies observed experimentally. However, the complexity of the model also limits the ability to systematically assess the relative roles of body mechanics and proprioception in coordination.   

{\red Other differences between our model and previous models \cite{Boyle:2012aa,Denham_2018} is the sign, directionality, and extent of nonlocal proprioception, the role of gap junctions, {\blue and the number of modules}.}  The directionality of proprioception in Boyle et al. \cite{Boyle:2012aa} and Denham et al. \cite{Denham_2018} is consistent with the directionality of undifferentiated processes extending posteriorly from the B-class neurons, which have postulated to be responsible for {\red proprioception \cite{NIEBUR199351,White:1986aa,Zhen:2015aa}. However, the extent of proprioception in these models needs to be much longer than this proposed biological mechanism \cite{Boyle:2012aa,Denham_2018}}.  We take the directionality of proprioception to be consistent with the functional directionality suggested by the experiments of Wen et al. \cite{Wen:2012aa} {\red and shown to produce locomotion in a neuromechanical model by Izquierdo and Beer \cite{Izquierdo_2018}}.  Note that symmetry arguments can be made that reversing both the sign and direction of the nonlocal proprioception will not change the behavior of the models, as Denham et al. \cite{Denham_2018} points out.  {\red When we reverse the sign and direction of the nonlocal proprioception in our model, we obtain qualitatively similar results (see Appendix \ref{ant_prop_section}). The full range of wavelength adapatation can still be achieved, and in particular the (two-module) weakly-coupled theory predicts that the change is only a small shift in the promoted phase-locked phase-difference of the proprioceptive coupling mode.}  The extent of proprioception in Boyle et al. \cite{Boyle:2012aa} is over half a bodylength, and Denham et al. \cite{Denham_2018} showed that
the larger the proprioceptive range, the longer the undulatory wavelength {\red in their model}.  We considered only nearest-neighbor proproception {\red because Wen et al. \cite{Wen:2012aa} found that proprioception likely extended anteriorly roughly 200 $\mu$m or $1/5$ of the bodylength, similar to the length of one module in our model.  This is} sufficient to achieve the long-wavelength undulations in water because of our inclusion of gap-junctional coupling that promotes synchrony between the modules and thus long wavelengths.  {\red This demonstration that gap-junctions provides a mechanism for long-wavelength coordination (in the absence of long-range proprioception) is a novel contribution of this model. {\blue Furthermore in \cite{johnson2020neuromechanical}, we explored the effect of changing the number of modules on coordination, and we showed that gait adaptation was robustly attainable with different neural coupling strengths.  Thus, the main conclusions of this paper do not depend on the number of modules considered.
}}

{\red Previous models which made use of more complicated mechanical and muscle models were also able to capture the swimming speed \cite{Boyle:2012aa,Izquierdo_2018,Denham_2018}.  Here, the main purpose of our work is to determine how gait adaptation emerges from the coordination of the neuromechanical modules.
%so we do not investigate the swimming speed.  %Other models examined the oscillation amplitudes and swimming speed like some of the previous models \cite{Boyle:2012aa,Izquierdo_2018,Denham_2018}. 
Capturing the right shapes and swimming speed may involve including some of the complexities of previous models.
% and other parameters that remained fixed in this work (e.g., the feedback strengths, which we explored more thoroughly in \cite{johnson2020neuromechanical}). 
% Computing the swimming speed requires both drag forces normal to the body and tangential, however, in the limit of small amplitude only normal forces show up in the curvature equations in the leading order, and tangential forces show up at higher orders.  The shape is determined, to leading order, by the normal forces.
Because we examine coordination using curvature in the small-amplitude limit, our model includes only the normal forces while tangential forces show up at higher order.  Thus the tangential forces, which are important for determining swimming speed, do not affect the curvature at leading order \cite{Thomases:2014aa}.}
% Given this focus on coordination, our fluid model includes only normal forces explicitly.  so we do not include them explicitly in our model, unlike some of the previous models \cite{Boyle:2012aa,Izquierdo_2018,Denham_2018}.}

%phrase this is terms of an even bigger question - Are there CPGs or emergent oscillations?
Our model assumes that the undulatory gait emerges from a chain of neuromechanical oscillators coupled by both body mechanics and neural connectivity.  However, there are several other hypotheses for how the undulatory gait is generated and coordinated \cite{Gjorgjieva_2014}: (1) a separate head circuit contains a CPG that drives the propogated bending wave along the body, and (2) a network of coupled CPGs generates and coordinates the bending wave in a feed-forward manner.  Modeling work by Olivares et al. \cite{Olivares_2018} shows that the anatomical structure of the neural circuitry of \textit{C. elegans} can be tuned to produce CPG-driven locomotion. However, there is no experimental evidence to date for such spontaneous isolated neural activity \cite{Cohen_2014,Zhen:2015aa}. Furthermore, recent experiments by \cite{Fouad:2018aa} showed that \textit{C. elegans} is capable of decoupled ``two-frequency undulations''.  By suppressing neural activity in the neck region, the head and body can undulate seemingly independent of one another at different frequencies (the head slower and the body faster).   This evidence supports the presence of multiple neural or neuromechanical oscillators.

In the present study, the theory of weakly coupled oscillators is used to identify the roles of the various coupling modalities in generating coordination for forward locomotion in \textit{C. elegans}.   
% Specifically, we explain how gait adaptation can arise from alterations in the mechanical (intermodular) coupling in response to changes in the viscosity of the fluid environment. 
The phase models derived by the theory of weakly coupled oscillators capture the influence of one oscillating module on another through the interaction functions $H(\phi)$, which are convolution-like integrals of the coupling input and the corresponding phase response function $Z(t)$ of the individual modules.
Therefore, our findings could be validated by experimentally measuring the phase-response curves of the neuromechanical circuit \cite{netoff_etal}.
This could be achieved using a combination of optogenetic techniques and mechanical stimuli to perturb the system \cite{Fouad:2018aa,hongfei_conf_abstract,Ji2020.06.22.164939,Wen:2012aa}.
Note also that the structure of the phase equations could be exploited to further dissect out the biophysical mechanisms that underlie coordination of the undulatory motion of \textit{C. elegans}.  Because the shapes of the PRCs and the coupling signals combine to determine the interaction functions, a systematic analysis of how cellular and synaptic dynamics \cite{Zhang_2013}, muscle properties, and body mechanics shape the PRCs and coupling signals would provide further insight into the integrated neuromechanical mechanisms underlying the generation and coordination of locomotion.

% By systematically examining how key processes -- of the neural circuit (cellular and synaptic dynamics), muscles properties, and body mechanics – shape the PRCs (response to input to the various modalities)  and the coupling signal (output that is impinging on other modules) on the neuromechanical modules.

% The theory of weakly coupled oscillators/structure of the phase equations could be exploited to obtain further insight into the biophysical mechanisms that underlie coordination of the undulatory motion of C. elegans. [theory provides a framework for dissecting out the mechanisms/roles of … key processes -- of the neural circuit (cellular and synaptic dynamics), muscles properties, proprioceptive feedback, and body mechanics (including the influence of the environment?).]

% The phase-locked states/coordination rely on the shapes of the PRCs and the coupling input of the NM oscillators …

% 
% … [concluding sentence linking to general statements made in intro] …

%look up gait adaptation in stick insects, cockroaches, and salamanders - INTRO?
%true neuromechanical oscillations

\appendix
% \section{Appendix}

% \subsection{Coupled Modules Schematic}

% {\red Figure \ref{wiring_diag_2modules_coupling} shows the coupling between two neuromechanical modules.  The modules are coupled mechanically through the body and neurally through both proprioceptive feedback and gap-junctions. }

% \begin{figure}
% \centering\includegraphics[width=.9\textwidth]{wiring_diagram_2modules.png}
% \caption{\red Coupling between two neuromechanical modules. The modules are coupled mechanically through the body and neurally through both proprioceptive feedback and gap-junctions.  The B-class motorneurons receive excitatory proprioceptive feedback from the anterior body segment. The VB (DB) neurons are also coupled symmetrically via gap-junctions with their nearest neighbors of the same class in the next modules.
% }
% \label{wiring_diag_2modules_coupling}
% \end{figure}

\section{Defining Wavelength}\label{appendix_defining_wvln}

\paragraph{Constant Wavespeed} For a wavelength of undulation in the neuromechanical model traveling front-to-back at constant speed, the phase is defined as 
\begin{equation}
\theta(x,t) = \qty(\dfrac{t}{T} - \dfrac{x}{\lambda}) \text{ mod 1}
\end{equation}
The phase corresponding to module $k$ ($k=1,\dots,6$) centered at body position $x = \ell(k - 1/2)$ is
\begin{equation}
\theta_k = \qty(\dfrac{t}{T} - \dfrac{\ell}{\lambda}\qty(k - \frac{1}{2})) \text{ mod 1},
\label{theta_k_defining_wvln}
\end{equation}
where $T$ is the oscillator period.
Thus, the constant phase difference $\phi^*$ is
\begin{equation}
\phi^* = \qty(\theta_{k+1} - \theta_k) \text{ mod 1} = \qty(- \dfrac{\ell}{\lambda}) \text{ mod 1} = 1-\dfrac{\ell}{\lambda}.
\label{phi_defining_wvln}
\end{equation}
and the constant wavelength is
\begin{equation}
\lambda  = \dfrac{\ell}{1-\phi^*}.
\label{wvln_defining_wvln}
\end{equation}
For the neuromechanical model, $\ell = L/6$, so the wavelength (normalized by  bodylength) is
\begin{equation}
\dfrac{\lambda}{L}  = \dfrac{1}{6\qty(1-\phi^*)}.
\label{app_wvln_per_bodylength_defining_wvln}
\end{equation}

\paragraph{Nonconstant Wavespeed}
The non-uniform phase differences $\phi_k = \theta_{k+1}-\theta_{k}$ ($k=1,\dots,5$) between modules are used to define an effective wavelength of undulation when the wavespeed is nonconstant.  The distance between the center of the first and center of the sixth module is $5/6L$, and the phase difference between them is $\sum_{k=1}^5(1-\phi_k)$.
This gives an effective wavelength (normalized by 5/6 bodylengths)
\begin{equation}
\dfrac{\lambda}{(5/6)L}  = \dfrac{1}{\sum_{k=1}^{5}\qty(1-\phi_k)},
\label{wvln_per_bodylength_defining_wvln_nonconstant_v1}
\end{equation}
so the wavelength (normalized by bodylength) is
\begin{equation}
\dfrac{\lambda}{L}  = \dfrac{1}{6\sum_{k=1}^{5}\qty(1-\phi_k)/5}.
\label{wvln_per_bodylength_defining_wvln_nonconstant}
\end{equation}
Note that this is equivalent to the constant phase difference wavelength (equation \ref{app_wvln_per_bodylength_defining_wvln}) using the \textit{average phase difference} between the modules as the constant phase difference $\phi^*$, i.e., with
\begin{equation}
1 - \phi^* = \dfrac{\sum_{k=1}^{5}1-\phi_k}{5}.
\end{equation}

For the neuromechanical model results, first the phase differences $\phi_k$ between the modules were computed, then the wavelength was computed according to equation \ref{wvln_per_bodylength_defining_wvln_nonconstant} above.

\section{Derivation of Mechanical Parameters}\label{appendix_deriv_mech_params}
First, the bending modulus $k_b = EI_c$ of the cuticle of the worm was determined, where $E$ is the Young's modulus and $I_c$ is the second moment of area of the cuticle. The nematode body can be thought of as a pressurized, fluid-filled tube or modeled as an annular cylinder as in Cohen and Ranner \cite{cohen_ranner_2017}, so the only elasticity in the body is that of the cuticle.  To approximate the second moment of area of the cuticle, $I_c$, note that the cuticle width $r_{\text{cuticle}} = 0.5$ $\mu$m is much smaller than the average worm radius $R=40$ $\mu$m.  Following Cohen and Ranner \cite{cohen_ranner_2017},

\begin{equation} I_c = 2\pi R^3 r_{\text{cuticle}} = 2.0 \times 10^{-7} \text{mm}^4. \label{moment_of_cuticle_area}\end{equation}

The Young's modulus $E$ has been estimated to be as small as $E = 3.77 \pm 0.62$ kPa \cite{Sznitman2010} or as large as $E = 13$ MPa \cite{Fang-Yen:2010aa}.  Backholm et al. \cite{Backholm_2013} gives a range of $110 \pm 30$ kPa $\leq E \leq 1.3 \pm 0.3$ MPa.
Using these estimates, we explore the range of bending moduli $k_b = EI_c = 7.53 \times 10^{-10} - 2.6 \times 10^{-6}$ N(mm)$^2$.

The \textit{cuticle} viscosity has been estimated as $5 \times 10^{-16}$ Nm$^2$s \cite{Fang-Yen:2010aa}.  
The internal \textit{tissue} viscosity has been estimated to be constant and negative (energy-generating) as $c_d = -177.1 \pm 15.2$ Pa s so that $\mu_b = c_d I = - 1.7 \times 10^{-11}$ N(mm)$^2$s  \cite{Sznitman2010} by a model fit, however this includes the active muscle components. 
Backholm et al. \cite{Backholm_2013} estimated the range $c_d \in \qty[1\times 10^{2}, 1\times 10^{4}]$ Pa s, so that the effective viscosity is $c_d I \in  \qty[2\times 10^{-11}, 2\times 10^{-9}]$ N(mm)$^2$s.  These experiments used different techniques and models for viscosity, so likely have different effects lumped into the viscosity parameter.  In order to explore the range of effective body mechanics timescales $\tau_f = \mu_b / k_b = 0.001 - 5$ s, we use the range of body viscosities $\mu_b = 5 \times 10^{-10} - 1.3 \times 10^{-7}$ N(mm)$^2$s in our model.

% In order to approximate the correct beating frequency in water and to maintain the separation of timescales described in Section 2.2, we set the internal viscosity in our model to be $\mu_b = 1.3\times10^{-7}$ N(mm)$^2$s, which we note is considerably larger than previous estimates.  However, unlike in the case of body elasticity, the viscosity of the whole-body is not wholly captured by the cuticle viscosity.  Previous estimates for internal viscosity have been for cuticle viscosity \cite{Fang-Yen:2010aa}, while we argue that the viscosity that effectively sets the timescale of the model here is a whole-body internal viscosity that includes the effects of the cuticle, muscles, tissue, and internal fluid.  For transparency, we also include results for our model for much smaller internal viscosities $\mu_b$, and note that this is a separate regime that does not display the same behavior as that described in the bulk of this paper.

Following previous modeling procedures \cite{Fang-Yen:2010aa,cohen_ranner_2017}, the normal drag coefficient $C_N$ of a slender body with length $L = 1$ mm and (average) radius $R = 40 \ \mu$m in a solution with viscosity $\mu_f$ is
\begin{equation}
C_N = \dfrac{ 4 \pi \mu_f}{\ln(L/R)+0.5} = \alpha \mu_f \approx 3.4 \mu_f.\label{drag_coeff_Eqn}\end{equation}

{\red 
\section{Posterior-to-Anterior Proprioceptive Coupling}\label{ant_prop_section}

When we reverse the sign and direction of the nonlocal proprioception in our model, we obtain qualitatively similar results.  That is, we replace the anterior-to-posterior proprioceptive coupling matrix $W_p$ (equation \ref{w_p_matrix}) with the opposite-signed, posterior-to-anterior proprioceptive coupling matrix
\begin{equation}
W_p = \qty(\begin{matrix}
0 & -1 &  & &  \\
 & 0 & -1 &  & \\
& &  \ddots & \ddots & \\
& & & 0 & -1\\
& & & & 0
\end{matrix}).\label{w_p_matrix_backwards}
\end{equation}
In the phase model, this yields the proprioceptive G-function $G_p(\phi) = H_p(\phi)$ (as opposed to $G_p(\phi) = H_p(-\phi)$ with opposite-signed anterior-to-posterior proprioception).

Figure \ref{gfns_plot_proprio}(a) shows that for particular parameters ($\mu_b = 1.3\times 10^{-8}$, $k_b = 2.6\times10^{-8}$, $\eps_p = 0.00068$, and $\eps_g = 0.00017$), the full range of wavelength adapatation can still be achieved.  Note that these are smaller coupling strengths than in our standard model (this is likely due to the nonlocal proprioception here being of the same sign as the local proprioception, so it must be weaker to have less of an effect on the dynamics).  In addition, the two-module weakly-coupled theory predicts that the change is only a small shift in the promoted phase-locked phase-difference of the proprioceptive coupling mode.
Figure \ref{gfns_plot_proprio}(b-c) shows the G-functions and corresponding phase-locked states of the proprioceptive coupling modalities.
For negative, anterior-to-posterior proprioceptive coupling, the stable phase-locked state is an intermediate phase-difference ($\phi^* \approx 0.75$, Figure \ref{gfns_plot_proprio}(b)), so the first oscillator leads the second (a front-to-back wave). Similarly, for positive, posterior-to-anterior proprioceptive coupling, the stable state is an intermediate phase-difference ($\phi^*\approx 0.82$,  Figure \ref{gfns_plot_proprio}(c)), so the first oscillator still leads the second (a front-to-back wave). Thus the change is only a small shift in the promoted phase-locked phase-difference but not a change in the direction of the traveling wave.  Based on this analysis, the full neuromechanical model is likely to behave qualitatively the same for either proprioceptive mode, though a full parameter study should be done to say so definitively.

\begin{figure}
\centering(a)\includegraphics[width=.7\textwidth]{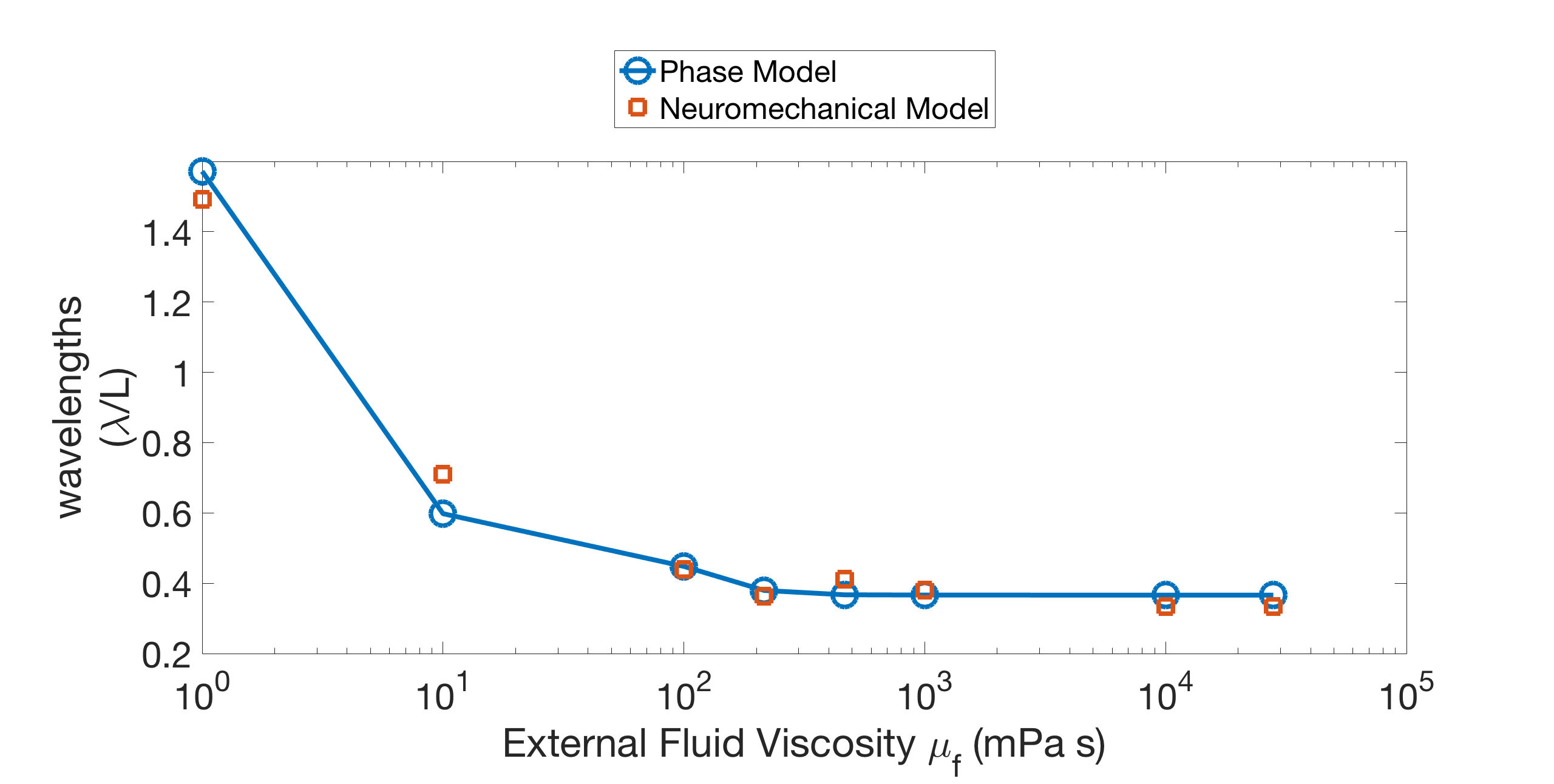}\\
\centering
(b)\includegraphics[width=.3\textwidth]{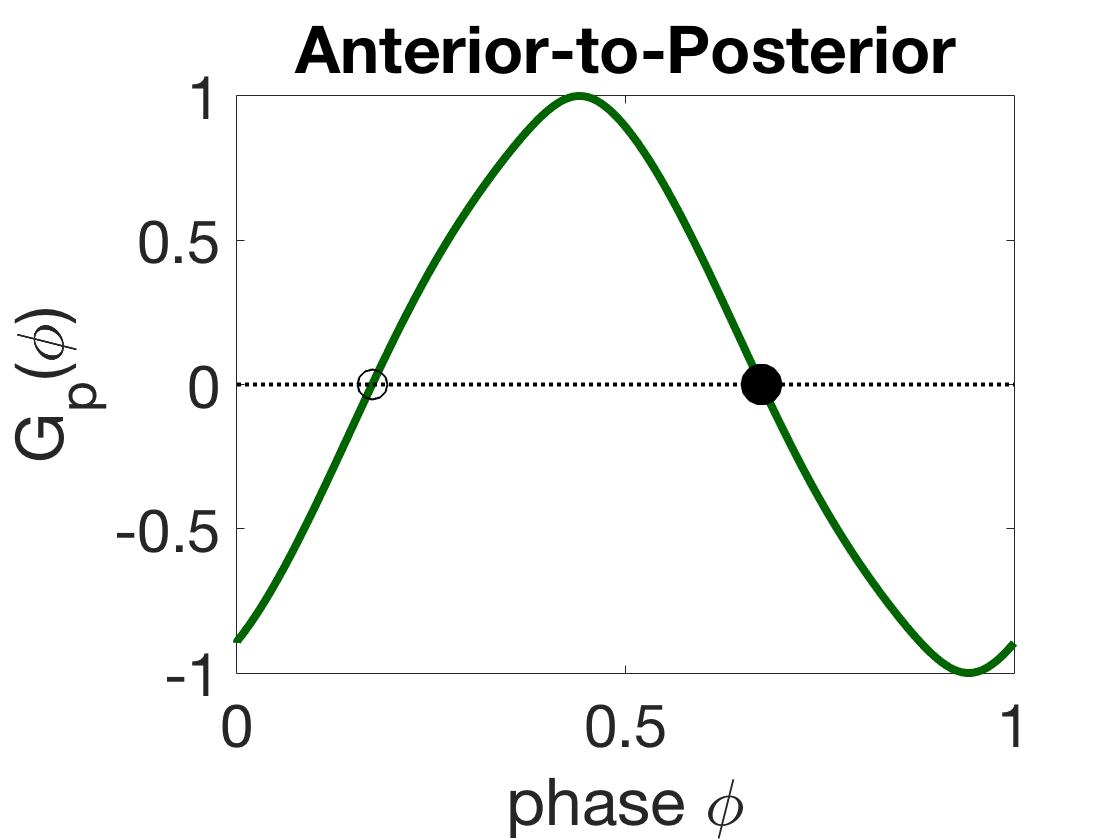}
(c)\includegraphics[width=.3\textwidth]{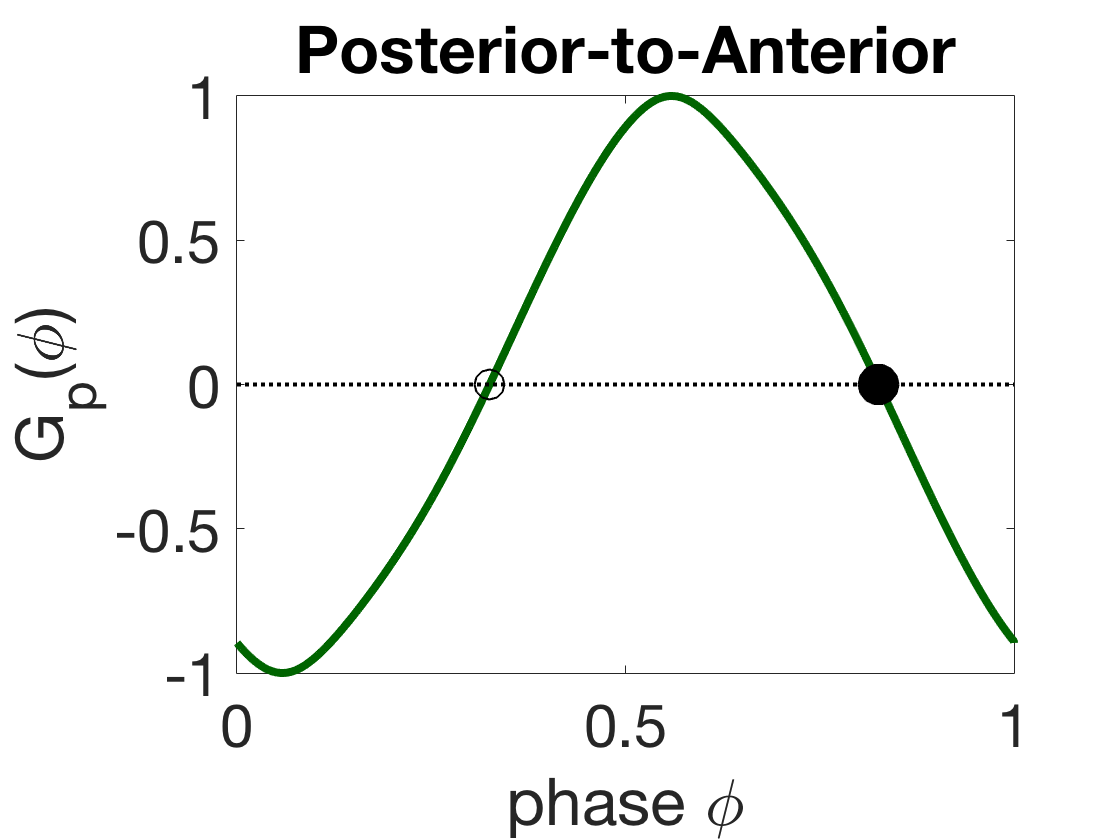}
\caption{(a) A sample wavelength trend for the six-module neuromechanical and phase model with positive, posterior-to-anterior proprioception. (b) The proprioceptive coupling G-function for negative, anterior-to-posterior nonlocal proprioception.  This mode promotes a phase-wave since $G_p(.75) = 0$ and $G_p'(.75)<0$; and (c) The proprioceptive coupling G-function for positive, posterior-to-anterior nonlocal proprioception.  This mode still promotes a phase-wave with the same directionality since $G_g(0.82) = 0$ and $G_g'(0.82)<0$.}
\label{gfns_plot_proprio}
\end{figure}
}

\section*{Acknowledgments}

The authors would like to thank Netta Cohen for helpful discussions related to this work, {\blue as well as the reviewers for their substantial comments and suggestions}. The work of RDG was partially supported by NSF grant DMS-1664679.

\bibliography{paper_1.bib}{}

\begin{thebibliography}{10}

\bibitem{ayali2015comparative}
{\sc A.~Ayali, A.~Borgmann, A.~B{\"u}schges, E.~Couzin-Fuchs, S.~Daun-Gruhn,
  and P.~Holmes}, {\em The comparative investigation of the stick insect and
  cockroach models in the study of insect locomotion}, Current Opinion in
  Insect Science, 12 (2015), pp.~1--10.

\bibitem{Backholm_2013}
{\sc M.~Backholm, W.~S. Ryu, and K.~Dalnoki-Veress}, {\em Viscoelastic
  properties of the nematode caenorhabditis elegans, a self-similar,
  shear-thinning worm}, Proceedings of the National Academy of Sciences, 110
  (2013), pp.~4528--4533, \url{https://doi.org/10.1073/pnas.1219965110},
  \url{http://dx.doi.org/10.1073/pnas.1219965110}.

\bibitem{Berri_2009}
{\sc S.~Berri, J.~H. Boyle, M.~Tassieri, I.~A. Hope, and N.~Cohen}, {\em
  Forward locomotion of the nematode c. elegans is achieved through modulation
  of a single gait}, HFSP Journal, 3 (2009), pp.~186--193,
  \url{https://doi.org/10.2976/1.3082260},
  \url{http://dx.doi.org/10.2976/1.3082260}.

\bibitem{Borgmann_2007}
{\sc A.~Borgmann, H.~Scharstein, and A.~B{\"u}schges}, {\em Intersegmental
  coordination: Influence of a single walking leg on the neighboring segments
  in the stick insect walking system}, Journal of Neurophysiology, 98 (2007),
  pp.~1685--1696, \url{https://doi.org/10.1152/jn.00291.2007},
  \url{http://dx.doi.org/10.1152/jn.00291.2007}.

\bibitem{Boyle:2012aa}
{\sc J.~H. Boyle, S.~Berri, and N.~Cohen}, {\em Gait modulation in c. elegans:
  An integrated neuromechanical model}, Front Comput Neurosci, 6 (2012), p.~10,
  \url{https://doi.org/10.3389/fncom.2012.00010}.

\bibitem{Bryden_2008}
{\sc J.~Bryden and N.~Cohen}, {\em Neural control of caenorhabditis elegans
  forward locomotion: the role of sensory feedback}, Biological Cybernetics, 98
  (2008), pp.~339--351, \url{https://doi.org/10.1007/s00422-008-0212-6},
  \url{http://dx.doi.org/10.1007/s00422-008-0212-6}.

\bibitem{chalfie1985neural}
{\sc M.~Chalfie, J.~E. Sulston, J.~G. White, E.~Southgate, J.~N. Thomson, and
  S.~Brenner}, {\em The neural circuit for touch sensitivity in caenorhabditis
  elegans}, Journal of Neuroscience, 5 (1985), pp.~956--964.

\bibitem{Cohen_1992}
{\sc A.~H. Cohen, G.~Bard~Ermentrout, T.~Kiemel, N.~Kopell, K.~A. Sigvardt, and
  T.~L. Williams}, {\em Modelling of intersegmental coordination in the lamprey
  central pattern generator for locomotion}, Trends in Neurosciences, 15
  (1992), pp.~434--438, \url{https://doi.org/10.1016/0166-2236(92)90006-t},
  \url{http://dx.doi.org/10.1016/0166-2236(92)90006-T}.

\bibitem{cohen_ranner_2017}
{\sc N.~Cohen and T.~Ranner}, {\em A new computational method for a model of c.
  elegans biomechanics: Insights into elasticity and locomotion performance},
  arXiv:1702.04988v1 [physics.bio-ph],  (2017).

\bibitem{Cohen_2014}
{\sc N.~Cohen and T.~Sanders}, {\em Nematode locomotion: dissecting the
  neuronal--environmental loop}, Current Opinion in Neurobiology, 25 (2014),
  pp.~99--106, \url{https://doi.org/10.1016/j.conb.2013.12.003},
  \url{http://dx.doi.org/10.1016/j.conb.2013.12.003}.

\bibitem{DengENEURO.0241-20.2020}
{\sc L.~Deng, J.~Denham, C.~Arya, O.~Yuval, N.~Cohen, and G.~Haspel}, {\em
  Inhibition underlies fast undulatory locomotion in c. elegans}, eNeuro,
  (2020), \url{https://doi.org/10.1523/ENEURO.0241-20.2020},
  \url{https://www.eneuro.org/content/early/2020/12/14/ENEURO.0241-20.2020},
  \url{https://arxiv.org/abs/https://www.eneuro.org/content/early/2020/12/14/ENEURO.0241-20.2020.full.pdf}.

\bibitem{Denham_2018}
{\sc J.~E. Denham, T.~Ranner, and N.~Cohen}, {\em Signatures of proprioceptive
  control in caenorhabditis elegans locomotion}, Philosophical Transactions of
  the Royal Society B: Biological Sciences, 373 (2018), p.~20180208,
  \url{https://doi.org/10.1098/rstb.2018.0208},
  \url{http://dx.doi.org/10.1098/rstb.2018.0208}.

\bibitem{Fang-Yen:2010aa}
{\sc C.~Fang-Yen, M.~Wyart, J.~Xie, R.~Kawai, T.~Kodger, S.~Chen, Q.~Wen, and
  A.~D.~T. Samuel}, {\em Biomechanical analysis of gait adaptation in the
  nematode caenorhabditis elegans}, Proc Natl Acad Sci U S A, 107 (2010),
  pp.~20323--8, \url{https://doi.org/10.1073/pnas.1003016107}.

\bibitem{Fouad:2018aa}
{\sc A.~D. Fouad, S.~Teng, J.~R. Mark, A.~Liu, P.~Alvarez-Illera, H.~Ji, A.~Du,
  P.~D. Bhirgoo, E.~Cornblath, S.~A. Guan, and C.~Fang-Yen}, {\em Distributed
  rhythm generators underlie caenorhabditis elegans forward locomotion}, Elife,
  7 (2018), \url{https://doi.org/10.7554/eLife.29913}.

\bibitem{fuchs2011intersegmental}
{\sc E.~Fuchs, P.~Holmes, T.~Kiemel, and A.~Ayali}, {\em Intersegmental
  coordination of cockroach locomotion: adaptive control of centrally coupled
  pattern generator circuits}, Frontiers in neural circuits, 4 (2011), p.~125.

\bibitem{Gjorgjieva_2014}
{\sc J.~Gjorgjieva, D.~Biron, and G.~Haspel}, {\em Neurobiology of
  caenorhabditis elegans locomotion: Where do we stand?}, BioScience, 64
  (2014), pp.~476--486, \url{https://doi.org/10.1093/biosci/biu058},
  \url{http://dx.doi.org/10.1093/biosci/biu058}.

\bibitem{Haspel14611}
{\sc G.~Haspel and M.~J. O{\textquoteright}Donovan}, {\em A perimotor framework
  reveals functional segmentation in the motoneuronal network controlling
  locomotion in caenorhabditis elegans}, Journal of Neuroscience, 31 (2011),
  pp.~14611--14623, \url{https://doi.org/10.1523/JNEUROSCI.2186-11.2011},
  \url{http://www.jneurosci.org/content/31/41/14611},
  \url{https://arxiv.org/abs/http://www.jneurosci.org/content/31/41/14611.full.pdf}.

\bibitem{doi:10.1137/S0036144504445133}
{\sc P.~Holmes, R.~J. Full, D.~Koditschek, and J.~Guckenheimer}, {\em The
  dynamics of legged locomotion: Models, analyses, and challenges}, SIAM
  Review, 48 (2006), pp.~207--304,
  \url{https://doi.org/10.1137/S0036144504445133},
  \url{https://doi.org/10.1137/S0036144504445133},
  \url{https://arxiv.org/abs/https://doi.org/10.1137/S0036144504445133}.

\bibitem{Izquierdo_2018}
{\sc E.~J. Izquierdo and R.~D. Beer}, {\em From head to tail: a neuromechanical
  model of forward locomotion in caenorhabditis elegans}, Philosophical
  Transactions of the Royal Society B: Biological Sciences, 373 (2018),
  p.~20170374, \url{https://doi.org/10.1098/rstb.2017.0374},
  \url{http://dx.doi.org/10.1098/rstb.2017.0374}.

\bibitem{hongfei_conf_abstract}
{\sc H.~Ji, A.~D. Fouad, and C.~Fang-Yen}, {\em A novel model of bending wave
  generation supports a threshold-switch mechanism in the c. elegans motor
  circuit}, in the Proceedings of the 22nd International C. elegans Conference,
  2019.

\bibitem{Ji2020.06.22.164939}
{\sc H.~Ji, A.~D. Fouad, S.~Teng, A.~Liu, P.~Alvarez-Illera, B.~Yao, Z.~Li, and
  C.~Fang-Yen}, {\em Phase response analyses support a relaxation oscillator
  model of locomotor rhythm generation in caenorhabditis elegans}, bioRxiv,
  (2020), \url{https://doi.org/10.1101/2020.06.22.164939},
  \url{https://www.biorxiv.org/content/early/2020/06/24/2020.06.22.164939},
  \url{https://arxiv.org/abs/https://www.biorxiv.org/content/early/2020/06/24/2020.06.22.164939.full.pdf}.

\bibitem{johnson2020neuromechanical}
{\sc C.~L. Johnson}, {\em Neuromechanical Mechanisms of Locomotion in C.
  elegans}, PhD thesis, University of California, Davis, 2020.

\bibitem{Kopell_1986}
{\sc N.~Kopell and G.~B. Ermentrout}, {\em Symmetry and phaselocking in chains
  of weakly coupled oscillators}, Communications on Pure and Applied
  Mathematics, 39 (1986), pp.~623--660,
  \url{https://doi.org/10.1002/cpa.3160390504},
  \url{http://dx.doi.org/10.1002/cpa.3160390504}.

\bibitem{Ludwar_2005}
{\sc B.~C. Ludwar, M.~L. G{\"o}ritz, and J.~Schmidt}, {\em Intersegmental
  coordination of walking movements in stick insects}, Journal of
  Neurophysiology, 93 (2005), pp.~1255--1265,
  \url{https://doi.org/10.1152/jn.00727.2004},
  \url{http://dx.doi.org/10.1152/jn.00727.2004}.

\bibitem{Marder_1996}
{\sc E.~Marder and R.~L. Calabrese}, {\em Principles of rhythmic motor pattern
  generation}, Physiological Reviews, 76 (1996), pp.~687--717,
  \url{https://doi.org/10.1152/physrev.1996.76.3.687},
  \url{http://dx.doi.org/10.1152/physrev.1996.76.3.687}.

\bibitem{Mellem:2008aa}
{\sc J.~E. Mellem, P.~J. Brockie, D.~M. Madsen, and A.~V. Maricq}, {\em Action
  potentials contribute to neuronal signaling in c. elegans}, Nat Neurosci, 11
  (2008), pp.~865--7, \url{https://doi.org/10.1038/nn.2131}.

\bibitem{Milligan2425}
{\sc B.~Milligan, N.~Curtin, and Q.~Bone}, {\em Contractile properties of
  obliquely striated muscle from the mantle of squid (alloteuthis subulata) and
  cuttlefish (sepia officinalis)}, Journal of Experimental Biology, 200 (1997),
  pp.~2425--2436, \url{https://jeb.biologists.org/content/200/18/2425},
  \url{https://arxiv.org/abs/https://jeb.biologists.org/content/200/18/2425.full.pdf}.

\bibitem{Mullins:2011aa}
{\sc O.~J. Mullins, J.~T. Hackett, J.~T. Buchanan, and W.~O. Friesen}, {\em
  Neuronal control of swimming behavior: comparison of vertebrate and
  invertebrate model systems}, Prog Neurobiol, 93 (2011), pp.~244--69,
  \url{https://doi.org/10.1016/j.pneurobio.2010.11.001}.

\bibitem{netoff_etal}
{\sc T.~Netoff, M.~Schwemmer, and T.~Lewis}, {\em Phase Response Curves in
  Neuroscience}, vol.~6 of Springer Series in Computational Neuroscience,
  Springer, 2012, ch.~Experimentally Estimating Phase Response Curves of
  Neurons: Theoretical and Practical Issues.

\bibitem{Niebur_1991}
{\sc E.~Niebur and P.~Erd{\"o}s}, {\em Theory of the locomotion of nematodes},
  Biophysical Journal, 60 (1991), pp.~1132--1146,
  \url{https://doi.org/10.1016/s0006-3495(91)82149-x},
  \url{http://dx.doi.org/10.1016/S0006-3495(91)82149-X}.

\bibitem{NIEBUR199351}
{\sc E.~Niebur and P.~Erdos}, {\em Theory of the locomotion of nematodes:
  Control of the somatic motor neurons by interneurons}, Mathematical
  Biosciences, 118 (1993), pp.~51 -- 82,
  \url{https://doi.org/https://doi.org/10.1016/0025-5564(93)90033-7},
  \url{http://www.sciencedirect.com/science/article/pii/0025556493900337}.

\bibitem{Nigon:2017aa}
{\sc V.~M. Nigon and M.-A. F{\'e}lix}, {\em History of research on c. elegans
  and other free-living nematodes as model organisms}, WormBook, 2017 (2017),
  pp.~1--84, \url{https://doi.org/10.1895/wormbook.1.181.1}.

\bibitem{Olivares_2018}
{\sc E.~O. Olivares, E.~J. Izquierdo, and R.~D. Beer}, {\em Potential role of a
  ventral nerve cord central pattern generator in forward and backward
  locomotion in caenorhabditis elegans}, Network Neuroscience, 2 (2018),
  pp.~323--343, \url{https://doi.org/10.1162/netn_a_00036},
  \url{http://dx.doi.org/10.1162/netn_a_00036}.

\bibitem{Pearson_2004}
{\sc K.~G. Pearson}, {\em Generating the walking gait: role of sensory
  feedback}, Brain Mechanisms for the Integration of Posture and Movement,
  (2004), pp.~123--129, \url{https://doi.org/10.1016/s0079-6123(03)43012-4},
  \url{http://dx.doi.org/10.1016/S0079-6123(03)43012-4}.

\bibitem{Schafer_2006}
{\sc W.~R. Schafer}, {\em Proprioception: A channel for body sense in the
  worm}, Current Biology, 16 (2006), pp.~R509--R511,
  \url{https://doi.org/10.1016/j.cub.2006.06.012},
  \url{http://dx.doi.org/10.1016/j.cub.2006.06.012}.

\bibitem{Schwemmer2012}
{\sc M.~A. Schwemmer and T.~J. Lewis}, {\em The Theory of Weakly Coupled
  Oscillators}, Springer New York, New York, NY, 2012, pp.~3--31,
  \url{https://doi.org/10.1007/978-1-4614-0739-3_1},
  \url{https://doi.org/10.1007/978-1-4614-0739-3_1}.

\bibitem{Sigvardt:1996aa}
{\sc K.~A. Sigvardt and T.~L. Williams}, {\em Effects of local oscillator
  frequency on intersegmental coordination in the lamprey locomotor cpg: theory
  and experiment}, J Neurophysiol, 76 (1996), pp.~4094--103,
  \url{https://doi.org/10.1152/jn.1996.76.6.4094}.

\bibitem{Skinner:1997aa}
{\sc F.~K. Skinner, N.~Kopell, and B.~Mulloney}, {\em How does the crayfish
  swimmeret system work? insights from nearest-neighbor coupled oscillator
  models}, J Comput Neurosci, 4 (1997), pp.~151--60.

\bibitem{Skinner:1998aa}
{\sc F.~K. Skinner and B.~Mulloney}, {\em Intersegmental coordination in
  invertebrates and vertebrates}, Curr Opin Neurobiol, 8 (1998), pp.~725--32.

\bibitem{Sznitman2010}
{\sc J.~Sznitman, X.~Shen, P.~K. Purohit, and P.~E. Arratia}, {\em The effects
  of fluid viscosity on the kinematics and material properties of c. elegans
  swimming at low reynolds number}, Experimental Mechanics, 50 (2010),
  pp.~1303--1311, \url{https://doi.org/10.1007/s11340-010-9339-1},
  \url{https://doi.org/10.1007/s11340-010-9339-1}.

\bibitem{Thomases:2014aa}
{\sc B.~Thomases and R.~D. Guy}, {\em Mechanisms of elastic enhancement and
  hindrance for finite-length undulatory swimmers in viscoelastic fluids},
  Physical Review Letters, 113 (2014),
  \url{https://doi.org/10.1103/PhysRevLett.113.098102}.

\bibitem{Tytell_2010}
{\sc E.~D. Tytell, C.-Y. Hsu, T.~L. Williams, A.~H. Cohen, and L.~J. Fauci},
  {\em Interactions between internal forces, body stiffness, and fluid
  environment in a neuromechanical model of lamprey swimming}, Proceedings of
  the National Academy of Sciences, 107 (2010), pp.~19832--19837,
  \url{https://doi.org/10.1073/pnas.1011564107},
  \url{http://dx.doi.org/10.1073/pnas.1011564107}.

\bibitem{Wen:2012aa}
{\sc Q.~Wen, M.~D. Po, E.~Hulme, S.~Chen, X.~Liu, S.~W. Kwok, M.~Gershow, A.~M.
  Leifer, V.~Butler, C.~Fang-Yen, T.~Kawano, W.~R. Schafer, G.~Whitesides,
  M.~Wyart, D.~B. Chklovskii, M.~Zhen, and A.~D.~T. Samuel}, {\em
  Proprioceptive coupling within motor neurons drives c. elegans forward
  locomotion}, Neuron, 76 (2012), pp.~750--61,
  \url{https://doi.org/10.1016/j.neuron.2012.08.039}.

\bibitem{White:1986aa}
{\sc J.~G. White, E.~Southgate, J.~N. Thomson, and S.~Brenner}, {\em The
  structure of the nervous system of the nematode caenorhabditis elegans},
  Philos Trans R Soc Lond B Biol Sci, 314 (1986), pp.~1--340.

\bibitem{Wiggins:1998aa}
{\sc C.~H. Wiggins and R.~E. Goldstein}, {\em Flexive and propulsive dynamics
  of elastica at low reynolds number}, Physical Review Letters, 80 (1998),
  pp.~3879--3882, \url{https://doi.org/10.1103/PhysRevLett.80.3879}.

\bibitem{Zhang_2014}
{\sc C.~Zhang, R.~D. Guy, B.~Mulloney, Q.~Zhang, and T.~J. Lewis}, {\em Neural
  mechanism of optimal limb coordination in crustacean swimming}, Proceedings
  of the National Academy of Sciences, 111 (2014), pp.~13840--13845,
  \url{https://doi.org/10.1073/pnas.1323208111},
  \url{http://dx.doi.org/10.1073/pnas.1323208111}.

\bibitem{Zhang_2013}
{\sc C.~Zhang and T.~J. Lewis}, {\em Phase response properties of half-center
  oscillators}, Journal of Computational Neuroscience, 35 (2013), pp.~55--74,
  \url{https://doi.org/10.1007/s10827-013-0440-1},
  \url{http://dx.doi.org/10.1007/s10827-013-0440-1}.

\bibitem{Zhen:2015aa}
{\sc M.~Zhen and A.~D.~T. Samuel}, {\em C. elegans locomotion: small circuits,
  complex functions}, Curr Opin Neurobiol, 33 (2015), pp.~117--26,
  \url{https://doi.org/10.1016/j.conb.2015.03.009}.

\end{thebibliography}
\bibliographystyle{siamplain.bst}

\end{document}